\documentclass[final,1p,times,number]{elsarticle}

\usepackage{amsmath}
\usepackage{amssymb}
\usepackage{amsfonts}
\usepackage{graphicx}
\usepackage{epstopdf}
\usepackage{color}
\usepackage{bm}
\usepackage{multirow}
\usepackage{natbib}
\usepackage{hyperref}

\usepackage{ifpdf}
\ifpdf%
\usepackage{pdflscape}
\else
\usepackage{lscape}
\fi

\def\be{\begin{equation}}
\def\ee{\end{equation}}
\def\bea{\begin{eqnarray}}
\def\eea{\end{eqnarray}}

\newcommand{\diag}{\mathop{\mathrm{diag}}}

\journal{Annals of Physics}
\biboptions{sort&compress}

\begin{document}

\begin{frontmatter}
\title{Transport in a disordered $\nu=2/3$ fractional quantum Hall junction}

\author[IVP1,IVP2]{I.\ V.\ Protopopov}
\cortext[cor]{Corresponding author. } 
\ead{ivan.protopopov@unige.ch}
\author[YG]{Yuval\ Gefen}
\author[ADM1,ADM2,YG,ADM4]{A.\ D.\ Mirlin}

\address[IVP1]{
Department of Theoretical Physics, University of Geneva,
24 quai Ernest-Ansermet, 1211 Geneva, Switzerland
}

\address[IVP2]{
 L.\ D.\ Landau Institute for Theoretical Physics RAS,
 119334 Moscow, Russia
}

\address[YG]{Dept. of Condensed Matter Physics, Weizmann Institute of
  Science, Rehovot 76100, Israel}

\address[ADM1]{
 Institut f\"ur Nanotechnologie, Karlsruhe Institute of Technology,
 76021 Karlsruhe, Germany
}

\address[ADM2]{
 Institut f\"ur Theorie der Kondensierten Materie and DFG Center for Functional
Nanostructures,
 Karlsruhe Institute of Technology, 76128 Karlsruhe, Germany
}

\address[ADM4]{
 Petersburg Nuclear Physics Institute,  188300 St.~Petersburg, Russia.
}

\begin{abstract}

Electric and thermal transport properties of a $\nu=2/3$ fractional quantum Hall junction are analyzed. We investigate the evolution of  the electric and thermal two-terminal conductances, $G$ and  $G^Q$, with  system size $L$ and temperature $T$. This is done both for the case of strong interaction between the 1 and 1/ 3 modes (when the low-temperature physics of the interacting segment of the device is controlled by the vicinity of the strong-disorder Kane-Fisher-Polchinski fixed point) and for relatively weak interaction, for which the disorder is irrelevant at $T=0$ in the renormalization-group sense.  The transport properties in both cases are similar in several respects. In particular, $G(L)$ is close to 4/3 (in units of $e^2/h$) and $G^Q$ to 2 (in units of $\pi T / 6 \hbar$) for small $L$, independently of the interaction strength. For large $L$ the system is in an incoherent regime, with $G$ given by 2/3 and $G^Q$ showing the Ohmic scaling, $G^Q\propto 1/L$, again for any interaction strength. The hallmark of the strong-disorder fixed point is the emergence of an intermediate range of $L$, in which the electric  conductance shows strong mesoscopic fluctuations and the thermal conductance is $G^Q=1$. The analysis is extended also to a device with floating 1/3 mode, as studied in a recent experiment [A. Grivnin et al, Phys. Rev. Lett. {\bf 113}, 266803 (2014)]. 

\end{abstract}


\end{frontmatter}

\section{Introduction}
\label{s1}

It is well understood that remarkable properties of 2D electron gas in the integer quantum Hall effect \cite{vonklitzing80}
are related to Anderson localization of electrons in the bulk of the system when the Fermi energy is away from the center of a Landau level. The only delocalized excitations at the Fermi energy are then edge states \cite{laughlin81}. Related physics occurs in the fractional quantum Hall effect (FQHE) \cite{tsui82}, albeit in a more complex setting. In this case, bulk excitations are fractionally charged quasiparticles \cite{laughlin83,haldane83,halperin84} that again become localized by disorder. As a result, the only low energy excitations that can contribute to transport are edge modes. It was shown by Wen \cite{wen} that these edge excitations can be described as one-dimensional bosonic systems in the framework of the chiral Luttinger liquid theory. Transport properties of various setups can then be evaluated with the appropriate generalization of the Landauer-B\"uttiker formalism \cite{landauer,buettiker}. This approach has allowed one to explain the conductance quantization for Laughlin fractional fillings $\nu = 1/(2m+1)$ and has also facilitated theoretical exploration of  transport in samples with constrictions, where edge states of opposite chirality are connected by point tunneling \cite{kane-fisher}. 

Properties of FQHE edges become more complicated for fractions that are different from Laughlin ones, in which case the edge hosts several branches of excitations. A prominent example of such a fraction is the hole-conjugate  $\nu=2/3$ state.  
It was shown  \cite{wen,macdonald90} on the basis of the Haldane-Halperin hierarchy of FQHE states \cite{haldane83,halperin84} (and supported by numerical simulations \cite{johnson91})
that the $\nu=2/3$ FQHE edge consists of two counterpropagating modes -- a ``downstream'' mode with $\delta\nu=1$, and an ``upstream''  one with $\delta\nu=-1/3$. More accurately, this applies when the confining potential is sufficiently steep; otherwise additional pair(s) of counterpropagating 1/3 modes may emerge \cite{chamon94,meir94,wang13}. We will restrict ourselves in the present paper to the ``minimal'' model of the 2/3 edge, with two counterpropagating modes (1 and 1/3). 

If the electrostatic interaction between the counterpropagating 1 and $-1/3$ modes is included, the actual eigenmodes of the Hamiltonian will be different. In general, there will be two counterpropagating eigenmodes as obtained by a Bogoliubov-type transformation. The effective fractional charge that may be associated with these modes will be non-universal and will depend on the interaction strength and the confining potential.  It was, however, shown in a seminal paper by 
Kane, Fisher and Polchinski \cite{kfp} that this picture is fundamentally modified if not only the interaction but also  tunneling between the modes 1 and $-1/3$ is taken into account. In a realistic sample, such tunneling will be induced by impurities, in full analogy with disorder-induced backscattering in more conventional realizations of quantum wires. It is thus assumed that the amplitude of tunneling is a random function of the coordinate along the wire; for simplicity it can be approximated by a Gaussian random variable with white-noise spatial correlations. In that situation, in analogy with the case of disordered Luttinger liquid with 1 and $-1$ original modes \cite{giamarchi88,giamarchi-book}, the action of the system shows a renormalization-group flow in the disorder-interaction plane \cite{kfp}. A remarkable property of this flow that was discovered by Kane, Fisher, and Polchinski is that, for a sufficiently strong bare interaction between the 1 and $-1/3$ modes, the flow terminates at the line corresponding to a downstream 2/3 mode and an upstream neutral mode. This determines the limiting behavior of the system at low temperatures. 
 It was further argued in Ref.~\cite{kfp} that this may explain the experimentally observed value $G = (2/3) e^2/h$ of the two-terminal conductance of $\nu = 2/3$ quantum-Hall devices. The original analysis of Ref.~\cite{kfp}
has been generalized to a variety of other FQHE filling fractions  \cite{kane-fisher95,moore98,ferraro10}.

A  extensive effort to observe FQHE edge neutral modes has taken almost two decades but finally turned up successful
\cite{bid09,bid10,yacoby12,gurman12,gross12,inoue14}. More specifically, an ``upstream'' energy flow---not accompanied by an upstream charge flow---has been experimentally observed at $\nu = 2/3$  as well as for several further FQHE filling fractions. 
These observations of the neutral modes demonstrated two facets of the dynamics involved: (i) the energy they carry causes heating, which may be detected directly \cite{yacoby12} or through a thermo-electric effect \cite{gurman12}; (ii) the annihilation of neutral excitations may be accompanied with the stochastic generation of quasi-particle/quasi-hole pairs \cite{bid10,gross12}, leading to shot-noise in the absence of average current.  It has been recently proposed \cite{park15} that the emergence of neutral modes in FQHE states may be responsible for the absence of experimental observation of anyonic interference.

Thus, experiments provide support to the emergence of the upstream energy (but not charge) flow at $\nu=2/3$ (as well as at several other fractions), consistent  with the theory  by Kane-Fisher-Polchinski. On the other hand, many important questions remain to be answered, in particular: Which of these observations are specific to the Kane-Fisher-Polchinski fixed point? In what respects is the system near this fixed point essentially different from that with weak interaction and what aspects of both scenarios are similar? What are the quantitative dependences of the physical observables on the parameters of the system (temperature, system size, interaction strength, etc.) and how do they compare to experiment? While some aspects of these questions have been addressed in  previous publication, the complete picture remains to be understood.

The goal of the present work is to analyze  electric and thermal transport properties  of a $\nu=2/3$ FQHE device. Specifically, we will study the dependence of the corresponding two-terminal conductances on temperature $T$, system size $L$, length of the  $\nu=1$ and $\nu=2/3$ ``leads'' $L_0$,  and the intermode  interaction strength. 
While some results for the zero-temperature electric conductance $G$ were presented in Ref.~\cite{kfp}, they were obtained by using the Kubo formula. 
It is known, however, that this approach yields an incorrect result for the dc conductance when one considers a finite-length Luttinger-liquid wire connected to two reservoirs \cite{Maslov,Safi,Ponomarenko}.  Specifically, a naive application of the Kubo formula yields  $G=K e^2/h$, where $K$ is the Luttinger-liquid parameter, while the correct value of the dc conductance is $G = e^2/h$. 
In this paper, we first explore a model where each  $\nu=2/3$ FQHE edge  consists of a ``wire''  attached to ``leads'' (noniteracting 1 and $-1/3$ modes). 
We consider both cases of strong and weak interaction between $1$ and $1/3$ modes. In the first case disorder is relevant and drives the interacting part of the device into the Kane-Fisher-Polchinskii fixed point, while in the second case the disorder is irrelevant and the eigenmodes in the interacting part of the system are non-universal. Despite this difference, the dependence of conductance on temperature in both cases bears considerable similarity. Specifically, at lowest temperatures the conductance is equal to $4/3$  (in units of $e^2/h$) both for weak and strong interaction. Furthermore, at high temperatures the system always enters an incoherent regime with a universal conductance $2/3$. The hallmark of the Kane-Fisher-Polchinskii fixed point turns out to be the regime of 
strong mesoscopic fluctuations in the intermediate temperature range, where the conductance  depends on configuration of disorder and can take an arbitrary value between 1/3 and 4/3. We also extend this analysis to a  setup where only the $1$ modes are coupled to external reservoirs. 
Furthermore, we  investigate the heat transport and present an explicit  comparison of the length and temperature dependence of the thermal conductance in the weak and strong interaction regimes. 

Various theoretical aspects of the transport through FQHE edges states in the presence of disorder were considered in a number of papers
\cite{Wen94,Chklovskii98,Chamon97,Zuelicke99,Naud00,Ponomarenko01,Zuelicke03,Sen08,Rosenow10}. Below, we will point out relations between approaches and results of the present paper and those of the above  works.
On the experimental side, our paper was largely motivated by the recent experimental work \cite{Grivnin14} where the conductance was studied for different lengths of the $\nu=2/3$ edge. It was found that the conductance has a very different behavior for short and long systems. At the end of the paper we  discuss  the connection of our results and existing experimental observations, in particular, those of  Ref.~\cite{Grivnin14} and of a very recent preprint on thermal transport \cite{Banerjee16}.

The structure of the paper is as follows. In Sec. \ref{s2} we formulate the model and discuss the general properties of the conductance matrix. In Sec. \ref{s3ALL}
we study the zero-temperature electric conductance for an exactly solvable model with interaction fine-tuned to the Kane-Fisher-Polchinski fixed point.  
Section \ref{s6} is devoted to a comprehensive analysis of temperature and length dependence of the conductance for a generic interaction strength, including both  cases of weak and strong interaction. In Sec. \ref{s7} we explore the thermal transport properties of the system. Finally, we extend the analysis to the case of a device with contacts attached only to mode 1 (but not to $1/3$) in Sec. \ref{s8}. Section \ref{s9} presents the summary of our results and an overview of future research directions.  

\section{Formulation of the problem}
\label{s2}

\subsection{Preliminaries}
\label{s2.1}

We consider the $\nu=2/3$ FQHE edge consisting of a left-moving mode $1$ and a right-moving mode $1/3$.
In the absence of both tunneling and interaction between these modes, the real-time action is given by \cite{wen,kfp}:
\begin{equation}
 S_0=\int dx dt \frac{1}{4\pi}\left[\partial_x \phi_{1}\left(\partial_t\phi_1-v_{1}\partial_x\phi_1\right)+3\partial_x \phi_{1/3}\left(-\partial_t\phi_{1/3}-v_{1/3}\partial_x\phi_{1/3}\right)\right].
  \label{e2.1}
\end{equation}
The imaginary-time (Euclidean) version of this action reads
\begin{equation}
 S_0 =\int dx d\tau \frac{1}{4\pi}\left[\partial_x \phi_{1}\left(-i\partial_\tau\phi_1+v_{1}\partial_x\phi_1\right)+3\partial_x \phi_{1/3}\left(i\partial_\tau\phi_{1/3}+v_{1/3}\partial_x\phi_{1/3}\right)\right].
  \label{e2.2}
\end{equation}
Calculating from this action fields canonically conjugated to $\phi_1$ and $\phi_{1/3}$, one finds the commutation relations 
\begin{eqnarray}
  \label{e2.3}
\left[\partial_x\phi_1(x), \phi_1(x^\prime)\right]=-2\pi i\delta(x-x^\prime);\\
\left[\partial_x\phi_{1/3}(x), \phi_{1/3}(x^\prime)\right]=\frac{2\pi i}{3}\delta(x-x^\prime).
  \label{e2.4}
\end{eqnarray}
It is easy to see that $\phi_1$  and $\phi_{1/3}$ are a  left-moving and a right-moving fields, respectively. 
The physical electric charge density is given by
\begin{equation}
 \rho(x)=\frac{1}{2\pi}\left(\partial_x\phi_1+\partial_x\phi_{1/3}\right).
   \label{e2.5}
\end{equation}
The operator for the intermode tunneling induced by disorder is given by
\begin{equation}
 S_{\rm dis}=\int dx d\tau \lambda(x)\exp\left(i\phi_1+3i\phi_{1/3}\right)+{\rm h.c.}
   \label{e2.6}
\end{equation}
Here the operator $e^{i\phi_1}$ creates an electron in the (left-moving) mode 1, while $e^{i\phi_{1/3}}$ annihilate a quasiparticle of charge 1/3 in the right-moving mode 1/3. The linear combination entering the exponent in Eq.~(\ref{e2.6}) is dictated by charge conservation. 

The electrostatic interaction between the modes 1 and 1/3 is described by the term proportional to the product of corresponding densities,
\be
\label{e2.7}
S_{\rm int} = \int dx d\tau \frac{1}{4\pi} \ 2 u \partial_x\phi_1 \partial_x\phi_{1/3},
\ee 
where $u$ is the intermode interaction strength. It is convenient to characterize the latter by a dimensionless parameter \cite{kfp}
\be
\label{e2.8}
c = \frac{2u}{\sqrt{3}(v_1 + v_{1/3})},
\ee 
with $c>0$ corresponding to the repulsive interaction. The condition of stability of the system amounts to $|c|<1$.  It is further useful to define \cite{kfp} a related dimensionless parameter $\Delta$ (which determines the scaling dimension of the tunneling operator) given by
\be
\label{e2.8a}
\Delta = \frac{2 - \sqrt{3} c}{\sqrt{1-c^2}} \,.
\ee 
The case of non-interacting 1 and 1/3 modes ($c=0$) corresponds to $\Delta = 2$. 
In the presence of interaction $u$, the bare modes 1 and 1/3 cease to be the eigenmodes of the problem. The new eigenmodes, which are linear combinations of  $\phi_1$ and $\phi_{1/3}$ can be straightforwardly obtained by a Bogoliubov transformation. For a particular interaction strength 
$c=\sqrt{3}/2$ (i.e., $\Delta=1$) they take the form 
\begin{eqnarray}
 \phi_\rho=\sqrt{3/2}\left(\phi_1+\phi_{1/3}\right)\, ,    \label{e2.9} \\
  \phi_{\sigma}=\sqrt{1/2}\left(\phi_1+3\phi_{1/3}\right).
 \label{e2.10}
\end{eqnarray}
The modes  $\phi_\rho$ and $\phi_{\sigma}$ play a central role for the problem under consideration. Indeed, $\phi_\rho$ is nothing but a total charge mode, as is clear from the comparison of Eq.~(\ref{e2.9}) with Eq.~(\ref{e2.5}). On the other hand,   $\phi_{\sigma}$ is a neutral mode, i.e., it  does not carry electric charge.

In the sequel, it will be convenient to normalize the  field $\phi_{1/3}$ differently, in order to get rid of the factor $1/3$ in the commutation relation   (\ref{e2.4}). We thus define the fields
\begin{equation}
 \phi_L\equiv \phi_1 \,, \qquad \phi_R=\sqrt{3}\phi_{1/3}
  \label{e2.11}
\end{equation}
and corresponding densities
\begin{equation}
 \rho_\eta=\frac{1}{2\pi}\partial_x \phi_\eta.
   \label{e2.11a}
\end{equation}
 They satisfy the standard commutation relations 
\begin{eqnarray}
   \label{e2.12}
\left[\rho_\eta(x), \phi_\eta(x^\prime)\right]=i\eta\delta(x-x'),
\end{eqnarray}
where $\eta = +1$ for the right-moving mode $\phi_R$ and $-1$ for the left-moving mode $\phi_L$. 
With these notations, the  Hamiltonian in the absence of interaction and tunneling between the $\phi_L$ and $\phi_R$ modes takes the standard form
\begin{equation}
   \label{e2.13}
H=\frac{1}{4\pi}\int dx \left[v_R \left(\partial_x \phi_R\right)^2+v_L \left(\partial_x \phi_L\right)^2\right],\qquad v_R=v_{1/3}, \qquad v_L=v_1.
\end{equation}

The modes $\phi_\rho$ and $\phi_\sigma$, with the corresponding densities defined by Eq.~(\ref{e2.11}), satisfy the same commutation relations    (\ref{e2.12}), with $\eta = +1$ for the right-moving mode $\phi_\sigma$ and $-1$ for the left-moving mode $\phi_\rho$. The two basis sets $(\phi_R,\ \phi_L)$ and $(\phi_\sigma,\ \phi_\rho)$ are related by a U(1,1) rotation
\begin{eqnarray}
   \label{e2.14}
\left(\begin{array}{c}
\phi_\sigma\\
\phi_\rho
\end{array}\right)=\frac{1}{{\cal T}}\left(\begin{array}{cc}
1 & {\cal R}\\
{\cal R} & 1
\end{array}\right)
\left(\begin{array}{c}
\phi_R\\
\phi_L
\end{array}\right)\,,
\\[0.3cm]
\left(\begin{array}{c}
\phi_R\\
\phi_L
\end{array}\right)=\frac{1}{{\cal T}}\left(\begin{array}{cc}
1 & -{\cal R}\\
-{\cal R} & 1
\end{array}\right)
\left(\begin{array}{c}
\phi_\sigma\\
\phi_\rho
\end{array}\right)\,,
\label{Eq:sigmarhotoRL}
\end{eqnarray}
where  the coefficients ${\cal R}$ and ${\cal T}$ are given by 
\begin{equation}
   \label{e2.15}
{\cal R} = 1 / \sqrt{3}; \qquad {\cal T} = \sqrt{2/3}; \qquad {\cal T}^2+{\cal R}^2=1.
\end{equation}

\subsection{Model and conductances}
\label{s2.3}

\begin{figure}
\centering
\includegraphics[width=250pt]{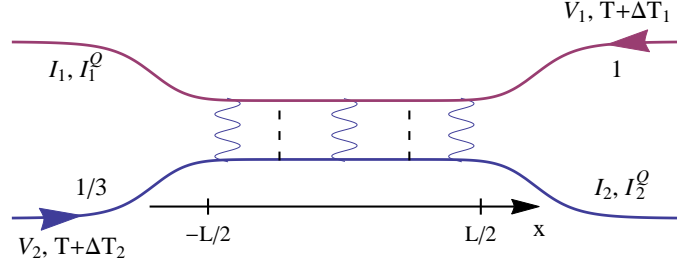}
\caption{Model of the $\nu=2/3$ FQHE edge as studied in this work.  In the middle region, $-L/2 < x < L/2$, the $(\phi_R,\ \phi_L)$ modes are coupled by both interaction and disorder while the leads are clean and non-interacting. }
\label{Fig:Setup}
\end{figure}

\begin{figure}
\centering
\includegraphics[width=250pt]{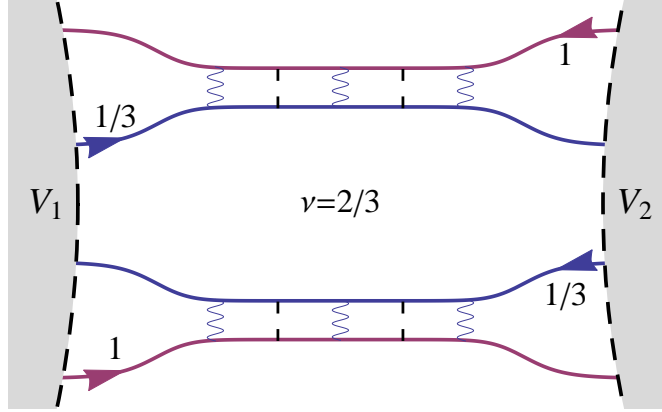}
\caption{Setup for investigation of the two-terminal conductance $G$ as studied in this work. Two opposite edges of a  $\nu=2/3$ FQHE system are shown; each of them is modelled as in Fig.~\ref{Fig:Setup}. On the left and right, the edges are coupled to metallic source and drain electrodes. Considered is the two-terminal conductance between these electrodes.}
\label{Fig:Setup-All}
\end{figure}

\begin{figure}
\centering
\includegraphics[width=250pt]{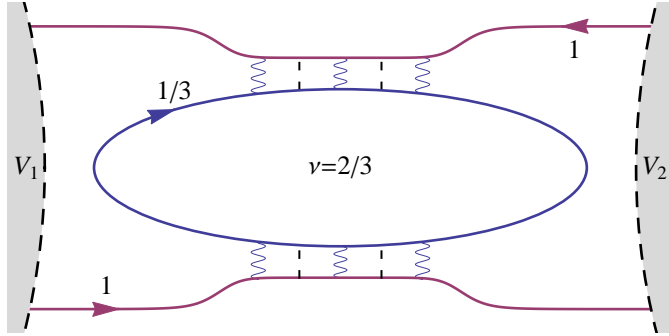}
\caption{Modified setup with floating $1/3$ mode as studied in experiment of Ref. \cite{Grivnin14}. This setup is analyzed in Sec.~\ref{s8} of the present work.}
\label{Fig:SetupFloating}
\end{figure}

We are now ready to formulate the problem to be studied in this paper. We consider portion of $\nu = 2/3$ FQHE edge of a length $L$ 
with certain interaction strength and disorder. This middle region of the setup is connected at points $ x = \pm L/2$ to ``leads'', which are modelled as non-interacting  $(\phi_R,\ \phi_L)$ edges, see Fig.~\ref{Fig:Setup}. Our goal will be to calculate the electric and thermal dc conductances  of this device. More accurately, the two-terminal conductance is defined in a FQHE system that contains two such edges, see Fig.~\ref{Fig:Setup-All}. We will also explore a related 
setup of Fig. \ref{Fig:SetupFloating} where only the mode 1 is contacted while the $1/3$ is floating.

\begin{figure}
\centering
\includegraphics[width=250pt]{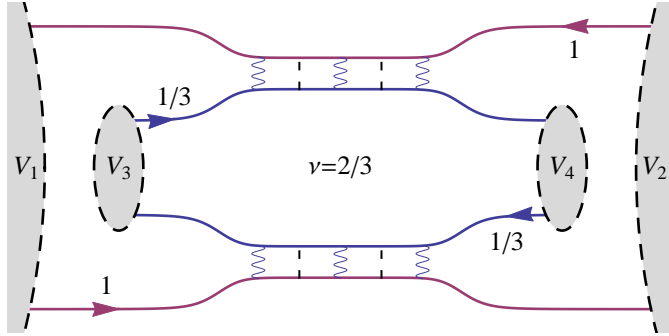}
\caption{Setup with four terminals that can in general have distinct voltages $V_1$, $V_2$, $V_3$, and $V_4$. It is characterized by a $4\times 4$ conductance matrix ${\cal G}_{ij}$, see Eqs.~(\ref{e2.22}--(\ref{e2.25}). The eigenvalues of the symmetric part of this conductance matrix are given by Eq.~(\ref{e2.26}). }
\label{Fig:Setup-Four}
\end{figure}

The single  $\nu=2/3$ FQHE edge shown in Fig.~\ref{Fig:Setup} is characterized by a $2\times 2$ conductance matrix $G_{ij}$ defined by $I_i = \sum_j (e^2/h) G_{ij} V_j$, where $V_1$ and $V_2$ are voltages characterizing the incoming 1 and 1/3 modes (i.e., $eV_1$ and $eV_2$ are electrochemical potentials of the reservoirs from which these modes emanate),  while $I_1$ and $I_2$ are currents in outgoing 1 and 1/3 modes, respectively. This matrix is subject to the following constraints:
\bea
   \label{e2.16}
G_{11} &+& G_{21} = 1 \,;\\
   \label{e2.17}
G_{12} &+& G_{22} = 1/3 \,;\\
   \label{e2.18}
G_{21} &=& G_{12} \,. 
\eea
The first two  constraints, Eqs.~(\ref{e2.16}) and  (\ref{e2.17}), {follow from} the standard condition in the theory of integer \cite{streda87,buettiker88} and fractional \cite{wen,beenakker90,Wen94} quantum-Hall edge states that the incident currents emanating from the reservoirs are completely determined by the potentials $V_1$ and $V_2$ of the corresponding reservoirs. Specifically, the current incident from the reservoir $V_1$ in the 1 mode is $(e^2/h)V_1$, while the current incident from the reservoir $V_2$ in the 1/3 mode is $(e^2/3h)V_2$.
The last condition, Eqs.~(\ref{e2.18}),  is the requirement that no current should flow between the 1 and 1/3 edge modes in equilibrium. Thus the conductance matrix of the edge is fully defined by a single parameter $G_{12}$,
\be
   \label{e2.19}
\hat G= \left( \begin{array} {cc} 1 - G_{12} \ & \ G_{12} \\ G_{12}  \ & \ 1/3- G_{12}
\end{array}
\right).
\ee
The two-terminal conductance $G$ of the  whole $\nu=2/3$ sample, as defined by Fig.~\ref{Fig:Setup-All}, is given by
\be
   \label{e2.20}
G = 4/3 - G_{12}^{(t)} - G_{12}^{(b)} \,,
\ee
where $G_{12}^{(t)}$ and $G_{12}^{(b)}$ are the off-diagonal elements of the matrices $\hat G$, Eq.~(\ref{e2.19}), characterizing the top and the bottom edges, respectively. Each of these matrix elements satisfies 
\be
\label{e2.21}
0 \le G_{12}^{(\alpha)} \le \frac{1}{2}.
\ee
Here the first inequality follows from the fact that $G_{12}^{(t)} + G_{12}^{(b)}$ is the conductance between the 1 and 1/3 modes and thus should be non-negative. (Note that $G_{12}^{(t)}$ and $G_{12}^{(b)}$ can be varied independently.) 
The second inequality is a consequence of the requirement that the matrix of conductances in a system with four different potentials applied in left and right parts of the system to  1 and 1/3 modes, see Fig. \ref{Fig:Setup-Four},  is positive semi-definite.  The condition $G \ge 0$, in view of    Eq.~(\ref{e2.20}), leads to $G_{12}^{(\alpha)} \le 2/3$ but the general requirement is more stringent, Eq.~(\ref{e2.21}), as we show now.

Let us consider a system shown in Fig.~\ref{Fig:Setup-Four}. We denote by  $V_1$, $V_2$, $V_3$, and $V_4$,  the potential of the left reservoir of the mode 1, left reservoir of the mode 1/3, right reservoir of the mode 1, and the right reservoir of the mode 1/3, respectively. Let $I_j$ with $j=1,2,3,4$ be the total currents flowing out of the respective reservoirs. We find, 
denoting $G_{12}^{(\alpha)}$ by $g_\alpha$, 
\bea
I_1 & = & V_1 - (1-g_t) V_{2} - g_t V_{3} \,,  \label{e2.22} \\
I_2 & = & V_{2} - (1-g_b) V_1 - g_b V_4 \,, \label{e2.24} \\
I_3 & = & \frac{1}{3}V_{3} - \left(\frac{1}{3}-g_b \right) V_4 - g_b V_1 \,, \label{e2.23} \\
I_4 & = & \frac{1}{3} V_4 - \left (\frac{1}{3}-g_t \right) V_{3} - g_tV_{2} \, .
\label{e2.25}
\eea
This yields a $4\times 4$ conductance matrix ${\cal G}_{ij}$. 
The dissipated energy is 
\be
\label{e2.27}
P = \sum_j V_j I_j = \sum_{ij} {\cal G}_{ij} V_iV_j \,,
\ee
and is thus determined by the symmetrized conductance matrix $({\cal G}_{ij}+{\cal G}_{ji})/2$. Diagonalizing this matrix,  we find the eigenvalues
\be
\label{e2.26}
0\,; \qquad g_b+g_t \,; \qquad \frac{1}{6}\left[8- 3 (g_b+g_t) \pm \sqrt{16 + 9 (g_b+g_t)^2 } \right]\,.
\ee
The requirement $P \ge 0$ implies that the symmetric part of the  matrix $\hat {\cal G}$ is positive semi-definite, i.e., all its eigenvalues are non-negative. Applying this condition to the eigenvalues (\ref{e2.26}), we find the constraint $0 \le g_b + g_t \le 1$. Since $g_b$ and $g_t$ can be varied independently, we get $0 \le g_\alpha \le 1/2$, which is the condition (\ref{e2.21}).   For zero temperature, the inequality  (\ref{e2.21}) can be also obtained by analyzing the energy currents \cite{Wen94,Sen08}. The proof that we have presented above is valid also for a non-zero temperature. 

Equation (\ref{e2.21}) implies that the two-terminal conductance $G$ satisfies
\be
\label{e2.22a}
\frac{1}{3} \le G \le \frac{4}{3}.
\ee
 Below we show, employing a microscopic approach (cf. Ref. \cite{Naud00}), that both limits of the inequality, Eq. (\ref{e2.22}), can be achieved in reality.

\section{Zero-temperature electric conductance of a $\Delta=1$ system}
\label{s3ALL}

In this section we study the zero-temperature conductance for the case when the interaction in the middle part of the edge is fine tuned to the value $c=\sqrt{3}/2$ corresponding to $\Delta=1$. In this situation the modes $\phi_\sigma$ and $\phi_\rho$ in the middle part of the device are completely decoupled. 
The reason for considering such a situation lies in the fact that $\Delta=1$ is an attractive infrared fixed point for a broad interval of bare interaction values. The analysis of this section will thus  serve as a starting point for the study of the dependence of the conductance on temperature, length and interaction strength in Sec. \ref{s6}.

\begin{figure}
\centering
\includegraphics[width=250pt]{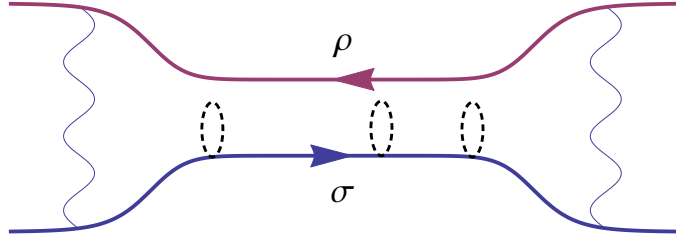}
\caption{Model of the $\nu=2/3$ FQHE edge with interaction strength $c=\sqrt{3}/2$ ($\Delta=1$): representation in terms of $(\phi_\sigma,\ \phi_\rho)$ modes. In the middle region of the edge, $-L/2 < x < L/2$, the interaction establishes eigenmodes  $(\phi_\sigma,\ \phi_\rho)$, while the non-interacting leads are characterized by $(\phi_R,\ \phi_L)$ modes. The disorder is present only in the middle part of the device and  affects the $\phi_\sigma$ (but not the  $\phi_\rho$) mode. In the leads, there is  interaction between the  $\phi_\sigma$ and $\phi_\rho$ modes. }
\label{Fig:SetupB}
\end{figure}

\subsection{Boundary between interacting and non-interacting sections}
\label{s3}

In order to evaluate the conductance of the system, we consider first the boundary between the noninteracting,  i.e., $(\phi_R,\ \phi_L)$, and the interacting, i.e., $(\phi_\sigma,\ \phi_\rho)$, parts of the system. We will consider this boundary as sharp, which means that the value of the interaction jumps abruptly at the boundary. This assumption is always justified in the considered dc limit $\omega \to 0$, since, independently of the specific profile of the interaction varying from $c=0$ to $c=\sqrt{3}/2$, this variation is sharp on the length scale set by frequency, $L_\omega \propto 1/ \omega \to \infty$.  

Assuming that the $(\phi_R,\ \phi_L)$ region is at $x<0$ and $(\phi_\sigma,\ \phi_\rho)$ at $x>0$, we obtain the following Hamiltonian of the non-uniform edge, cf. Eq. (\ref{Eq:sigmarhotoRL}):
\begin{eqnarray}
H &=& \frac{1}{4\pi}\int_{-\infty}^0 dx  \left[v_R \left(\partial_x \phi_R\right)^2+v_L \left(\partial_x \phi_L\right)^2\right]+
\frac{1}{4\pi}\int_{0}^\infty dx  \left[v_\sigma \left(\partial_x \phi_\sigma\right)^2+v_\rho \left(\partial_x \phi_\rho\right)^2\right]
\nonumber \\ &=&
\frac{1}{4\pi}\int_{-\infty}^0 dx  \left[v_R \left(\partial_x \phi_R\right)^2+v_L \left(\partial_x \phi_L\right)^2\right]+
\nonumber
\\
&&+\frac{1}{4\pi}\int_{0}^\infty dx  \left[\frac{v_\sigma}{{\cal T}^2}
 \left(\partial_x \phi_R +{\cal R} \partial_x\phi_L\right)^2+\frac{v_\rho}{{\cal T}^2} \left({\cal R} \partial_x \phi_R+\partial_x\phi_L\right)^2\right]
 \nonumber  \\ &=&
 \frac{1}{4\pi} \int dx
 \left(\begin{array}{cc}
 \partial_x\phi_R , & \partial_x\phi_L
 \end{array}\right){\cal H}(x)
  \left(\begin{array}{c}
 \partial_x\phi_R \\ \partial_x\phi_L
 \end{array}\right),
    \label{e3.1}
\end{eqnarray}
where 
\begin{equation}
    \label{e3.2}
{\cal H}(x)=
\Theta(-x)\left(\begin{array}{cc}
v_R & 0\\
0 & v_L
\end{array}
\right)+\frac{\Theta(x)}{{\cal T}^2}\left(\begin{array}{cc}
v_\sigma+{\cal R}^2 v_\rho & {\cal R}(v_\sigma+v_\rho)\\
{\cal R}(v_\sigma+v_\rho) & v_\rho+{\cal R}^2 v_\sigma
\end{array}
\right)
\end{equation}
and $\Theta(x)$ is the Heaviside theta function. The coefficients ${\cal R}$ and ${\cal T}$ are given by Eq. (\ref{e2.15}). The four velocities $v_R$, $v_L$, $v_\sigma$, and $v_\rho$ can be in general considered as independent parameters. 

\begin{figure}
\centering
\includegraphics[width=380pt]{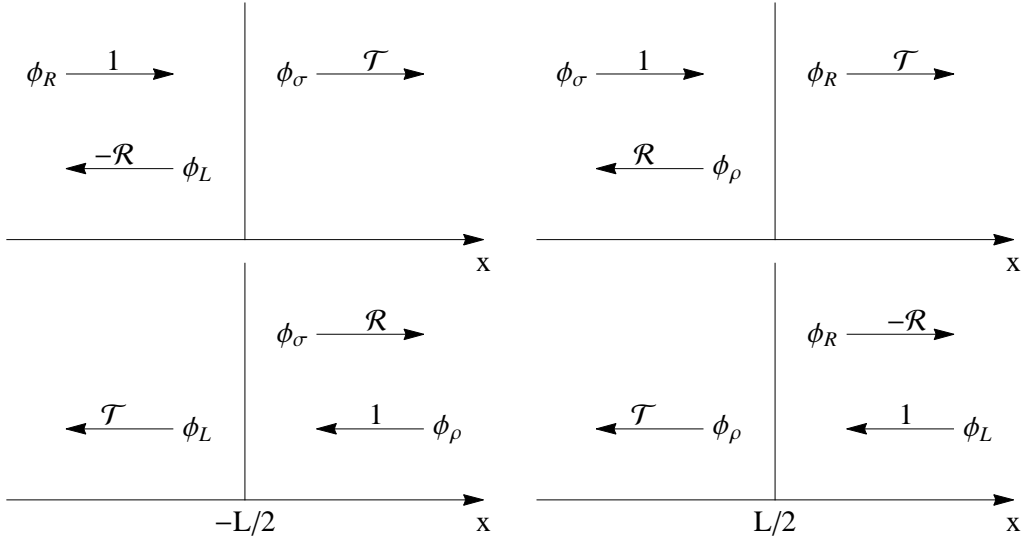}
\caption{Bosonic scattering states. {\it Left:}  Scattering states at the boundary between the left lead and the middle section ($x=-L/2$). {\it Right:}  Scattering states at the boundary between the  right lead  and the middle section ($x=L/2$). }
\label{Fig:Scattering}
\end{figure}

The Hamiltonian (\ref{e3.1}) is quadratic in the bosonic fields.  Its eigenstates are one-boson scattering states. Solving the equations of motion,
\begin{equation}
\partial_t \left(\begin{array}{c}
 \phi_R \\ \phi_L
 \end{array}\right)+\partial_x \sigma_z {\cal H}\left(\begin{array}{c}
 \phi_R \\ \phi_L
 \end{array}\right)=0,
\end{equation}
we obtain the in-scattering states corresponding to an incoming $\phi_R$ wave:
\begin{eqnarray}
    \label{e3.3}
 \left(\begin{array}{c}
 \phi_R \\ \phi_L
 \end{array}\right)=\left\{
 \begin{array}{cc} 
 \left(\begin{array}{c}
 e^{i\epsilon x/v_R} \\  -{\cal R}\, e^{-i\epsilon x/v_L}
 \end{array}\right), &\quad  x<0\,;  \\[0.4cm]
 e^{i\epsilon x/v_\sigma}
 \left(\begin{array}{c}
 1 \\  -{\cal R}
 \end{array}\right), &\quad  x>0 \,,
 \end{array} \right.
 \end{eqnarray}
 as illustrated in the top left panel of Fig.~\ref{Fig:Scattering}. Here, $\epsilon$ denotes the energy of the scattering state.  Similarly, the in-scattering states corresponding to an incoming $\Phi_\rho$ wave are (see  the top right panel in Fig.~\ref{Fig:Scattering})
\begin{eqnarray}
    \label{e3.4}
  \left(\begin{array}{c}
 \phi_R \\ \phi_L
 \end{array}\right)=\left\{
 \begin{array}{cc} 
 {\cal T}e^{-i\epsilon x/v_L}
 \left(\begin{array}{c}
  0\\ 1
 \end{array}\right), &\quad  x<0\,;  \\[0.4cm]
 \frac{1}{{\cal T} }e^{-i\epsilon x/v_\rho}
 \left(\begin{array}{c}
 -{\cal R} \\  1
 \end{array}\right)
+\frac{{\cal R}}{{\cal T} }e^{i\epsilon x/v_\sigma}
 \left(\begin{array}{c}
 1 \\  -{\cal R}
 \end{array}\right), &\quad  x>0 \,.
 \end{array} \right.
 \end{eqnarray}
The coefficients ${\cal T}$ and ${\cal R}$  defined in Eq.~(\ref{e2.15}) have ths the meaning of the transmission and reflection amplitudes at the boundary between the  $(R, L)$ and the $(\rho, \sigma)$ regions. 
The scattering states for the case where the $(R, L)$ region is to the right and the $(\rho, \sigma)$ region is to the left of the boundary are obtained in  full  analogy, cf. the  bottom panels of Fig.~\ref{Fig:Scattering}. 

In the next subsection, Sec.~\ref{s4}, we will study the effect of disorder in the middle part of the setup. After this, we will be able to study the whole device by combining the analysis of the middle region with that of two boundaries.

\subsection{Middle segment: interaction and disorder}
    \label{s4}

The Hamiltonian in the middle part of the system can be expressed as a sum of Hamiltonians for the charge mode and the neutral mode. The latter is given by
\begin{equation}
    \label{e4.1}
H_{\sigma}=\int_{-L/2}^{L/2} dx\left[ \frac{1}{4\pi}v_\sigma (\partial_x\phi_\sigma)^2+\frac1{2\pi a}\left(\lambda(x)e^{i\sqrt{2}\phi_\sigma}+{\rm h.c.}\right) \right],
\end{equation}
where the second term accounts for disorder, cf.   Eq.~(\ref{e2.6}). Here $a$ is the ultraviolet cutoff with dimension of length, and $\lambda(x)$ has dimension of energy. 
The field $\phi_\sigma(x)$ is a chiral boson field with compactification radius $1/\sqrt{2}$ and the mode expansion (see \ref{a2})
\begin{equation}
\phi_\sigma(x)=\frac{2\pi}{L}\frac{N_\sigma}{\sqrt{2}}x-\sqrt{2}\chi_\sigma +\frac{1}{i}\sum_{q> 0}\sqrt{\frac{2\pi}{L q}}\left[e^{iq x}b_q-e^{-iq x}b_q^+\right] .
\end{equation}
 
We shall now employ the bosonic language and demonstrate that the disorder can be ``gauged out''.  Here, we will do so by considering the Hamitonian     (\ref{e4.1}) on the entire $x$ axis (with disorder restricted to the region $-L/2 < x < L/2$); in the next subsection, Sec.~\ref{s5}, we will generalize this analysis to a system with leads hosting R and L eigenmodes. The theory defined by Eq.~(\ref{e4.1}) is a chiral version of the random sine-Gordon theory. In the case of non-random $\lambda(x)$ such a theory was considered also in other related contexts, including bilayer quantum Hall systems \cite{Naud00} and reconstructed quantum Hall edges in the presence of Umklapp scattering \cite{Zuelicke99}. Our analysis in this section employs  the  methods of Ref.~\cite{Naud00}.  The difference with Ref.~\cite{Naud00} is in the randomness of $\lambda(x)$ as well as in the presence of leads, see 
also Refs.~\cite{Zuelicke03, Rosenow10}.

To proceed, we introduce the operators
\begin{eqnarray}
   \label{e4.2}
 J^z (x)&=& \frac{1}{2\pi \sqrt{2}}\partial_x \phi_\sigma(x) \,,\\
 J^{\pm}(x) \equiv J^x(x) \pm i J^y(x)&=& \frac{1}{2\pi a}e^{\pm i\sqrt{2}\phi_\sigma(x)} \,.
    \label{e4.3}
\end{eqnarray}
The commutation relations for the Fourier components of the operators $J^a(x)$ (with $a = x,y,z$) read (see \ref{App:CommRelations})
\begin{equation}
   \label{e4.4}
 \left[J^{a}_q, J^b_{q^\prime}\right]=\frac{Lq}{4\pi}\delta_{q+q^\prime}\delta^{ab}+i\epsilon^{a bc}J^{c}_{q+q^\prime},
\end{equation}
which is the level-1 su(2) Kac-Moody algebra. 
In the real space representation, the commutation relations take the form 
\begin{equation}
   \label{e4.5}
 \left[J^{a}(x), J^b(x^\prime)\right]=-\frac{i}{4\pi}\delta^{ab}\delta^\prime(x-x^\prime)+i\epsilon^{abc}J^c(x)\delta(x-x^\prime).
\end{equation}
 
In order to gauge out the disorder, we will use the fact that the algebra of the operators $J^a$ is invariant under transformations
\begin{equation}
   \label{e4.6}
 \tilde{J}^{a}(x)=S^{ab}(x)J^{b}(x)+h^a_{S}(x)\,, \qquad J^{a}(x)=S^{ba}(x)\left[\tilde{J}^b(x)-h^b_{S}(x)\right]
\end{equation}
parametrized by a real orthogonal matrix $S(x)$. Here the vector $h^a_{S}(x)$ is related to $S(x)$ as follows:
\begin{equation}
   \label{e4.7}
 h^{c}_{S}(x)=\frac{1}{8\pi}\epsilon^{abc}\Omega_S^{ab}\,, \qquad
 \Omega_S\equiv S(x)\partial_x S^T\,, \qquad \Omega^{ab}_S=4\pi \epsilon^{abc}h^c_S.
\end{equation}
Indeed,  we have
\begin{eqnarray}
\left[\tilde{J}^{a}(x), \tilde{J}^{b}(x^\prime)\right]&=&S^{a\mu}(x)S^{b\nu}(x^\prime)\left[J^{\mu}(x), J^{\nu}(x^\prime)\right]\nonumber \\
&=&S^{a\mu}(x)S^{b\nu}(x^\prime)\left[-\frac{i}{4\pi}\delta^{\mu\nu}\delta^\prime(x-x^\prime)+i\epsilon^{\mu\nu \gamma}J^\gamma(x)\delta(x-x^\prime)\right]
\nonumber
\\&=&
-\frac{i}{4\pi}\left[S^{a\mu}(x)S^{b\mu}(x)\delta^\prime(x-x^\prime)-S^{a\mu}(x)\left(\partial_x S^{b\mu}(x)\right)\delta(x-x^\prime)\right]+
\nonumber\\
&&
+\, i S^{a\mu}(x)S^{b\nu}(x)S^{\lambda\gamma}(x)
S^{\lambda\gamma^\prime}(x)\epsilon^{\mu\nu\gamma}J^{\gamma^\prime}(x)\delta(x-x^\prime)
\nonumber\\
&=&
-\frac{i}{4\pi}\left[\delta^{ab}\delta^\prime(x-x^\prime)-\Omega_S^{ab}(x)\delta(x-x^\prime)\right]
+i 
S^{\lambda\gamma^\prime}(x)\epsilon^{ab\lambda}J^{\gamma^\prime}(x)\delta(x-x^\prime)
\nonumber\\
&=&
-\frac{i}{4\pi}\delta^{ab}\delta^\prime(x-x^\prime)
+i\epsilon^{abc}\delta(x-x^\prime)\left[h_S^c(x)
+i 
S^{c\gamma}(x)J^{\gamma}(x)\right]
\nonumber\\
&=&-\frac{i}{4\pi}\delta^{ab}\delta^\prime(x-x^\prime)
+i\epsilon^{abc}\delta(x-x^\prime)\tilde{J}^c(x) \,,
   \label{e4.8}
\end{eqnarray}
which proves the invariance stated above.

In terms of the operators $J^a(x)$,  the Hamiltonian (\ref{e4.1}) can be expressed as follows (see \ref{App:CommRelations}):
\begin{equation}
 \label{e4.9}
 H_\sigma=\int dx \left[\frac{2\pi v_\sigma}{3}J^2(x)+2\lambda^a(x) J^a(x)\right]\,, \qquad \lambda(x)\equiv\lambda^x(x)-i\lambda^y(x)\,,
\end{equation}
where $J^2 \equiv (J^x)^2 + (J^y)^2 + (J^z)^2$.
We now look for a transformation of the operators $J^a$ that would remove the linear-in-$J$ term in Eq.~(\ref{e4.9}) or  at least would make it as simple as possible. 
The transformation has a form    [see Eq.~(\ref{e4.6})]
\begin{equation}
 \label{e4.10}
 J^{a}(x)=S^{ab}(x)\tilde{J}^{b}(x)+h^a_{S}(x),
\end{equation}
where  $h^a_S$ is related to the matrix $S$ by Eq.~(\ref{e4.7}).
The transformed Hamiltonian reads
\begin{equation}
 \label{e4.11}
H_\sigma=\int dx \left[ \frac{2\pi v_\sigma}{3}\tilde{J}^2+2\tilde{\lambda}^a \tilde{J}^a(x)\right] \,, \qquad \tilde{\lambda}^a=\lambda^b(x)S^{ba}+ \frac{2\pi v_\sigma}{3}S^{ba}h_S^b(x).
\end{equation}
The condition for the cancelation of the linear terms in the transformed Hamiltonian yields the equation
\begin{equation}
 \label{e4.12}
 \frac{4\pi v_\sigma}{3}h^a_S=-2\lambda^a.
\end{equation}
Using the definition of $h_S$,  Eq.~(\ref{e4.7}), we find the equation for the rotation matrix $S$:
\begin{equation}
 \label{e4.13}
\frac{v_\sigma}{12}\epsilon^{abc}\Omega_S^{ab}=-\lambda^c \,,
 \end{equation}
 or, equivalently,  
 \begin{equation}
  \partial_x S=\frac{6}{v_\sigma}U_\lambda S\,, \qquad {\rm where} \qquad U^{ab}_\lambda=\epsilon^{abc}\lambda^c.
\label{SEq}
 \end{equation}
 Since $U_\lambda$ is a real antisymmetric matrix, this equation is consistent with the orthogonality of the matrix $S$.  The explicit solution can be written in terms of the path-ordered exponent:
\begin{equation}
S_\lambda (x) =T_x\exp\left[\frac{6}{v_\sigma}\int_{-\infty}^x dx' U_\lambda(x') \right]S(-\infty) \,,
\label{Eq:SLambda}
\end{equation}
with a constant matrix $S(-\infty)$. 

\begin{figure}
\centering
\includegraphics[width=250pt]{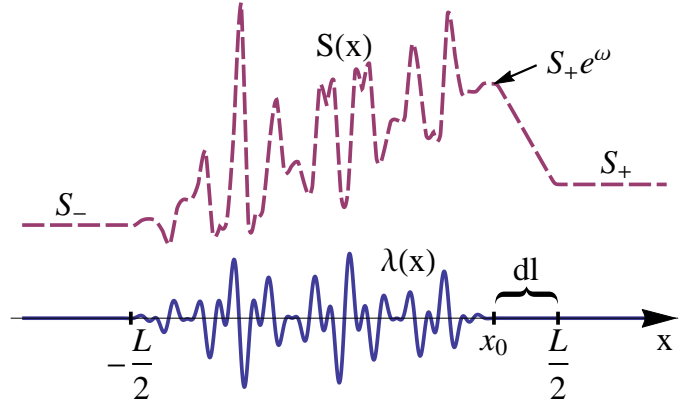}
\caption{Schematic view of the disorder $\lambda(x)$ and of the gauge transformation $S(x)$, Eq. (\ref{Eq:SFinal}), used to collect the effect of disorder into a single point $x_0$. Outside the interval $(-L/2,\, L/2)$ the transformation $S(x)$ is constant and given by rotations $S_-$ and $S_+$ around the $z$ axis.  }
\label{Fig:SigmaGauge}
\end{figure}

As it will be clear in Sec. \ref{s5}, in a system with leads, where the neutral and charge modes interact, it is important to restrict the possible gauge transformations $S(x)$ to those not modifying the operator $\partial_x\phi_\sigma\sim J^z$ outside the middle part of the system, $-L/2<x<L/2$.  Equivalently, 
$S(x)$ should rotate around $z$ axis for all $|x|>L/2$. To construct such a gauge transformation, let us assume that  the disorder $\lambda(x)$ is non-zero only in the region $-L/2 < x < x_0=L/2-dl$, where $dl$ is a small interval width that will be later sent to zero,   see Fig. \ref{Fig:SigmaGauge}. 
For a given realisation of disorder, we  compute now the path-ordered exponent
\begin{equation}
 \label{e4.14}
  T_x\exp\left[\frac{6}{v_\sigma}\int_{-\infty}^{L/2} dx\, U_\lambda(x) \right]\equiv T_x\exp\left[\frac{6}{v_\sigma}\int_{-\infty}^{+\infty} dx\, U_\lambda(x) \right]
\end{equation}
and establish its decomposition in terms of Euler angles
\begin{equation}
 \label{e4.15}
 T_x\exp\left[\frac{6}{v_\sigma}\int_{-\infty}^{+\infty} dx\, U_\lambda(x) \right]=R_z(\psi)R_x(\theta)R^{-1}_z(\phi)  \equiv S_+e^{\hat{\omega}}
S_-^{-1} \,.
\end{equation}
As the last equality indicates, we  denote the two rotation around the $z$ axis by $S_+$ and $S_-$ and the $x$-axis rotation by~$e^{\hat{\omega}}$,
\begin{equation}
 \label{e4.16}
R_x(\theta)=e^{\hat{\omega}}\,, \qquad {\rm with} \qquad  \hat{\omega}=\theta\cdot 
\left(\begin{array}{ccc}
0 & 0 & 0\\
0& 0 & 1\\
0& -1 &0
\end{array}\right)
\,, \qquad 0 < \theta < \pi \,.
\end{equation}
 We chose $S(x)$ according to
 \begin{equation}
S(x)=\left\{\begin{array}{ll}
S_\lambda(x), & \ \  x<x_0 \,;\\
S_+\exp\left(\frac{\hat{\omega}(L/2-x)}{dl} \right), & \ \  x_0<x<L/2\,;\\
S_+, & \ \  x>L/2 \,,
\end{array}\right.
\label{Eq:SFinal}
\end{equation}
where $S_\lambda(x)$ is given by Eq. (\ref{Eq:SLambda}) with $S(-\infty)=S_-$.  The corresponding $\Omega_S$ and $h_S$  are given by
\begin{equation}
 \label{e4.17}
\Omega_S(x)=S\partial_x S^T=\left\{\begin{array}{ll}
-\frac{6}{v_\sigma}U_\lambda, & \ \  x<x_0 \,;\\[0.2cm]
\frac{1}{l}S_+\hat{\omega} S_+^T, & \ \ x_0<x<L/2\,;\\[0.2cm]
0, & \ \ x>L/2 \,,
\end{array}\right.
\end{equation}
and
 \begin{equation}
  \label{e4.18}
h_S^a=\left\{\begin{array}{cc}
-\frac{3}{2\pi v_\sigma}\lambda^a, & \ \ x<x_0\,;\\[0.2cm]
\frac{1}{8\pi l}\epsilon^{rpq}S_+^{ar}\omega^{pq}\equiv \frac{1}{8\pi l} \epsilon^{rpq}\left[S_+e^{\hat{\omega}}\right]^{ar}\omega^{pq}, & \ \ x_0<x<L/2\,;\\[0.2cm]
0, & \ \ x>L/2 \,.
\end{array}\right.
\end{equation}
 Using Eq.~(\ref{e4.11}), we find the vector function $\tilde{\lambda}^a(x)$ representing the disorder in the transformed Hamiltonian: 
\begin{equation}
 \label{e4.19}
\tilde{\lambda}^a=
\left\{\begin{array}{cc}
0, & x<x_0\,;\\[0.2cm]
\frac{v_\sigma}{12 l}\epsilon^{abc}\omega^{bc}, & x_0<x<L/2\,;\\[0.2cm]
0, & x>L/2.
\end{array}\right.
\end{equation}
Substituting here $\omega$ from Eq.~(\ref{e4.16}), we thus obtain the Hamiltonian expressed in terms of transformed operators,
\begin{equation}
 \label{e4.20}
 H_\sigma=\int dx \: \frac{2\pi v_\sigma}{3} \tilde{J}^2+\frac{v_\sigma\theta}{3dl}\int_{x_0}^{L/2}dx\, \tilde{J}^x(x)\,,
\end{equation}
or, in the limit $dl\rightarrow 0$,
\begin{equation}
 \label{e4.21}
 H_\sigma=\int dx \: \frac{2\pi v_\sigma}{3}\tilde{J}^2+\frac{v_\sigma\theta}{3}\tilde{J}^x(x_0).
\end{equation}

Equation (\ref{e4.21}) represents the main result of this subsection. It shows that the effect of disorder can be accumulated in a single point $x_0$. It is easy to show by generalizing the above derivation that the role of $x_0$ can be played by any point, also by one inside the disordered region. In a certain sense, this is similar to the Aharonov-Bohm effect, where, depending on a choice of a gauge,  the effect of magnetic flux  amounts to a phase jump that can be shifted to an arbitrary point on the contour.  We note that  sophisticated choice of the gauge transformation $S(x)$, Eq. ({\ref{Eq:SFinal}}), allows us to prevent the appearance of $J^z(0)$ and $J^y(0)$ terms in the transformed Hamiltonian.

Let us emphasize that the transformation we have used satisfies the boundary conditions
\begin{equation}
S(x< -L/2)=S_-\,, \qquad S(x> L/2)=S_+ \,,
 \label{e4.21a}
\end{equation}
with two non-trivial (but constant) matrices $S_+$ and $S_-$. 
Thus, generally this transformation modifies the operators even outside the disordered region. Moreover, for $S_-\neq S_+$ this modification is different  at $x < -L/2$ and $x > L/2$.  However, due to our requirement that $S_+$ and $S_-$ are rotations around the $z$ axis, the operator $\partial_x \phi_\sigma$ is not affected by these rotations.  This will be of key importance for the analysis of the whole system in the next subsection, Sec.~\ref{s5}. 

\subsection{Adding the leads: analysis of the entire system}
\label{s5}

\subsubsection{Transformation of the Hamitonian: Exploiting the $SU(2)$ symmetry }
\label{s5.1}

We are now ready to study the whole system consisting of a disordered middle section whose eigenmodes are  neutral and charge (2/3) and  of clean leads with eigenmodes R and L. 
The Hamiltonian of the total system has the form 
\begin{equation}
\label{e5.1}
 H=\int_{-L/2}^{+L/2}\frac{1}{4\pi}\left[v_{\sigma}(\partial_x \phi_\sigma)^2+v_\rho(\partial_x \phi_\rho)^2\right]+\int_{|x|>L/2}
 \frac{1}{4\pi}\left[v_R(\partial_x \phi_R)^2+v_L(\partial_x \phi_L)^2\right]+\int_{-L/2}^{L/2}{\rm disorder}.
 \end{equation}
Splitting off the Hamiltonian of an infinite $\sigma$--$\rho$ system, we rewrite $H$ as follows:
\begin{eqnarray}
\label{e5.2}
 H &=&
 \int_{-\infty}^{+\infty}\frac{1}{4\pi}\left[v_{\sigma}(\partial_x \phi_\sigma)^2+v_\rho(\partial_x \phi_\rho)^2\right]+\int_{-L/2}^{L/2}{\rm disorder}+\nonumber\\&&+\int_{|x|>L/2}
 \frac{1}{4\pi}\left[v_R(\partial_x \phi_R)^2+v_L(\partial_x \phi_L)^2-v_{\sigma}(\partial_x \phi_\sigma)^2-v_\rho(\partial_x \phi_\rho)^2\right]\,,
  \end{eqnarray}
or, expressing everything through the $\phi_\rho$ and $\phi_\sigma$ fields (cf. Sec. \ref{s3}),
 \begin{eqnarray}
 \label{e5.3}
 H&=& \int_{-\infty}^{+\infty}\frac{1}{4\pi}\left[v_{\sigma}(\partial_x \phi_\sigma)^2+v_\rho(\partial_x \phi_\rho)^2\right]+\int_{-L/2}^{L/2}{\rm disorder}\, +\nonumber\\&+&\int_{|x|>L/2}
 \frac{1}{4\pi}\left[\frac{v_R}{{\cal T}^2}(\partial_x \phi_\sigma-{\cal R}\partial_x\phi_\rho)^2+\frac{v_L}{{\cal T}^2}(\partial_x \phi_\rho-{\cal R}\partial_x \phi_\sigma)^2-v_{\sigma}(\partial_x \phi_\sigma)^2-v_\rho(\partial_x \phi_\rho)^2\right].\qquad
 \end{eqnarray}
 Now we apply to this Hamiltonian the transformation $S(x)$ (which acts on the $\phi_\sigma$ field only) defined in Sec.~\ref{s4}. 
The key point here is that this transformation acts trivially on the $\partial_x\phi_\sigma$ outside the interval $(-L/2, \; L/2)$ and thus does not affect the last line in Eq.~(\ref{e5.3}). Therefore, as  in Sec.~\ref{s4}, we can gauge out the disorder---collecting its effect onto a single point $x_0=0$---without affecting the form of the Hamiltonian in the leads. Choosing for definiteness $x_0=0$ and omitting the tilde in the symbols for transformed fields, we get the Hamiltonian
\begin{equation}
 \label{e5.4}
H=\int_{-L/2}^{+L/2}\frac{1}{4\pi}\left[(v_{\sigma}(\partial_x \phi_\sigma)^2+v_\rho(\partial_x \phi_\rho)^2\right]+\int_{|x|>L/2}
 \frac{1}{4\pi}\left[v_R(\partial_x \phi_R)^2+v_L(\partial_x \phi_L)^2\right]+\frac{ v_\sigma\theta }{3} J^x(0),
\end{equation}
or, equivalently, with   $h=\theta/6\pi $, 
\begin{equation}
 \label{e5.5}
H=\int_{-L/2}^{+L/2}\frac{1}{4\pi}\left[v_{\sigma}(\partial_x \phi_\sigma)^2+v_\rho(\partial_x \phi_\rho)^2\right]+\int_{|x|>L/2}
 \frac{1}{4\pi}\left[v_R(\partial_x \phi_R)^2+v_L(\partial_x \phi_L)^2\right]+\frac {v_\sigma h}{a} \cos \sqrt{2}\phi_\sigma(0).
\end{equation}

\subsubsection{Renormalization group near $\theta=0$}
\label{s5.2}

We are now going to develop a renormalization-group (RG) treatment of the last (disorder-induced) term  in Eq.~(\ref{e5.4}) [or, equivalently, Eq.~(\ref{e5.5})]. This RG will describe renormalization of this disorder-induced term by interaction between the $\phi_\rho$ and $\phi_\sigma$ modes in the leads. From symmetry considerations, it is clear that the points $\theta=0$ and $\theta=\pi$ are fixed points of the RG. We first consider the
problem near the clean fixed point $\theta=0$. We begin by integrating out  all the degrees of freedom apart from $\phi_\sigma(x=0, t)\equiv \phi(t)$. 
This is done in a standard way by introducing a Lagrange multiplier $\xi(t)$, after which the intergral over $\phi(t)$ becomes Gaussian: 
\begin{eqnarray}
\nonumber
e^{-S[\phi]} &=& \int {\cal D}\xi(t){\cal D}\phi_R{\cal D}\phi_L e^{-S[\phi_R, \phi_L]+i\int dt \xi(t)\left(\phi(t)-\phi_\sigma(0, t)\right) } \\
&=& \int{\cal D} \xi(t) e^{-\frac{1}{2}\int dt dt^\prime D_{\sigma\sigma}(0, t-t^\prime) \xi(t)\xi(t^\prime) +i\int dt \xi(t)\phi(t)+\frac{v_\sigma h}{a}\int dt \cos{\sqrt{2}\phi}} \,,
 \label{e5.6}
\end{eqnarray}
where $D_{\sigma\sigma}$ is the Green function of the field $\phi_\sigma$,
\begin{equation}
 \label{e5.7}
D_{\sigma\sigma}(0, t-t^\prime)=\langle \phi_\sigma(0, t)\phi_{\sigma}(0, t^\prime)\rangle.
\end{equation}
Integrating now out the $\xi$-field, we get the action for the field  $\phi(t)$ at the position of the ``impurity'' ($x=0$):
\begin{equation}
 \label{e5.8}
S=\frac{1}{2}\int \frac{d\omega}{2\pi} \frac{|\omega|}{\pi K(\omega)}|\phi(\omega)|^2+ \frac{v_{\sigma}h}{a}\int d\tau \cos \sqrt{2}\phi \,,
\end{equation}
where
\begin{equation}
 \label{e5.9}
\frac{\pi K(\omega)}{|\omega|}=D_{\sigma\sigma}(x-x^\prime=0,\omega) \,.
\end{equation}
Evaluating the propagator entering  Eq.~(\ref{e5.9}), we obtain
\begin{equation}
K(\omega)=\frac{2\omega}{\pi}(1-{\cal R}^4)\int_0^\infty d\Omega \frac{1}{\omega^2+\Omega^2}\frac{1}{1+{\cal R}^4-2{\cal R}^2\cos2\Omega\tau} \,,
\label{Eq:K}
\end{equation}
where $\tau=L/\bar v$ is the mean flight time through the middle part of the system, with $\bar v^{-1} = (v_\rho^{-1}+v_\sigma^{-1})/2$.
Equation (\ref{Eq:K}) stems from the expansion of the field $\phi_\sigma(0, t)$ in terms of the incoming right and left fields   (see Fig. \ref{Fig:Scattering}),
\begin{equation}
 \label{e5.10}
\phi_\sigma(0, t)={\cal T}\sum_{n=0}^{\infty}{\cal R}^{2n}\Phi_R(0,   t-2n \tau) +{\cal T}\sum_{n=0}^{\infty}{\cal R}^{1+2n}\Phi_L( 0, t -2 n \tau ),
\end{equation}
and the Green functions of the incoming fields,
\begin{equation}
 \label{e5.11}
D_{RR}(\omega, q)=\frac{2\pi}{q}\frac{1}{i\omega-v_R q}\,, \qquad D_{LL}(\omega, q)=-\frac{2\pi}{q}\frac{1}{i\omega+v_L q}.
\end{equation}
Indeed, it follows from  Eq.~(\ref{e5.10}) that
\begin{multline}
 \label{e5.12}
D_{\sigma\sigma}(0, t)={\cal T}^2\sum_{n, m=0}^{\infty}{\cal R}^{2n+2m}D_{RR}(0, t+2n\tau-2m\tau)+{\cal T}^2{\cal R}^2\sum_{n, m=0}^{\infty}{\cal R}^{2n+2m}D_{LL}(0, t+2n\tau -2m\tau).
\end{multline}
Performing  the Fourier transformation, using Eq.~(\ref{e5.11}), and substituting the result in  Eq.~(\ref{e5.9}), we obtain Eq.~(\ref{e5.10}).
The function $K(\omega)$ has the asymptotic behavior (we recall that ${\cal R}=1/\sqrt{3}$)
\begin{equation}
 \label{e5.13}
K(0)=\frac{1+{\cal R}^2}{1-{\cal R}^2}=2\,, \qquad K(\infty)=1 \,;
\end{equation}
the crossover between them takes place at frequencies $\omega \sim 1/\tau$ determined by the flight time through the middle part of the device. 

The RG equation that describes the flow of $h$ with the frequency $\omega$ reads
\begin{equation}
 \label{e5.14}
d h/d{\cal L}=[1-K(\omega)]h \,,
\end{equation}
where ${\cal L} = \ln (1/\omega)$. 
Equivalently, one can view Eq.~(\ref{e5.14}) as describing the flow of $h$ with the length $L_0$ of the leads.
 For this purpose,  the frequency $\omega$ in Eq. (75) should be understood as a running energy scale corresponding to the running length $L_0$
of the leads. It can be checked that the non-trivial renormalization of $h$ comes from the fluctuations of the bosonic  filed in the $1/3$ lead. Thus, the relevant velocity is that of the $1/3$ mode, i.e., $\omega\sim v_L/L_0$.  The flow becomes non-trivial when $K(\omega)$ differs essentially from unity, which is the case  for $\omega \lesssim 1/\tau$, or, in terms of the length scale, for $L_0 \gtrsim (v_L/{\bar v})L$.
Below we will drop for brevity  the dimensionless ratio of velocities in this condition, writing it simply as $L_0 \gtrsim L$.
 In this regime $1-K$ is negative, so that $h$ decreases under the RG flow. Thus, the clean ($h=0$) fixed point is stable (attractive). 

\subsubsection{Renormalization group near $\theta=\pi$}
\label{s5.3}

Let us now describe the RG near the second fixed point of the Hamiltonian (\ref{e5.4}), (\ref{e5.5}), $\theta = \pi$ (or, equivalently, $h = 1/6$). 
When $\theta$ is exactly equal to $\pi$, the factor $R_x(\theta)$ in Eq.~(\ref{e4.15})  is a rotation by angle $\pi$ around the $x$ axis. Such a rotation transforms the $z$ axis into $-z$.   In this situation, we can gauge out the disorder completely by using the transformation generated by $S_\lambda(x)$ and still have a quadratic Hamiltonian.
The resulting fixed-point Hamiltonian reads:
\begin{eqnarray}
 H&=&\int_{-L/2}^{+L/2}\frac{1}{4\pi}\left[v_{\sigma}(\partial_x \phi_\sigma]^2+v_\rho(\partial_x \phi_\rho)^2\right)+\int_{x<-L/2}
 \frac{1}{4\pi}\left[v_R(\partial_x \phi_R)^2+v_L(\partial_x \phi_L)^2\right]\nonumber+\\&&+\int_{x> L/2}
 \frac{1}{4\pi}\left[v_R(\partial_x \tilde{\phi}_R)^2+v_L(\partial_x \tilde{\phi}_L)^2\right] \,,
 \label{Eq:HNFix}
 \end{eqnarray}
where 
\begin{equation}
\left(\begin{array}{c}
\tilde{\phi}_R\\
\tilde{\phi}_L
\end{array}\right)=\frac{1}{{\cal T}}\left(\begin{array}{cc}
1 & {\cal R}\\
{\cal R} & 1
\end{array}\right)
\left(\begin{array}{c}
\phi_\sigma\\
\phi_\rho
\end{array}\right).
\label{Eq:sigmarhotoRtLt}
\end{equation}
Equation (\ref{Eq:sigmarhotoRtLt}) is obtained from Eq.~({\ref{Eq:sigmarhotoRL}}) by a transformation ${\cal R} \to -{\cal R}$ associated with the sign change $\partial_x \phi_\sigma \to  - \partial_x \phi_\sigma$ generated by $S_\lambda(+\infty)$. 

A small deviation from this fixed point, $\tilde \theta = \pi - \theta \ll 1$, will generate a perturbation of the Hamiltonian (\ref{Eq:HNFix}) by a cosine term
 analogous to the last term in Eq.~(\ref{e5.4}) with a prefactor given by $\tilde \theta $. 
 Proceeding in full analogy with the analysis in Sec.~\ref{s5.2}, we can derive an RG equation for $\tilde h = \tilde \theta / 6\pi$,
 \begin{equation}
  \label{e5.15}
d \tilde h/d{\cal L}=[1-\tilde K(\omega)]\tilde h \,.
\end{equation}
The resulting expression for $\tilde K(\omega)$ can be obtained from  that for $K(\omega)$ by a replacement 
${\cal R}^2\rightarrow -{\cal R}^2$ in Eqs. (\ref{Eq:K}), yielding
\begin{equation}
\tilde{K}(\omega)=\frac{2\omega}{\pi}(1-{\cal R}^4)\int_0^\infty d\Omega \frac{1}{\omega^2+\Omega^2}\frac{1}{1+{\cal R}^4+2{\cal R}^2\cos2\Omega\tau} \,,
\label{Eq:K2}
\end{equation}
with asymptotic values 
\begin{equation}
\tilde{K}(0)=\frac{1-{\cal R}^2}{1+{\cal R}^2}=\frac{1}{2}\,, \qquad \tilde{K}(\infty)=1 \,.
  \label{e5.16}
\end{equation}
Since $\tilde K(0) < 1$, this fixed point is unstable (repulsive). Thus, the infrared RG flow is directed from the $\theta = \pi$ fixed point towards the $\theta = 0$ fixed point. 

\subsection{Conductance at the fixed points}
\label{s5.4}

It remains to calculate the value of the (two-terminal) conductance at the fixed points with $\theta=0$ and $\theta=\pi$. 
We consider a quantum Hall sample with two opposite edges, see Fig.~\ref{Fig:Setup-All}. In the top-edge ``leads'' the mode 1 propagates to the left and the mode 1/3 to the right, as shown in Fig.~\ref{Fig:Scattering}. In the bottom edge the situation is opposite. We bias the incoming modes in the left leads (which are the 1/3 mode in the top edge and the 1 mode in the bottom edge)  by a small voltage $V$ as compared to the incoming modes in the right leads. The two-terminal conductance $G$ is defined as $G = I / V$ where $I$ is the resulting current from left to right, see Sec.~\ref{s2.3}.  According to Eq.~(\ref{e2.20}), the two-terminal conductance is determined by the parameters $G_{12}^{(\alpha)}$  characterizing the top ($\alpha = t$) and the bottom ($\alpha=b$) edges. A detailed derivation of the values of $G_{12}^{(\alpha)}$ for each of two saddle points is presented in \ref{App:Conductance}; here we present a brief sketch of the argument and the result.

For the case when each of two edges is characterized by the trivial ($\theta=0$) fixed point, the zero-frequency transmission amplitudes for both the $1/3$ and the  $1$ incoming modes are equal to unity, which yields the conductance (in units of $e^2/h$)
\be
  \label{e5.17}
G = 1 + \frac{1}{3} = \frac{4}{3}.
\ee
In terms of the notations introduced in Sec.~\ref{s2.3}, this corresponds to $G_{12}^{(t)} = G_{12}^{(b)} = 0$ (no backscattering between the two channels).  This situation realizes the largest possible value of the two-terminal conductance $G$, see Eq.~(\ref{e2.22}).

If each of the edges is characterized by the nontrivial fixed point ($\theta=\pi$), the transmission amplitudes of  both  incoming modes  are equal to 
\begin{equation}
{\cal T}^2-{\cal T}^2 {\cal R}^2+{\cal T}^2{\cal R}^4+\ldots =\frac{{\cal T}^2}{1+{\cal R}^2}=\frac12
\label{e5.18}
\end{equation}
where we have take into account that ${\cal R}=1/\sqrt{3}$ and ${\cal T}=\sqrt{2/3}$. 
 In terms of  the fields $\tilde{\phi}_R$ and $\tilde{\phi}_L$ the physical charge density reads
\begin{equation}
\label{e5.18a}
{  \rho(x)= \frac{1}{2\pi}\left(\partial_x\tilde{\phi}_L-{\cal R}\partial_x\tilde{\phi}_R\right). }
\end{equation}
Thus, the field $\tilde{\phi}_R$ carries charge of $-1/3$, and the conductance is equal to
\begin{equation}
  \label{e5.19}
-\frac{1}{2}\times\frac{1}{3} + \frac{1}{2}\times 1=\frac{1}{3}.
\end{equation}
In terms of the notations of Sec.~\ref{s2.3}, this corresponds to $G_{12}^{(t)} = G_{12}^{(b)} = 1/2$. This situation realizes the lowest possible value of the two-terminal conductance $G$, see Eq.~(\ref{e2.22}).  The value $G_{12}= 1/2$ corresponds to the limit of the strongest local scatterer as found in Ref.~\cite{Chamon97}. The emergence of this value in the context of line junction between counterpropagating 1 and 1/3 modes was also pointed out in Refs.~\cite{Ponomarenko01,Rosenow10}. 

When the renormalization discussed in Sections \ref{s5.2} and \ref{s5.3} is inefficient (i.e., the leads are relatively short, or else, the frequency is sufficiently high), the two-terminal conductance of the junction can  take any value between 1/3 and 4/3, depending on the specific configuration of disorder (regime of mesoscopic fluctuations, see Sec. \ref{s6}).  When the infrared cutoff is lowered (i.e., the length of the leads is increased or the frequency is lowered), the renormalization yields a flow of the conductance towards the value 4/3. 

\section{Dependence of conductance on system size, temperature, and bare interaction strength}
\label{s6}

The above arguments predict that for the case of relatively short leads,  $L_0 \ll L$, the two-terminal conductance of a  $\nu=2/3$ FQHE junction can take any value between $1/3$ and $4/3$ (in units of $e^2/h$). When the  leads are long, an additional renormalization takes place, and the conductance is renormalized towards $G=4/3$ for almost any realization of disorder.

Clearly, the model considered above contains a number of assumptions that are idealizations as compared to a realistic experimental situation. Specifically, we have assumed
\begin{enumerate}
\item[(i)] ``ideal contacts'': the segments of the edge to which the bias voltage is applied can be modelled as decoupled 1 and 1/3 modes, and the modes leaving the reservoir are in equilibrium with this reservoir;
\item[(ii)] that the interaction in the central region of the device is fine-tuned to the value $c = \sqrt{3}/2$ for which the neutral and 2/3 modes are the eigenmodes;
\item[(iii)] zero temperature, $T = 0$.
\end{enumerate}

The importance of the assumption (i) in the context of transport through FQHE devices has been analysed in the literature \cite{kane-fisher95a,Chamon97} and we will not discuss it here. In the analysis of the length and temperature dependence of the conductance below we will follow the ideal-contact assumption (i).

Let us now discuss  the implications of  relaxing the remaining two assumptions, which is crucial for understanding the dependence of the conductance on temperature and on the size of the device, as studied experimentally. 

\subsection{Zero temperature, strong interaction}
\label{s6.1}

Imagine first that the assumption (ii) is relaxed [but (iii) still holds: the temperature is $T=0$]. For a broad range of bare values of the interaction and disorder, the theory will be  in the basin of attraction of the ``neutral plus 2/3'' fixed-point theory, cf.  Ref.~\cite{kfp}.  Under these conditions, the above results should retain their validity, up to small corrections. (We assume, of course, that the size $L$ of the central region is much larger than the ultraviolet scale $a$.)  In other words, at zero temperature the assumption (ii) can be substantially weakened: it is sufficient that the initial parameters are in the above basin of attraction, which is rather broad \cite{kfp}. For  weak bare disorder, the requirement is that the parameter $\Delta$ is in the range $1 < \Delta < 3/2$, which means that the repulsive interaction between the 1 and 1/3 modes is neither too weak nor extremely strong, $0.34 < c < 0.98$. 

\subsection{Zero temperature, weak interaction}
\label{s6.2}

What happens if the interaction is weaker, $\Delta > 3/2$, i.e., $c < 0.34$?  The weak disorder is then RG-irrelevant (i.e., it renormalizes to zero in the infrared), and the renormalization of $\Delta$ is not particularly important. The infrared limit of the theory is then on a line of fixed points with $\Delta > 3/2$ and no disorder, which is characteristic for the Berezinskii-Kosterlitz-Thouless transition.
The zero-temperature conductance will be then essentially the same (up to small corrections) as that of a clean system. Clearly, for a clean system we have $G_{12}^{(t)} = G_{12}^{(b)} = 0$.
For the two-terminal conductance $G$ this implies  
\be
\label{e6.1}
G = 4/3 \,. 
\ee
It is worth emphasizing that this value is independent of the interaction strength (i.e., on $\Delta$).

Thus, there is an important difference between the $T=0$ value of the conductance  in the cases of strong interaction (disordered fixed point with $\Delta=1$) and weak interaction (clean fixed point with $\Delta>3/2$). In the first case the conductance $G$ shows strong mesoscopic fluctuations bounded between $1/3$ and $4/3$ if the leads are not too long, $L_0 \lesssim L$, and renormalizes to 4/3 in the limit $L_0 \gg L$. In the second case the conductance is equal to 4/3, independently of the relation between $L_0$ and $L$.  The existence of strong mesoscopic fluctuations is thus a hallmark of the $\Delta=1$ fixed point, i.e., essentially of the neutral-mode physics. As we discuss below, the difference between the strong-interaction and weak-interaction regimes shows up also in the limit of long leads, $L_0 \to \infty$, as one considers the conductance either as function of temperature or as  function of the length of the interacting segment of the edge. 

\subsection{Finite temperature}
\label{s6.3}

Finally, let us relax the assumption (iii), i.e., consider the problem at hand at finite temperature $T$. As for the zero-$T$ limit above, we consider separately the two cases of strong and weak interaction between the 1 and 1/3 modes.

\subsubsection{Strong interaction: $\Delta < 3/2$}
\label{s6.3.1}

\begin{figure}
\centering
\includegraphics[width=270pt]{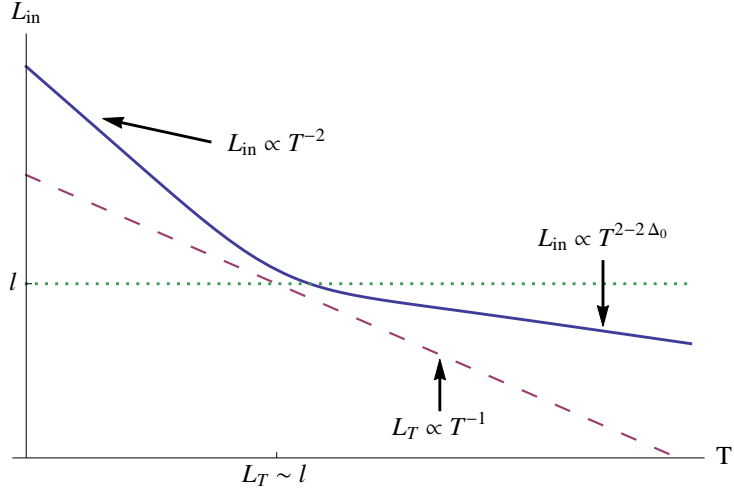} 
\caption{Schematic plot (on the log-log scale) of the temperature dependence of the inelastic equilibration length $L_{\rm in}$ for the case of strong interaction, $\Delta_0 < 3/2$.  The equilibration length is given by Eqs.~(\ref{e6.2}) and (\ref{e6.2-high}) for low and higher temperatures, respectively. (In the figure, only the corresponding temperature scaling is indicated.)  Also shown are the thermal length $L_T$ as well as the temperature-independent length $\ell$, Eq.~(\ref{e-length2}), at which the disorder becomes strong at low temperature. 
}
\label{Fig:L-in}
\end{figure}

We assume that the bare value of the interaction between the 1 and 1/3 modes is sufficiently strong ($\Delta_0 < 3/2$), so that the system flows under RG towards the ``neutral plus 2/3'' fixed point  ($\Delta=1$) in the infrared. For lowest temperatures (and at given system size $L$) this flow is cut off by $L$. In this situation, the temperature is of no particular importance and can be safely set to zero---which is the case considered above. In the opposite situation, it is the temperature that stops the RG flow of interaction and disorder at the corresponding thermal length scale $L_T \sim v_\sigma/T \ll L$. The system thus does not reach the fixed point characterized by the neutral and 2/3 modes: the eigenmodes remain slightly different. As a result, tunneling couples these counterpropagating eigenmodes, establishing a length $L_{\rm in}(T)$ of backscattering between them. It is important that this scattering is a genuine inelastic process. {  In view of this, we expect} that the scattering establishes equilibration between the counterpropagating modes at scales  $L \gg L_{\rm in}(T)$. As is clear from the above discussion, the length $L_{\rm in}(T)$ diverges in the limit $T\to 0$.  An explicit estimate of this equilibration length  $L_{\rm in}(T)$ (as well as of other characteristic scales for the finite-temperature behavior of transport characteristics) is presented in \ref{App:Length-scales} where we closely follow Ref.~\cite{kfp}. The result is [see Eq.~(\ref{e-length8})]
\be 
\label{e6.2}
L_{\rm in}(T) \sim \frac{L_T^2}{\ell W_0^{\rm in}} \sim \frac{1}{\Delta_0-1} \frac{1}{\ell} \left( \frac{v_\sigma}{T} \right)^2 \,, \qquad L_T \gtrsim \ell \,,
\ee
where $\ell$ is the length of disorder-induced mixing between the bare modes, Eq.~(\ref{e-length1}). Equation (\ref{e6.2}) is valid at sufficiently low temperature, when $L_T \gtrsim \ell$, so that the RG flow has taken the system to the strong-disorder regime and the renormalized $\Delta$ is close to unity.   In this low-temperature regime the hierarchy of scales is $\ell \ll L_T \ll L_{\rm in}(T)$. 
In the opposite case of higher temperatures, $L_T \lesssim \ell$, the equilibration length is given by 
\be
\label{e6.2-high}
L_{\rm in}(T) \sim \frac{\ell}{\Delta_0-1} \left( \frac{\ell}{L_T}\right)^{2-2\Delta_0} \,,  \qquad L_T \lesssim \ell \,,
\ee
see Eq.~(\ref{e-length8d}).
Thus, the temperature scaling of $L_{\rm in}$ changes from $T^{-(2\Delta_0-2)}$ (with $0 < 2\Delta_0-2 < 1$) at higher temperatures to $T^{-2}$ at lower temperatures, see Fig.~\ref{Fig:L-in}. The crossover temperature is determined by the condition $L_T \sim \ell$ and thus depends on disorder strength.

Let us consider the evolution of the conductance $G$ with increasing system size $L$ at a given (non-zero) temperature $T$ and given length of the leads $L_0$. 
The physics is particularly rich in the case of low temperatures, $L_T \gg \ell$. 
 In view of the characteristic length scales identified above, we can  then distinguish between the following four transport regimes:

\begin{itemize}

\item[(i)] Almost decoupled bare modes: $L \ll \ell$. In this situation the disorder is of no influence at all. In particular, the conductance is given by Eq.~(\ref{e6.1}), up to small corrections.

\item[(ii)] Nearly decoupled neutral and 2/3 modes with renormalization by leads:  $\ell \ll L \ll \min(L_0,L_T)$.  The interacting part of our device is now described in terms of decoupled neutral and 2/3 modes, with the former subject to strong disorder. However, the renormalization of the conductance by the leads again yields the result (\ref{e6.1}), see Secs.~\ref{s5} and \ref{s6.1}.

\item[(iii)] Mesoscopic regime: $ \min(L_0,L_T)\ll L \ll L_{\rm in}(T)$. Now the renormalization by leads becomes inefficient, and the conductance shows mesoscopic fluctuations between the limiting values 1/3 and 4/3, see Secs.~\ref{s5} and \ref{s6.1}.

\item[(iv)] Incoherent regime: $L \gg L_{\rm in}(T)$. In this situation the parameter $G_{12}^{(\alpha)}$ characterizing each of the two edges (top and bottom) is equal to 1/3, and the two-terminal conductance is 
\be
\label{e6.2a}
G = 2/3 \,,
\ee
up to corrections that are exponentially small in $L_{\rm in}(T) / L \ll 1$.  This result for the conductance in the incoherent regime was obtained in Refs.~\cite{kfp,Sen08,Rosenow10,Gefen-unpub}. We present a brief analysis of the transport in the incoherent regime with a derivation of Eq.(\ref{e6.1}) in Sec.~\ref{s6.3.3} below.

\end{itemize}

\begin{figure}
\centering
\includegraphics[width=180pt]{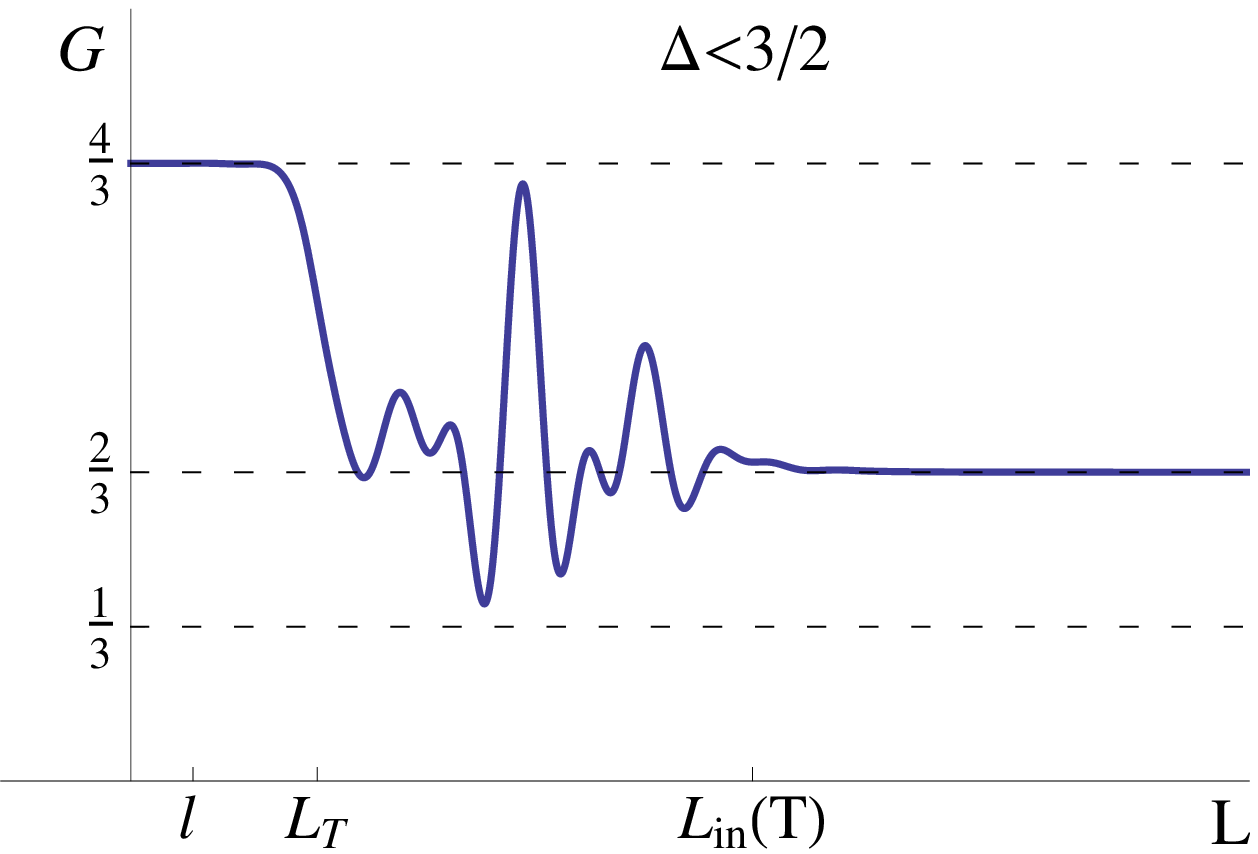} \hspace*{0.5cm}
\includegraphics[width=180pt]{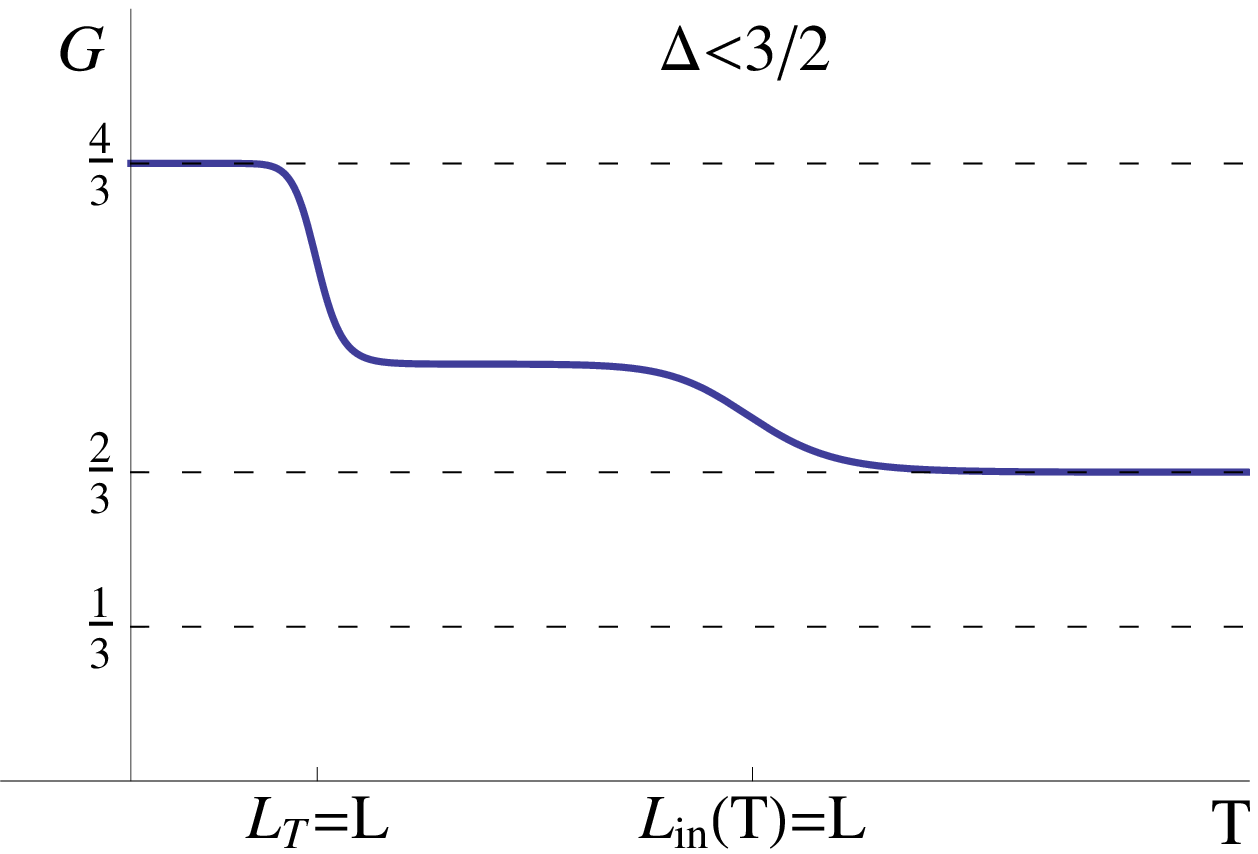}
\caption{ Schematic plot of the temperature and length dependence of the two-terminal conductance $G$ for the case of strong interaction (bare value of $\Delta$ below 3/2), when the renormalization drives the system towards the strong-disorder $\Delta=1$ fixed point.
{\it  Left panel:} Evolution of $G$ with the length $L$ of the central part of the system, at fixed (non-zero) temperature $T$.  It is assumed that the temperature is sufficiently low, so that $L_T \gg \ell$.
{\it  Right panel:} Evolution of $G$ with the temperature $T$ at fixed $L \gg \ell$. 
The characteristic length scales are: the length $\ell$ of tunneling between the 1 and 1/3 modes (which is essentially $T$-independent at low $T$), the thermal length $L_T \propto T^{-1}$, and the equilibration length $L_{\rm in}(T) \propto T^{-2}$, which satisfy the hierarchy $\ell \ll L_T \ll L_{\rm in}(T)$.  For $L \ll L_T$,  the renormalization drives the system towards the fixed point with  $G = 4/3$.
For $L_T \lesssim L \lesssim L_{\rm in}(T)$ the system is in the regime of strong mesoscopic fluctuations.
It is assumed here that the leads are sufficiently long $L_0 \gg L_T$; otherwise, the lower border of the regime of strong mesoscopic fluctuations will be at $L_0$ rather than at $L_T$. For $L \gg L_{\rm in}(T)$ the system is in the incoherent regime with $G = 2/3$.  
}
\label{Fig:Evolution}
\end{figure}

The evolution of the conductance $G$ of a given sample with system size length $L$ at fixed temperature $T$ in the strong-interaction regime is illustrated in the left panel of Fig.~\ref{Fig:Evolution}. The right panel of the figure shows the same results plotted as a function of $T$ for fixed $L$. The mesoscopic-fluctuation regime at $ \min(L_0,L_T)\ll L \ll L_{\rm in}(T)$ shows up in this plot in the form of a sample-specific value of the conductance satisfying $1/3 < G < 4/3$. 

In the case of higher temperatures, $L_T \lesssim \ell$, the regime (i) extends up to  $L \sim L_{\rm in}(T)$, with the equilibration length given by Eq.~(\ref{e6.2-high}), where a crossover directly  to the incoherent regime (iv) takes place. The behavior of the conductance in this situation is essentially the same as for the case of weak interaction $\Delta > 3/2$, see Sec.~\ref{s6.3.2} and the left panel of Fig.~\ref{Fig:Evolution1} below.

\subsubsection{Weak bare interaction: $\Delta > 3/2$}
\label{s6.3.2}

Now we turn to the case of weak interaction, $\Delta_0 > 3/2$, when the disorder is RG-irrelevant and the renormalization of interaction is inessential. 
{  In this regime of weak interaction,} the only length scale where the character of transport changes is the inelastic equilibration length $L_{\rm in}$ given by [see Eq.~(\ref{e-length10}) in \ref{App:Length-scales}]
\be
L_{\rm in}(T) \sim W_0^{-1} a  \left( \frac{L_T}{a} \right)^{2\Delta-2}\sim W_0^{-1} a  \left(\frac{v_{1/3}}{a T} \right)^{2\Delta-2} \,,
\label{e6.2b}
\ee

Now there are only two different transport regimes:

\begin{itemize}

\item[(i)] Almost decoupled bare modes: $L \ll L_{\rm in}(T)$, which is fully analogous to the regime (i) of the strong-interaction case, Sec.~\ref{s6.3.1}.
Disorder is of no importance, and the conductance is given by Eq.~(\ref{e6.1}), up to small corrections.

\item[(ii)] Incoherent regime: $L \gg L_{\rm in}(T)$, which is analogous to   the regime (iv) of the strong-interaction case, Sec.~\ref{s6.3.1}. 
The parameter $G_{12}^{(\alpha)}$ characterizing each of the two edges  is equal to 1/3, and the two-terminal conductance is given by Eq.~(\ref{e6.2a}).

\end{itemize}

\begin{figure}
\centering
\includegraphics[width=180pt]{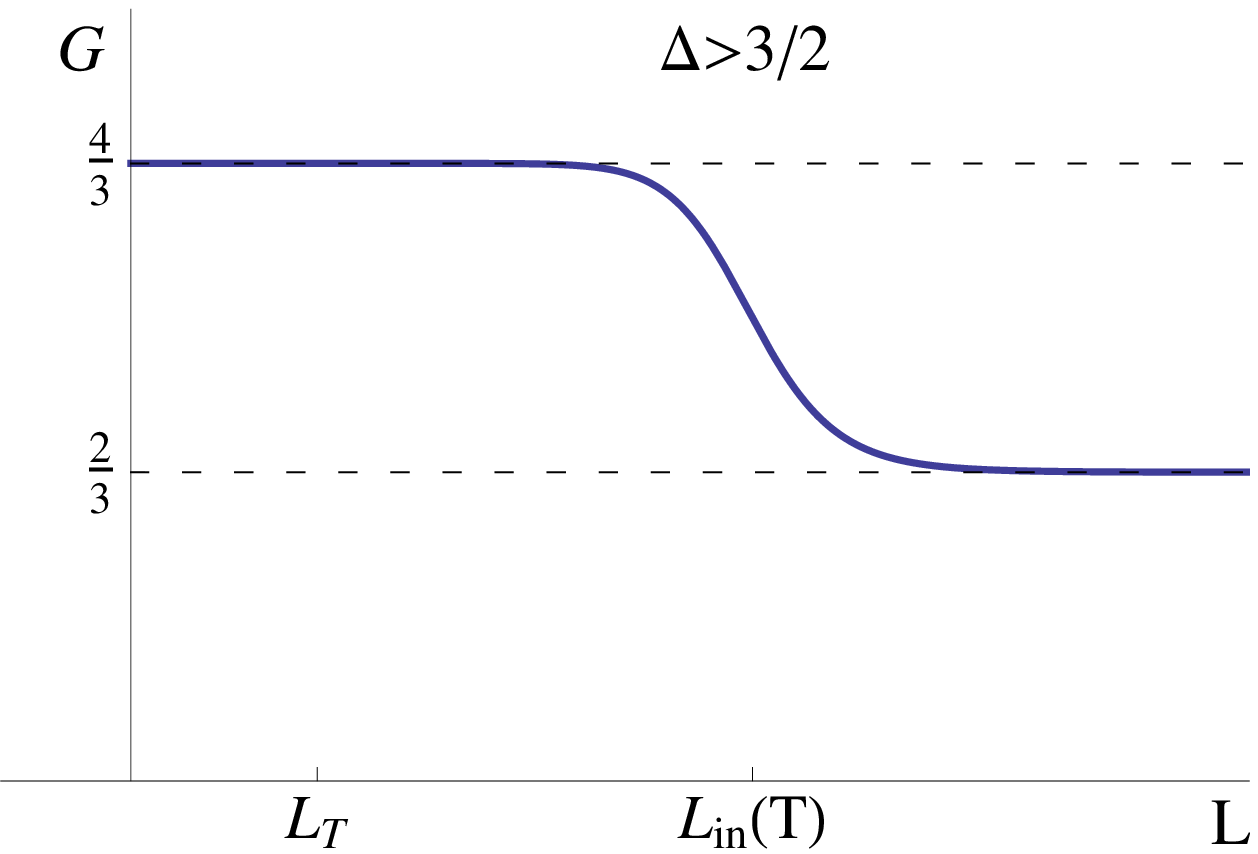} \hspace*{0.5cm}
\includegraphics[width=180pt]{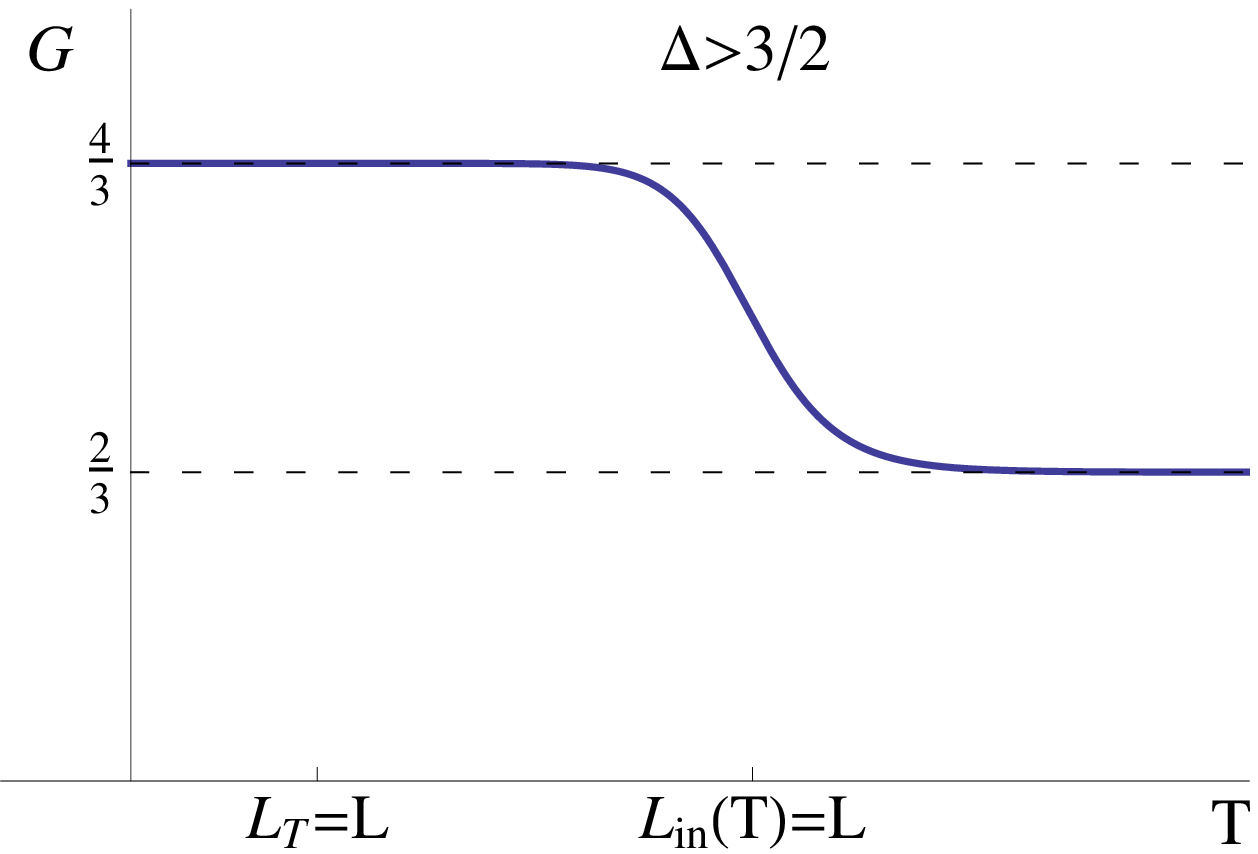}
\caption{Schematic plot of the temperature and length dependence of the two-terminal conductance $G$ for the case of weak interaction, $\Delta>3/2$. 
{\it  Left panel:} Evolution of $G$ with the length $L$ of the central part of the system, at fixed (non-zero) temperature $T$.  
{\it  Right panel:} Evolution of $G$ with the temperature $T$ at fixed $L$. 
The characteristic length scales are:  the thermal length $L_T \propto T^{-1}$ and the equilibration length $L_{\rm in}(T) \propto T^{-(2\Delta-2)}$, which satisfy $L_T \ll L_{\rm in}(T)$.  For $L \ll L_{\rm in}(T)$,  the renormalized tunneling between the eigenmodes is weak, and the conductance is $G = 4/3$. For $L \gg L_{\rm in}(T)$ the system is in the incoherent regime with $G = 2/3$. 
}
\label{Fig:Evolution1}
\end{figure}

These results are illustrated in Fig.~\ref{Fig:Evolution1}.  It is instructive to compare the behavior for the case of weak interaction shown (Fig.~\ref{Fig:Evolution1}) with that for the strong-interaction case (Fig.~\ref{Fig:Evolution}). The behavior for small $L$ (where $G = 4/3$) and for large $L$ (incoherent regime with $G = 2/3$) is similar in both cases. The difference is, however, in the scaling of the borders of these regimes with temperature.  Another difference, which is arguably the most dramatic one, is in the emergence of the mesoscopic regime in the case of strong interaction, for $ \min(L_0,L_T)\ll L \ll L_{\rm in}(T)$, with the conductance showing strong mesoscopic fluctuations within the range $1/3 < G < 4/3$.

\subsubsection{Incoherent transport, $L \gg L_{\rm in}$}
\label{s6.3.3}

Here we present an analysis of the incoherent regime, $L \gg L_{\rm in}$, the key results for which were already stated in Sec.~\ref{s6.3.2}. To do this, we rewrite the conductance matrix  (\ref{e2.19}) of an interacting disordered edge in terms of a transfer matrix expressing currents to the left (in the outgoing 1 channel and incoming 1/3 channel)  in terms of those to the right (incoming 1 channel and outgoing 1/3 channel):
\be
\label{e6.3}
\left( 
\begin{array}{c}
I_1^{\rm out} \\ I_{1/3}^{\rm in} 
\end{array}
\right) =
\frac{1}{1-3g} \left( 
\begin{array}{cc} 
1-4g & 3g \\
-g & 1
\end{array}
\right) 
\left( 
\begin{array}{c}
I_1^{\rm in} \\ I_{1/3}^{\rm out} 
\end{array}
\right) \,,
\ee
where we used again the notation $g= G_{12}$. 
In the incoherent regime such transfer-matrices for adjacent segments of the wire will be simply multiplied. It is easy to check that multiplication of two matrices of the type (\ref{e6.3}) with the reflection coefficients $g_1$ and $g_2$ yields a matrix of the same type with the reflection coefficient $g$ given by 
\be
\label{e6.4}
g = \frac{g_1 + g_2 - 4 g_1 g_2}{1 - 3 g_1 g_2} \,.
\ee
Setting here $g_1 = g_2 = g$, we obtain the  equation
\be 
g = \frac{2g - 4 g^2}{1-3g^2} \,,
\label{e6.5}
\ee
which has an attractive fixed point $g=1/3$. This is the limiting value of $g(L)$ at $L/L_{\rm in} \to \infty$. In order to see how this value is approached with increasing $L$, we consider attaching a segment of the wire of length $\sim L_{\rm in}$ to a wire of a length $L \gg L_{\rm in}$. Using Eq.~(\ref{e6.4}), we get the evolution equation for $g(L)$ (more precisely, for its typical value):
\be
\label{e6.6}
\frac{dg}{dy} = (1-4g + 3g^2) \,,
\ee
where $y \sim L/L_{\rm in}$.
The solution $g(y)$ of Eq.~(\ref{e6.6}) with the initial condition $g(0)=0$ is
\be
\label{e6.7}
g(y) = \frac{1-e^{-2y}}{3-e^{-2y}} \,,
\ee
which shows that the incoherent limiting value $g=1/3$ is approached exponentially fast, $\delta(y) \equiv g(y) - 1/3 \sim e^{-2y}$, in agreement with Refs.~\cite{Sen08,Rosenow10,Gefen-unpub}. It is easy to check that this conclusion is not essentially modified by fluctuations.
Indeed, using Eq.~(\ref{e6.4}) for small $\delta_i = g_i - 1/3$, we get $\delta_{12} = - \frac{9}{2} \delta_1\delta_2$, which implies that $\ln \delta$ is a Gaussian-distributed quantity, with the average $\langle \ln \delta \rangle = - 2y$ and fluctuations ${\rm var} (\ln \delta) \sim y$.  

Substituting Eq.~(\ref{e6.7}) into Eq.~(\ref{e2.20}), we find for the two-terminal conductance in the incoherent regime
\be
\label{e6.8}
G  = \frac{2}{3} \: \frac{1 + \frac{1}{3}e^{-2y}}{1 - \frac{1}{3}e^{-2y}} \simeq  \frac{2}{3} \left(1 +  \frac{2}{3} e^{-2y}\right) \,, \quad y \sim L / L_{\rm in} \gg 1 \,.
\ee
Thus, the limiting incoherent value of the  two-terminal conductance at $L / L_{\rm in} \to \infty$ is 2/3; a correction at finite (but large) $L / L_{\rm in}$ is exponentially small and positive.

Let us emphasize that these results for the incoherent regime apply equally to the cases of strong and weak interactions discussed in Sections \ref{s6.3.1} and \ref{s6.3.2}, respectively. The only difference is in the value of the equilibration length $L_{\rm in}(T)$, which is given by Eqs.~(\ref{e6.2}) and (\ref{e6.2-high}) in the first case and by  Eq.~(\ref{e6.2b}) in the second case.

\section{Thermal transport}
\label{s7}

In this Section we discuss the thermal transport through the $\nu=2/3$ FQH edge states. 

\subsection{General consideration}
\label{s7.1}

 Similarly to the above study of the charge transport, we consider the four-terminal setup  of Fig.~\ref{Fig:Setup-Four}, where now the electrodes are kept at zero potential but have slightly different temperatures $T_i=T+\Delta T_i$, with $i=1, 2, 3, 4$. 
Within the linear response approximation, the system is characterized by the matrix of thermal conductances, ${\cal  G}^{Q}_{ij}$, relating the energy currents 
$I^{Q}_i$ flowing out of the electrodes to the  temperature differences, $I^{Q}_i={\cal  G}^{Q}_{ij}\Delta T_j$.

In full analogy with the analysis of the electric transport carried out in Sec. \ref{s2.3}, the matrix ${\cal  G}^{Q}_{ij}$ can be constructed out of the two matrices 
$\hat{G}^{Q}_{t}$ and $\hat{G}^Q_{b}$, characterizing the response of the top and bottom edges of the the sample to temperature variations. 
Specifically, each of the matrices $\hat{G}^Q_{t, b}$ relates the energy currents $I_1^{Q}$ and $I_{2}^{Q}$ in the outgoing 1 and $1/3$ channels 
of the respective edge to the temperatures of the incoming $1$ and $1/3$ channels (see Fig. \ref{Fig:Setup}), and has the form [cf. Eq. (\ref{e2.19})]
\be
   \label{Eq:GTh}
\hat{G}^Q= \left( \begin{array} {cc} 1 - G^Q_{12} \ & \ G^Q_{12} \\ G^Q_{12}  \ & \ 1- G^Q_{12}
\end{array}
\right).
\ee
Here, we measure the thermal conductance in units of the thermal-conductance quantum $\pi T/6\hbar$; the sum of the matrix elements in each column (unity) reflects the fact that the $1$ and $1/3$  modes are indistinguishable from the point of view of  thermal transport, and the thermal current emanating  via each of the incoming modes from the corresponding  reservoir is given by $I_i^{Q, {\rm in}}= \pi T\Delta T_i/6\hbar$. 

Constructing now the matrix ${\cal  G}^Q_{ij}$ [cf. Eqs. (\ref{e2.22})--(\ref{e2.25})] and analysing  the eigenvalues of its symmetric part, one readily finds that thermodynamics dictates the inequality 
\begin{equation}
0\leq G^Q_{12}\leq 1.
\label{Eq:G12ThLimits}
\end{equation}
For the two-terminal thermal conductance   of the $\nu=2/3$ sample as measured in the setup of Fig. (\ref{Fig:Setup-All}) (with voltage bias replaced by  temperature bias),
\begin{equation}
\label{e7.1a}
G^{ Q}=2-G_{12}^{{Q, (t)}}-G_{12}^{{Q, (b)}},
\end{equation}
the inequality (\ref{Eq:G12ThLimits}) implies
\begin{equation}
\label{e7.1}
0\leq G^{Q}\leq 2.
\end{equation}
Thus, in contrast to the the case of electric transport, thermodynamics does not guarantee ballistic transport of energy in the system. 
As is shown below, thermal transport remains nevertheless ballistic in the coherent regime, $L<L_{\rm in}(T)$, and crosses over to diffusion --- which implies  the standard Ohmic behavior of thermal conductance, $G^{Q}\sim 1/L$ --- on larger length scales,  $L > L_{\rm in}(T)$.   

\subsection{Coherent regime}

In this section we analyse the thermal transport in the coherent regime when the length of the system is smaller then the inelastic scattering length, $L<L_{\rm in}(T)$, with the later given by Eq. (\ref{e6.2}) and Eq.  (\ref{e6.2b}) for the cases of strong ($\Delta<3/2$) and weak ($\Delta>3/2$) interaction respectively.  For the clarity of presentation we assume that the length $L_0$ of the leads exceeds all other length scales of the problem. 

\subsubsection{Weak interaction}
\label{Sec:CoherentWeak}

In the case of weak interaction ($\Delta_0>3/2$) and for $L<L_{\rm in}(T)$, the effect of impurities on the system can be fully neglected. 
One deals then with a system of $1$ and $1/3$ modes non-interacting for $|x|>L/2$ and  with interaction $c$ (non-universal, not renormalized by impurities up to small corrections)  in the middle part of the wire, $|x|<L/2$. 
Such a structure is characterized by an (energy-dependent) transmission amplitude 
\begin{equation}
{\cal T}_{\rm tot}(\omega)=\frac{\tilde{{\cal T}}^2}{1-\tilde{{\cal R}}^2 e^{2i\omega\tilde{\tau}}}.
\label{calTGeneric}
\end{equation}
Here,  $\tilde{{\cal R}}$ and  $\tilde{{\cal T}}=\sqrt{1-\tilde{{\cal R}}^2}$  are the bosonic  reflection and transmission amplitudes at the boundary of the interacting part given by
\begin{equation}
\tilde{{\cal R}}=\frac{1-\sqrt{1-c^2}}{c} \,,
\label{eRtilde}
\end{equation}
and $\tilde{\tau}$ is the mean flight time through the interacting region determined by its length and by the velocities $\tilde{v}_R$ and $\tilde{v}_L$ of the local eigenmodes,
\begin{equation}
\tilde{\tau}=\frac{L}{2}\left(\tilde{v}_R^{-1}+\tilde{v}_L^{-1}\right).
\end{equation}
Computing the energy current in the outgoing $1$-mode of Fig. \ref{Fig:Setup}, we find  a general expression for the parameter $G^Q_{12}$ characterizing the thermal transport through the edge: 
\begin{equation}
\label{e7.2}
G_{12}^Q=1-\frac{6}{\pi^2}\int_{0}^{\infty}\frac{\Omega d\Omega}{e^{\Omega}-1}\left|{\cal T}_{\rm tot}\left(\Omega T\right)\right|^2.
\end{equation}
The integration in Eq.~(\ref{e7.2}) goes over a dimensionless variable $\Omega = \omega/T$. 
The value of $G_{12}^Q$ depends on the relation between the temperature $T$ and the characteristic flight time $\tilde{\tau}$ proportional to the system size $L$. 
For low temperatures,  $T\tilde{\tau} \ll 1$, the   transmission coefficient is given, up to small corrections, by ${\cal T}_{\rm tot}(\omega=0)=1$, which  leads to 
 $G_{12}^{Q}$=0. Thus, in this regime the two-terminal thermal conductance
attains a universal value (again up to small corrections),
\begin{equation}
\label{e7.3}
 G^{Q}=2, \qquad T\tilde{\tau}\ll 1.
\end{equation}
On the other hand, for higher temperatures, $T\tilde{\tau}\gg1$  [but still under the condition $L\ll L_{\rm in}(T)$ with  the inelastic length given by Eq. (\ref{e6.2b})] 
the thermal conductance $G_{12}^{Q}$ (and, correspondingly, $G^Q$) is non-universal and given by
\begin{equation}
G_{12}^{Q}=\frac{2\tilde{{\cal R}}^2}{1+\tilde{{\cal R}^2}}=1-\sqrt{1-c^2}\,, \qquad \qquad T\tilde{\tau}\gg 1.
\label{G12WeakInt}
\end{equation}
 When the interaction $c$ is varied, $G_{12}^{Q}$ can in principle take arbitrary values between $0$ and $1$, corresponding to the two-terminal conductance taking values in the range $0\leq G^{Q}\leq 2$. However, if we restrict ourselves to relatively weak repulsive interactions satisfying the condition $\Delta>3/2$, i.e., to the interval $0 < c < 0.34$, we find a much narrower range $0 \le G_{12}^{Q} \le 0.06$, and thus 
 \be
 \label{e7.3a}
 1.88\leq G^{Q}\leq 2 \,, \qquad L_T\ll L\ll L_{\rm in}(T). 
 \ee

 We thus conclude that the thermal conductance shows a crossover from the universal value, $G^{Q}=2$, to 
a non-universal regime as the length of the system exceeds  $L_T=\tilde{v}_L\tilde{v}_R/T(\tilde{v}_L+\tilde{v}_R)$. By contrast, the electric conductance discussed in Sec. \ref{s6}  retains its universal value $G=4/3$ up to the scale $L_{\rm in}(T)$. The difference can be traced back to the fact that  thermal conductance is determined by bosonic excitations within energy window of the order of temperature, $\omega \sim T$, while the electric transport (in the dc limit) is solely due to bosons with zero energy. 
 
\subsubsection{Strong interaction}
\label{Sec:CoherentStrong}

Let us now consider  the case of strong interaction in the middle part of the edge ($\Delta_0<3/2$), still under the assumption of a coherent regime, $L < L_{\rm in}(T)$, with the inelastic length given by Eq. (\ref{e6.2}).  At the same time, let us assume that the length of the system, $L$,  is much larger than the (temperature-independent) length $\ell$ of the disorder-induced mixing between the bare modes, Eq.~(\ref{e-length1}). 
The interacting part of the edge is then at the Kane-Fisher-Polchinski fixed point.
In this situation, the neutral and charge modes are decoupled and, generally,  the former  is  subject to strong disorder.
However, at lowest temperatures,  $L_T=v_\sigma/T\gg L$,  the disorder in the neutral mode  is renormalized to zero by the leads.  At such temperatures the bosonic transmission amplitude ${\cal T}_{\rm tot}(\omega)$ is approximately equal to unity for all $\omega\lesssim T$, cf. Eq.~(\ref{calTGeneric}). Thus, in this regime $G^Q_{12} = 0$,  yielding the two-terminal thermal conductance
\begin{equation}
\label{e7.4}
G^{Q}=2 \,, \qquad L \ll L_T \,,
\end{equation}
up to small corrections. 

\begin{figure}
\centering
\includegraphics[width=180pt]{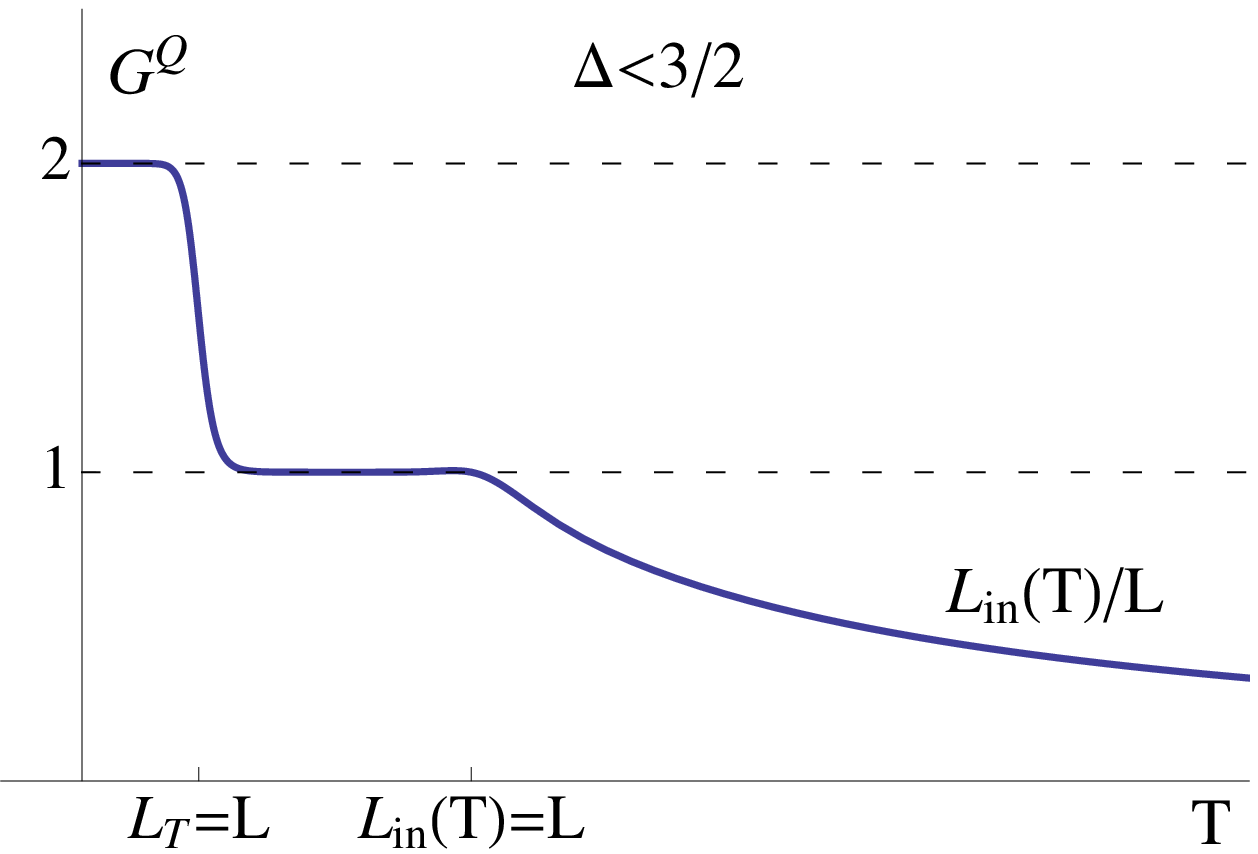}\hspace*{0.5cm}
\includegraphics[width=180pt]{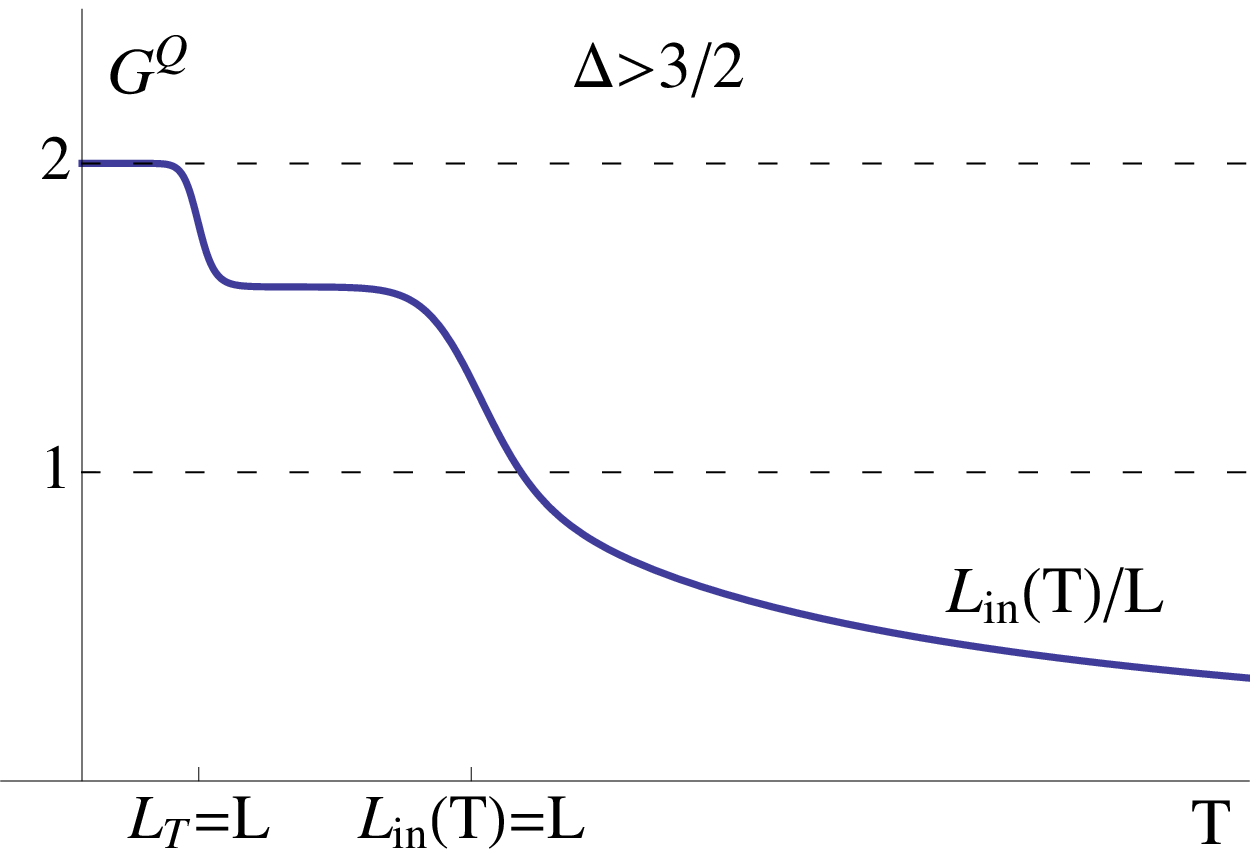}
\caption{
Two-terminal thermal conductance of a $\nu=2/3$ FQH sample for the case of strong ($\Delta <3/2$, left panel) and weak ($\Delta >3/2$, right panel) interaction between 1 and $1/3$ modes in the middle part of the edges. In both cases, the conductance at low temperatures (such that $L < L_T$) is $G^Q = 2$. 
At high enough temperatures [such that $L > L_{\rm in}(T)$], the system crosses over to the incoherent regime with Ohmic scaling of thermal conductance. In this regime, the difference between the weakly interacting  and the strongly interacting situations manifests itself in the different scaling of $L_{\rm in}(T)$ with temperature. Specifically, while $L_{\rm in}(T)$ scales according to Eq.~(\ref{e6.2b}) for $\Delta >3/2$, its behavior for $\Delta < 3/2$ is given by Eqs. (\ref{e6.2}) and (\ref{e6.2-high}) with a crossover at $L_T \sim \ell$ .  In the intermediate temperature range, $L_T < L < L_{\rm in}$,  a  system with $\Delta<3/2$ is characterized by universal thermal conductance $G^{Q} = 1$, while a system with 
$\Delta>3/2$ develops an interaction-dependent value of the thermal conductance. }
\label{ThermalCondPlot}
\end{figure}

At higher temperatures, $L_T<L$, the disorder manifest in the neutral mode survives the renormalization by the leads, giving rise to strong mesoscopic fluctuations of the electric conductance discussed in Sec. \ref{s6}. These fluctuations originate from the dependence of the electric conductance on the parameter $\theta$ that characterizes the disorder configuration and make take any value in the range between 0 and $\pi$. We argue now that, in contrast to the electric conductance,  the thermal conductance of the edge remains insensitive to disorder in this regime due to thermal averaging. Indeed, for the two exactly solvable cases of $\theta=0$ and $\theta=\pi$, the edge is characterized by the transmission amplitude [cf. Eq.~(\ref{calTGeneric})]
\begin{equation}
{\cal T}_{\rm tot}(\omega)=\frac{{\cal T}^2}{1\mp {\cal R}^2 e^{2i\omega\tau}}\,, \qquad \tau=\frac{L(v_\sigma+v_\rho)}{2v_\sigma v_\rho} \,,
\label{calTGeneric1}
\end{equation}
where ${\cal R}=1/{\sqrt{3}}$; the plus and minus signs in the denominator correspond to  $\theta=\pi$ and $\theta=0$, respectively. 
Substituting Eq.~(\ref{calTGeneric1})  into Eq.~(\ref{e7.2}), we find [cf. Eq. (\ref{G12WeakInt})]
\begin{equation}
\label{e7.6}
G_{12}^{Q}=\frac{2 {\cal R}^2}{1+{\cal R}^2}=1/2 \,,
\end{equation}
in both cases of $\theta=0$ and $\theta=\pi$. 
This yields the two-terminal conductance
\begin{equation}
G^{Q}=1\,, \qquad L_T\ll L\ll L_{\rm in}(T).
\label{GThAv}
\end{equation}
While we have proven this only for the limiting cases of the weakest ($\theta=0$) and strongest ($\theta=\pi$) disorder, we expect that the 
analogous thermal averaging washes out the disorder in the neutral mode for arbitrary value of $\theta$. If this conjecture is correct,  the 
thermal conductance assumes the universal value (\ref{GThAv}), which is a half of the maximal thermal conductance [which is found for the lowest temperatures, see Eq.~(\ref{e7.4})].
This universal behavior, along with mesoscopic fluctuations of electric conductance in the same temperature range, is then a hallmark of the proximity of the system to the Kane-Fisher-Polchinski fixed point. 

This analysis of the case of strong interaction assumed that the temperature remains sufficiently low, $L_T \gg \ell$, so that the interaction is renormalized by disorder from an initial value $1 < \Delta_0<3/2$ to a vicinity of the Kane-Fisher-Polchinski fixed point, $\Delta=1$.
In the opposite case of higher temperatures, $L_T \ll \ell$, the analysis is similar to that in the case of weak interaction (Sec.~\ref{Sec:CoherentWeak}), see the comment  in the end of Sec.~\ref{s6.3.1}.  Under the condition $L_T<L<L_{\rm in}(T)$, the coefficient $G^Q_{12}$   determining the thermal conductance  is then given by Eqs. (\ref{G12WeakInt}) and (\ref{eRtilde}) with the unrenormalized  interaction, $0.34<c<0.98$. This implies that the two-terminal thermal conductance  $G^Q$ may take a value in the range
\begin{equation}
0.41<G^Q<1.88 \,.
\end{equation}

\subsection{Incoherent regime}

Let us now consider  the system in the incoherent regime, $L>L_{\rm in}(T)$, with the equilibration length given by Eq.~(\ref{e6.2}),  (\ref{e6.2-high}), or (\ref{e6.2b}), depending on the interaction strength and the temperature range.
The behaviour of the thermal conductance in the incoherent regime can be studied in close analogy to the analysis of the electric conductance in Sec. \ref{s6.3.3}. 
To this end, we describe each segment of the system of length $L \gtrsim L_{\rm in}$ by a transfer matrix relating incoming and outgoing currents [see Eq. (\ref{Eq:GTh})]:
\be
\left( 
\begin{array}{c}
I_1^{Q, {\rm out}} \\ I_{1/3}^{Q,  {\rm in}} 
\end{array}
\right) =
\frac{1}{1-g^Q} \left( 
\begin{array}{cc} 
1-2g^Q & g^Q \\
-g^Q & 1
\end{array}
\right) 
\left( 
\begin{array}{c}
I_1^{Q, {\rm in}} \\ I_{1/3}^{Q, {\rm out}} \,
\end{array}
\right),
\label{e7.8}
\ee
where we used a shorthand notation $g^Q\equiv G^Q_{12}$. Joining two such segments of length $L_1$ and $L_2$ characterized by $g^Q_1$ and $g^Q_2$, we get a segment of length $L_1+L_2$ having the transfer matrix of the same structure with
\begin{equation}
g^Q=\frac{g^Q_1+g^Q_2-2g^Q_1 g_2^Q}{1-g^Q_1 g^Q_2} \: .
\label{gThIt}
\end{equation}
The iterative scheme (\ref{gThIt})  has a stable fixed point $g^Q=1$. Indeed, it can be equivalently be rewritten as
\begin{equation}
\label{e7.10}
\frac{g^Q}{1-g^Q}=\frac{g^Q_1}{1-g^Q_1}+\frac{g^Q_2}{1-g^Q_2} \: .
\end{equation}
The parameter $g$ characterizing  a segment of length $L\gg L_{\rm in}(T)$ thus satisfies
\begin{equation}
\left\langle \frac{g^Q(L)}{1-g^Q(L)}\right\rangle=C \frac{L}{L_{\rm in}(T)} \,,
\label{gAv}
\end{equation}
where the angular brackets denotes the average with respect to disorder and $C$ is a numerical constant of order unity, 
\begin{equation}
\label{e7.11}
C=\left\langle \frac{g^Q(L_{\rm in}(T))}{1-g^Q(L_{\rm in}(T))}\right\rangle \sim 1.
\end{equation}
It follows from Eq. (\ref{gAv}) that in the incoherent regime 
\begin{equation}
\label{e7.12}
\left\langle g^Q(L)\right\rangle=1-\frac{1}{C} \frac{L_{\rm in}(T)}{L},
\end{equation}
so that the two-terminal thermal conductance shows Ohmic scaling with the system size,
 \begin{equation}
 G^Q= \frac{2}{C} \frac{L_{\rm in}(T)}{L}\,, \qquad L \gg  L_{\rm in}(T),
 \label{GThInc}
 \end{equation}
 in agreement with Ref.~\cite{Gefen-unpub}.
 As follows from Eq.~(\ref{e7.10}), the mesoscopic fluctuations of  $G^Q$ scale in this regime as 
 \be
  \label{e7.13}
 \frac{\left\langle \left( \delta G^Q \right)^2 \right \rangle}{ \left[ G^Q \right]^2} \sim \frac{L_{\rm in}(T)}{L} \,.
 \ee
 In the time domain the $1/L$ scaling  of the thermal conductance in the incoherent regime, Eq.~(\ref{GThInc}), corresponds to diffusive propagation of energy along the edge  found in Ref. \cite{KaneFisher97}. This should be contrasted to ballistic propagation of charge. The $2/3$ edge with random tunneling  is thus, for $L > L_{\rm int}$,  a system with a strong charge-energy separation. In fact, this separation would become even more dramatic if the dominant disorder would be not the random tunneling but rather a random interaction between the modes. In this case, the heat transport would be suppressed still stronger due to Anderson localization of bosonic modes, see a discussion in Sec.~\ref{s9}. 
 
Let us emphasize once more that Eq. (\ref{GThInc}) describes thermal transport in the incoherent regime independently of the strength of the interaction.   The only  dependence on the latter is in the value of the equilibration length $L_{\rm in}(T)$, which is given by Eqs.~(\ref{e6.2}) and (\ref{e6.2-high}) in the  case $\Delta<3/2$ and by  Eq.~(\ref{e6.2b}) in the case $\Delta>3/2$. 
The dependence of the thermal conductance on temperature is summarized in Fig. \ref{ThermalCondPlot}.

It is worth mentioning that we have discarded a contribution to thermal conductivity which originates from transport via localized quasiparticle states in the 2D bulk of the quantum-Hall system due to long-range Coulomb interaction \cite{gutman16}. The corresponding 2D heat conductivity was found to scale as $T^3$ at low temperatures \cite{gutman16}, i.e., as  $T^2$ when measured in units of the thermal conducance quantum $\pi T/6\hbar$. Thus, the contribution of the bulk thermal transport is small in comparison with the edge contribution in the regime of coherent edge transport, when $G^Q$ is of order unity. On the other hand, the situation becomes more intricate for larger systems (or higher temperatures), when the edge transport is in the incoherent regime (which, as we discuss in Sec.~\ref{s9}, is the case for the majority of experiments). Indeed, the edge thermal conductance then decreases with increasing system size as $1/L$. On the other hand, the two-terminal conductance via the 2D bulk scales with the distance $L$ between the terminals only logarithmically, $G^Q_{\rm bulk} \propto 1/\ln(L/b)$, where $b$ is the size of the contact. Therefore, with increasing $L$, the bulk contribution will become progressively more sizeable in comparison with the edge one, and for large enough $L$ the bulk thermal  transport will dominate. 

\section{Transport properties of a device with floating $1/3$ mode}
\label{s8}

\begin{figure}
\centering
\includegraphics[width=250pt]{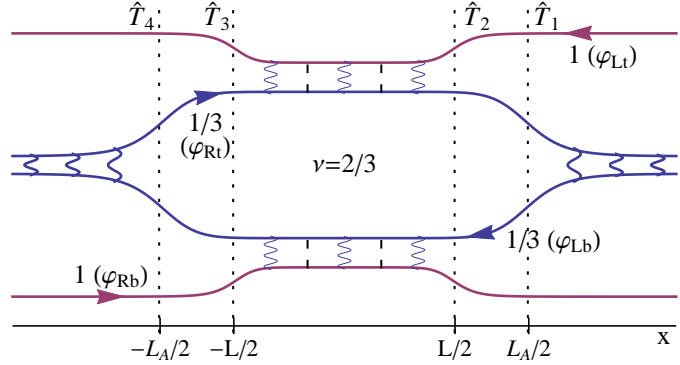}
\caption{Equivalent representation of a setup with floating $1/3$ mode. The top and bottom $1/3$ modes are formally extended 
to infinity. Strong interaction between them at $|x|>L_A/2$ prevents the plasmons from leaving the loop in the middle of the 
sample trough  these artificially added channels, thus making this setup equivalent to that of Fig.~\ref{Fig:SetupFloating}. }
\label{Fig:SetupFloatingEquivalent}
\end{figure}

In the previous Sections, we have analysed in detail transport properties of a $\nu=2/3$ FQH sample where both $1$ and $1/3$ channels are contacted by external electrodes, see Figs. \ref{Fig:Setup-All} and   \ref{Fig:Setup-Four}. In those setups the coherence between the top and bottom edges of the sample was fully broken by the ohmic contacts, irrespectively of temperature or system size.   This has allowed us to describe the properties of the system in terms of two parameters, 
$g_t$ and $g_b$, characterizing the top and bottom edges separately.  

In the present Section we explore a setup of different kind---the one with  the ``floating'' $1/3$ mode, i.e., not connected to any metallic contact, see Fig. \ref{Fig:SetupFloating}. The charge transport in such a device was recently studied experimentally in Ref. \cite{Grivnin14}. Throughout this Section we will denote the two-terminal 
electric and thermal conductances of this device by $G$ and $G^Q$, respectively.

\subsection{Incoherent regime, $L > L_{\rm in}(T)$}
\label{s8.1}

In the incoherent regime, when the length $L$ of the interacting parts of the edges exceeds the inelastic length $L_{\rm in}(T)$ 
[given by Eq. (\ref{e6.2}), (\ref{e6.2-high}), or Eq. (\ref{e6.2b}) depending on the value of $\Delta$],  the setup of Fig. \ref{Fig:SetupFloating} 
is equivalent to the four-terminal setup, Fig. \ref{Fig:Setup-Four}, where now the voltages $V_3$ and $V_4$ are adjusted 
such that the currents flowing from reservoirs $3$ and $4$ vanish, $I_3 = I_4 =0$.  This gives us for the two-terminal conductance of the device
\begin{equation}
\label{e8.1}
G=\frac{g_t+g_b-4 g_t g_b}{g_t+g_b-3g_t g_b}\,.
\end{equation}
Using the asymptotic behavior (\ref{e6.7}) of $g_b(y) = g_t(y) = g(y)$ with $y \sim L / L_{\rm in} \gg 1$, we obtain
\be
\label{e8.2}
G   \simeq  \frac{2}{3} \left(1 +  \frac{2}{3} e^{-2y} \right), \qquad y \sim L / L_{\rm in} \gg 1 \,.
\ee
Thus, the two-terminal conductance of the device with a floating 1/3 mode has exactly the same asymptotic behavior in the incoherent regime 
(the limiting value 2/3, with a positive exponentially small correction) as in a system where both 1 and 1/3 modes are contacted, see Eq.~(\ref{e6.8}).

An analogous consideration yields the ohmic scaling of thermal conductance in the incoherent regime:
\begin{equation}
\label{e8.3}
G^Q=\frac{g^Q_t+g^Q_b-2 g^Q_t g^Q_b}{g^Q_t+g^Q_b-g^Q_t g^Q_b}\sim \frac{L_{\rm in}(T)}{L}.
\end{equation}
In the last equality we have used the asymptotic behavior (\ref{e7.12}) of $g^Q(L)$. 

\subsection{Coherent regime, $L<L_{\rm in}(T)$}
\label{s8.2}

Upon lowering the  temperature, the system goes over into the coherent regime, $L<L_{\rm in}(T)$, and 
the equivalence to the four-terminal setup breaks down. To study the transport properties under these conditions, it is convenient 
to present the setup with the floating mode in the equivalent form shown in Fig. \ref{Fig:SetupFloatingEquivalent}. 
Here we have extended the top and bottom $1/3$ modes to $\pm \infty$ but simultaneously added interaction between them 
for $|x|>L_A/2$ with $2L_A$ being the length of the $1/3$ loop in Fig. \ref{Fig:SetupFloating}. We take the interaction between 
$1/3$ modes at $|x|> L_A/2$ to be infinitely strong such that the bosonic reflection coefficient $r$ at the junction 
points $x=\pm L_A/2$ is equal to unity. The quadratic part of the Hamiltonian density of our system then reads 
[cf. Eq. (\ref{e3.1}); see also Fig. \ref{Fig:SetupFloatingEquivalent} for notations]
\begin{equation}
\pi{\cal H}(x)=\left\{
\begin{array}{cc}
\displaystyle
v_L \left[\rho_{Lt}^2+\rho_{Rb}^2\right]
+\frac{v_R}{t^2}\left[\left(\rho_{Rt}-r\rho_{Lb}\right)^2+\left(\rho_{Lb}-r\rho_{Rt}\right)^2\right],
& \hspace{-0.4cm}|x|>L_A/2;\\[0.5cm]
v_L\rho_{Lt}^2+v_R\rho_{Rt}^2+v_R\rho_{Lb}^2+v_L\rho_{Rb}^2\,, &\hspace{-1.4cm} L/2<|x|< L_A/2;\\[0.5cm]
\displaystyle 
\hspace{-0.1cm} \frac{v_\sigma}{\tilde{{\cal T}}^2}\left[\left(\rho_{Rt} +\tilde{\cal R} \rho_{Lt}\right)^2+\left(\rho_{Lb} +\tilde{{\cal R}} \rho_{Rb}\right)^2\right]+\frac{v_\rho}{\tilde{{\cal T}}^2} 
 \left[\left(\tilde{{\cal R}} \rho_{Rt}+\rho_{Lt}\right)^2+ \left(\tilde{{\cal R}} \rho_{Lb}+\rho_{Rb}\right)^2\right],  & \\ 
 &\hspace{-0.2cm} |x|< L/2. 
\end{array}
\right.
\label{HFloating}
\end{equation}
Here $t=\sqrt{1-r^2}$, and $r$ is eventually sent  to unity.   
{  Further, $\tilde{{\cal R}}$ and $\tilde{{\cal T}}=\sqrt{1-\tilde{{\cal R}}^2}$ are the bosonic reflection and transmission coefficients at the boundary between the interacting and non-interacting parts of the 2/3 edge.}
In the case of weak interaction, $\Delta>3/2$, the coefficient $\tilde{{\cal R}}$ is determined by the interaction in the middle part of the edge, 
Eq. (\ref{eRtilde}), while for $\Delta<3/2$ (and for sufficiently low temperatures, such that $L_T \gg \ell$)  we have $\tilde{{\cal R}}={\cal R}\equiv1/\sqrt{3}$ due to renormalization of  interaction by disorder. 

\subsubsection{Weak interaction, $\Delta > 3/2$}
\label{s8.2.1}

In the weak interaction regime and for $L< L_{\rm in}(T)$ the disorder can be neglected and the system is fully characterized by the bosonic scattering matrix.  The latter can be reconstructed from the transfer matrix 
\begin{equation}
\hat{T}=\hat{T}_4 \hat{T}_{L-L_A}   \hat{T}_3\hat{T}_L\hat{T}_2 \hat{T}_{L-L_A} \hat{T}_1 \,,
\label{TFull}
\end{equation}
where $\hat{T}_1$, $\hat{T}_2$, $\hat{T}_3$, and $\hat{T}_4$ describe the bosonic scattering  at $x=L_A/2, \, L/2,\, -L/2$, and $-L_A/2$ respectively, while $\hat{T}_L$ and  $\hat{T}_{L-L_A}$ describe the propagation of bosons
on the intervals $|x|<L/2$ and $L/2<|x|<L_A/2$, see Fig.~\ref{Fig:SetupFloatingEquivalent}.
The elementary transfer-matrices have the form (we set $\Delta L=L_A-L$)
\begin{eqnarray}
\label{e8.4}
\hat{T}_1=\hat{T}_4^{-1}=\frac{1}{t}\left(\begin{array}{cccc}
1 & 0 &0 &0\\
0 & 1& r & 0\\
0 & r & 1 & 0\\
0 &0&0 &1
\end{array}\right), \qquad 
\hat{T}_2=\hat{T}_3^{-1}=\frac{1}{\tilde{\cal T}}
\left(\begin{array}{cccc}
1 & \tilde{{\cal R}} &0 &0\\
\tilde{{\cal R}} & 1& 0 & 0\\
0 & 0 & 1 & \tilde{{\cal R}}\\
0 &0&\tilde{{\cal R}} &1
\end{array}\right),\label{T1234}\\[0.3cm]
\label{e8.4a}
\hat{T}_{L-L_A}=\diag
\left(\exp\left[i\omega\Delta L/2v_L\right], \, \exp\left[-i\omega\Delta L/2v_R\right], \,  \exp\left[i\omega\Delta L/2v_R\right], \, \exp\left[-i\omega\Delta L/2v_L\right]\right),\\[0.3cm]
\label{e8.4b}
\hat{T}_{L}=
\diag \left(\exp\left[i\omega L/v_\rho\right], \,\exp\left[-i\omega L/v_\sigma\right],\,
\exp\left[i\omega L/v_\sigma\right], \,\exp\left[-i\omega L/v_\rho\right]
\right).
\end{eqnarray}
A straightforward algebra leads now to the transmission amplitude  for the $1$ modes (in the limit $r\rightarrow 1$)
\begin{equation}
{\cal T}_{\rm tot}(\omega)=\frac{2i \tilde{{\cal T}}^2(\tilde{{\cal R}}^2\sin\omega\tau_1+\sin\omega\tau_2)}{\tilde{{\cal T}}^4-[e^{-i\omega\tau_2}-e^{i\omega\tau_1}\tilde{{\cal R}}^2]^2}, 
\qquad {\cal {\cal T}}_{\rm tot}(\omega=0)=1 \,,
\label{Eq:TransWeakGrivnin}
\end{equation}
where $\tau_1=L/v_\rho-\Delta L/v_R$ and $\tau_2=L/v_\sigma+\Delta L/v_R$.  

It follows from Eq. (\ref{Eq:TransWeakGrivnin}) that  
the electrical conductance of the system, determined by the transmission amplitude at zero frequency, 
equals unity  in the coherent regime  for any 
$\Delta>3/2$ (cf. Sec. \ref{s6.3.2}). On the other hand, the thermal conductance in the considered setup with a floating 1/3 mode is given by 
\begin{equation}
\label{e8.5}
G^Q=\frac{6}{\pi^2}\int_{0}^{\infty}\frac{\Omega d\Omega}{e^{\Omega}-1}\left|{\cal T}_{\rm tot}\left(\Omega T\right)\right|^2.
\end{equation} 
At  lowest temperatures, $T< \min(1/\tau_1,\, 1/\tau_2)$,  the thermal conductance $G^Q$ is universal and equals unity. At higher temperatures,
 (but still in the coherent regime), $G^Q$ crosses over to a 
non-universal interaction-dependent value, cf. Sec. \ref{Sec:CoherentWeak} where a similar behavior was found for the case of setup with both 1 and 1/3 modes contacted.  For the repulsive interaction that is sufficiently weak that the system is outside the basin of attraction of the Kane-Fisher-Polchinski fixed point, $0<c<0.34$, the thermal conductance $G^Q$ is then found to be in a numerically narrow  range
\begin{equation}
\label{e8.5a}
0.97<G^Q<1 \,.
\end{equation}
Note that if an attractive interaction ($c<0$, which corresponds to $\Delta>2$) between the $1$ and $1/3$ modes is allowed, the thermal conductance can take in the coherent regime with RG-irrelevant disorder an arbitrary value from the interval $0<G^Q<1$. 

\subsubsection{Strong interaction, $\Delta < 3/2$}
\label{s8.2.2}

Let us now turn to the case of strong interaction when  $\Delta<3/2$ in the middle part of the edges.  
It is clear that the gauge transformation used in Sec.  \ref{s3ALL} to collect the effect of disorder onto a single point 
can be applied to the bottom and top edges in Fig. \ref{Fig:SetupFloatingEquivalent} independently.
The system  is then characterized---in analogy with a setup with both 1 and 1/3 modes coupled to reservoirs, see Sec.~\ref{s5.4}---by two parameters $0<\theta_t,\, \theta_b<\pi$ describing the disorder in the 
top and bottom edges, respectively.  
At lowest temperatures, when the length of the interacting parts of the edge is much smaller then the temperature length, $L \ll L_T$, 
the resulting disorder is further renormalized down by the leads. As a result, we end up with the quadratic Hamiltonian (\ref{HFloating})
 where now $\tilde{{\cal R}}={\cal R}=1/\sqrt{3}$. Accordingly,  the electric and thermal  conductances are both equal to unity. 
 
 As the temperature becomes higher, the thermal length $L_T$ drops below the length $L$. In this situation,  the renormalization of 
 disorder by the leads  is absent and the system enters the regime of mesoscopic fluctuations (cf. Sec.~\ref{s6.3.1}). 
To understand the boundaries for the fluctuations, we study the exactly solvable limits when 
$\theta_{t}$ and $\theta_b$ are equal to 0 or $\pi$. 

In the case $\theta_t=\theta_b=0$ the bosonic transmission coefficient is given by Eq. (\ref{Eq:TransWeakGrivnin}) and the 
electrical conductance obviously equals unity.  As we are now at 
 relatively high temperatures, $L>L_T$, the thermal conductance of the device is however no longer determined by the 
 zero-frequency transmission coefficient but rather by its average value, see Eq.~(\ref{e8.5}).
 Assuming that the times $\tau_1$ and $\tau_2$ are 
 incommensurate, we get
 \begin{equation}
 \label{e8.6}
 G^Q\simeq 0.63 \,.
 \end{equation}

Suppose now that  the  configuration of disorder is  characterized by $\theta_t=\pi$ and 
$\theta_b=0$. Inspecting the action of the gauge transformation of Sec. \ref{s5.3} on the Hamiltonian (\ref{HFloating}), we find 
that in this situation the transfer-matrices  $\hat{T}_1$ and $\hat{T}_2$ in Eq. (\ref{TFull}) are given by [cf. Eq. (\ref{T1234})]
\begin{equation}
 \label{e8.7}
\hat{T}_1=\frac{1}{t}\left(\begin{array}{cccc}
1 & 0 &0 &0\\
0 & 1& -r & 0\\
0 & -r & 1 & 0\\
0 &0&0 &1
\end{array}\right), \qquad 
\hat{T}_2=\frac{1}{{\cal T}}
\left(\begin{array}{cccc}
1 & -{\cal R} &0 &0\\
-{\cal R} & 1& 0 & 0\\
0 & 0 & 1 & {\cal R}\\
0 &0&{\cal R} &1
\end{array}\right),
\end{equation}
while other transfer-matrices involved are not modified in comparison with Eqs.~(\ref{e8.4})--(\ref{e8.4b}). {  Let us note that the modes of the top edge that are used here as basis in the interval $[L/2, L_A/2]$ are $\tilde\phi_L$ and $\tilde\phi_R$, with the latter one carrying the charge $-1/3$, see Eq.~(\ref{e5.18a}). }
The transmission amplitude for the system is now found to be 
\begin{equation}
{\cal T}_{\rm tot}(\omega)=\frac{2 {\cal T}^2(-i{\cal R}^2\sin\omega\tau_1+\cos\omega\tau_2)}
{1-e^{2i\omega(\tau_1+\tau_2)}{\cal R}^4+e^{2i\omega\tau_2}{\cal T}^4}, 
\qquad {\cal T}_{\rm tot}(\omega=0)=1.
\label{Eq:TransPi0Grivnin}
\end{equation}
The electrical conductance is $G=1$ and the thermal conductance in this situation is
\begin{equation}
 \label{e8.8}
G^Q\simeq 0.75.
\end{equation}

Finally, an analogous consideration for the case when both edges realise the case of the strongest possible impurity, 
$\theta_t=\theta_b=\pi$ yields
\begin{equation}
{\cal T}_{\rm tot}(\omega)=-\frac{2i {\cal T}^2({\cal R}^2\sin\omega\tau_1-\sin\omega\tau_2)}{{\cal T}^4-[e^{-i\omega\tau_2}+e^{i\omega\tau_1}{\cal R}^2]^2}, 
\qquad {\cal T}_{\rm tot}(\omega=0)=0 \,.
\label{Eq:TranspPiPiGrivnin}
\end{equation}
The electric conductance in this case equals $0$ while the thermal conductance is given by Eq.~(\ref{e8.6}),
$G^Q\simeq 0.63$.

We thus conclude that in the case of strong interaction, 
$\Delta<3/2$, the electric conductance of the system experiences in the intermediate  temperature range, $L_T<L<L_{\rm in}(T)$,  strong mesocopic fluctuations within the (maximally wide) interval $0<G<1$. Under the same conditions, the thermal conductance of the device, 
$G^Q$, also shows mesoscopic fluctuations. The above solution of the limiting cases implies that all values satisfying $0.63 < G^Q < 0.75$ are allowed; it is plausible that this is exactly the interval of fluctuations.   

In this analysis of the case of strong interaction, we have assumed that the temperature remains sufficiently low, $L_T \gg \ell$, so that the interaction is renormalized by disorder from an initial value $1 < \Delta_0<3/2$ to a vicinity of the Kane-Fisher-Polchinski fixed point, $\Delta=1$.
In the opposite case of higher temperatures, $L_T \ll \ell$, the same analysis as in the case of weak interaction applies, see the analogous comment  
in the end of Sec.~\ref{s6.3.1} and \ref{Sec:CoherentStrong}.  The thermal conductance is then given by 
Eqs. (\ref{Eq:TransWeakGrivnin}),  (\ref{e8.5}), and (\ref{eRtilde}) with unrenormalized value of the interaction, $0.34<c<0.98$. In the intermediate temperature range, $L_T<L<L_{\rm in}(T)$, this implies
\begin{equation}
0.28<G^Q<0.97 \,.
\end{equation}
{  The electric conductance remains equal unity, $G\simeq 1$, in this regime.}

\begin{figure}
\includegraphics[width=180pt]{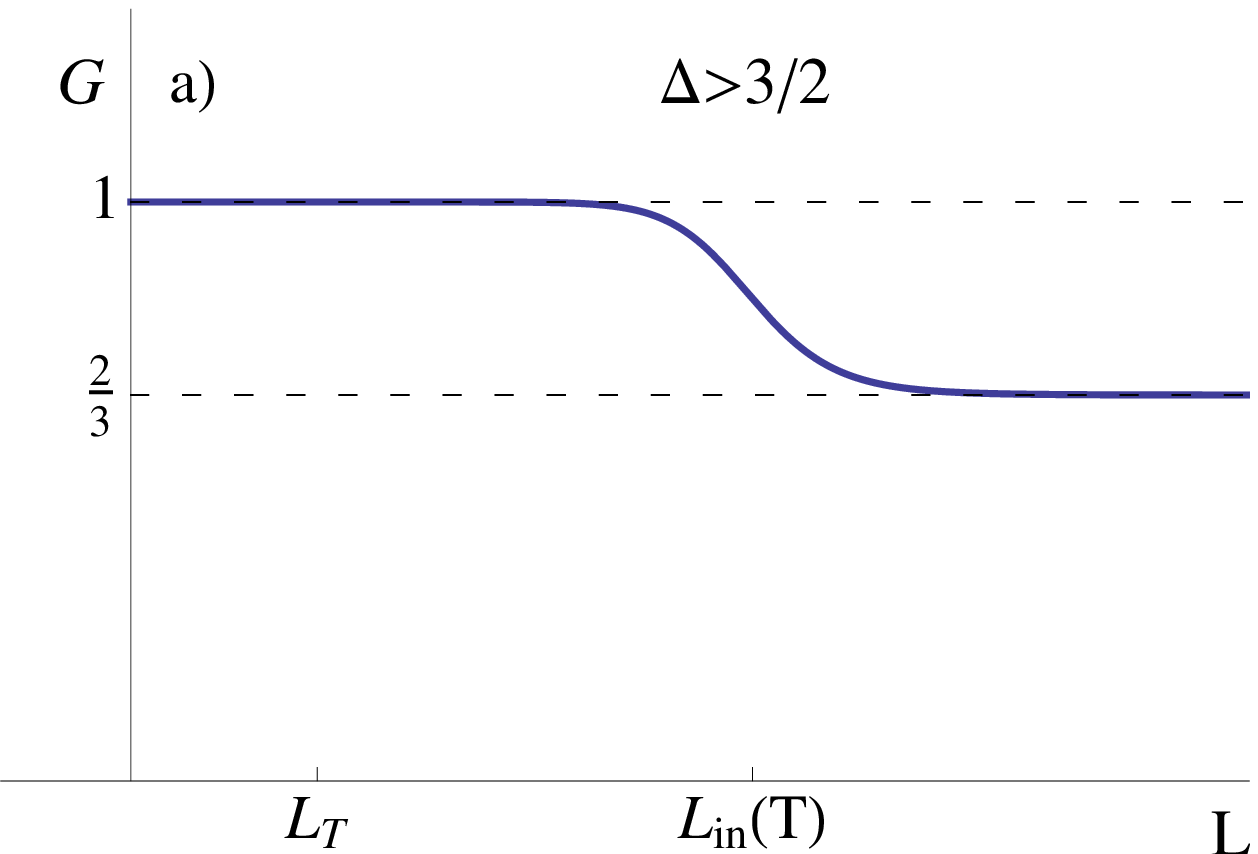}\hspace*{0.5cm}
\includegraphics[width=180pt]{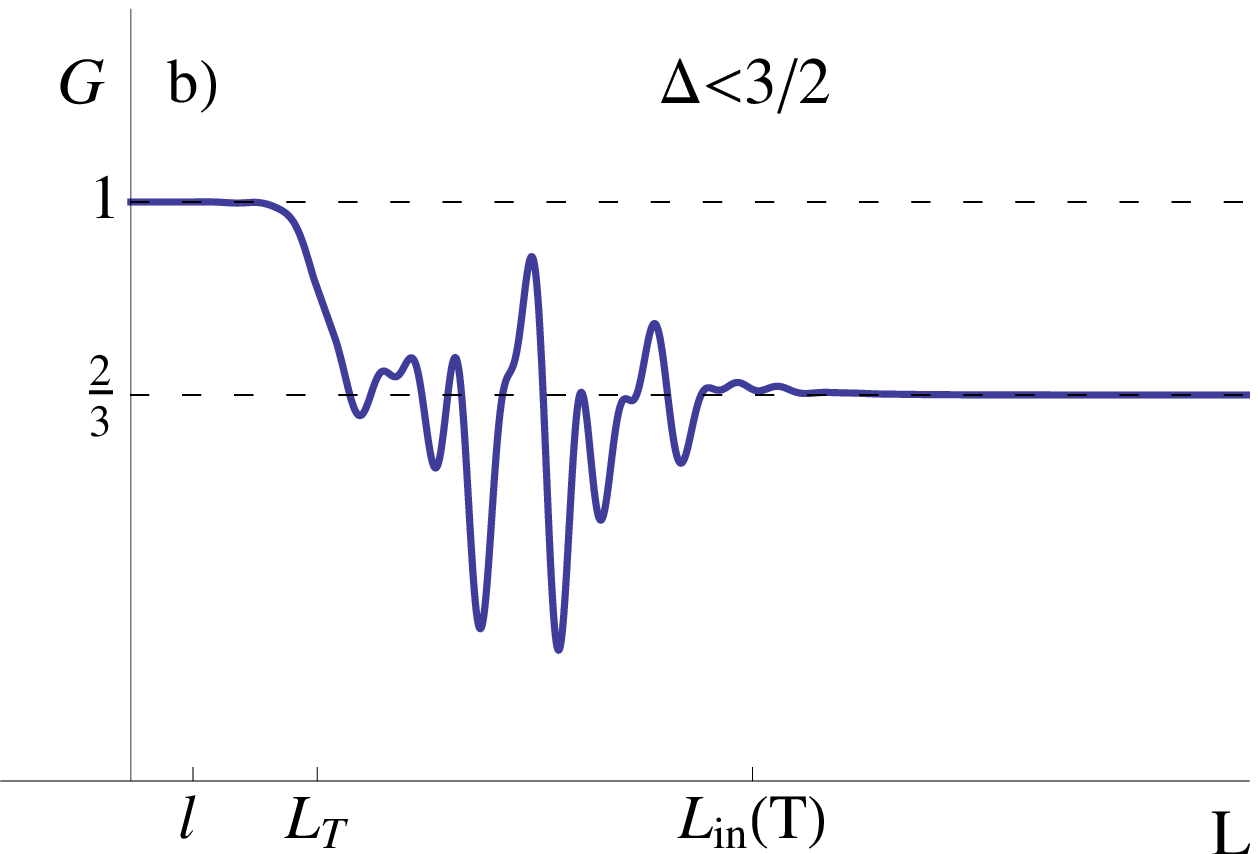}\\
\includegraphics[width=180pt]{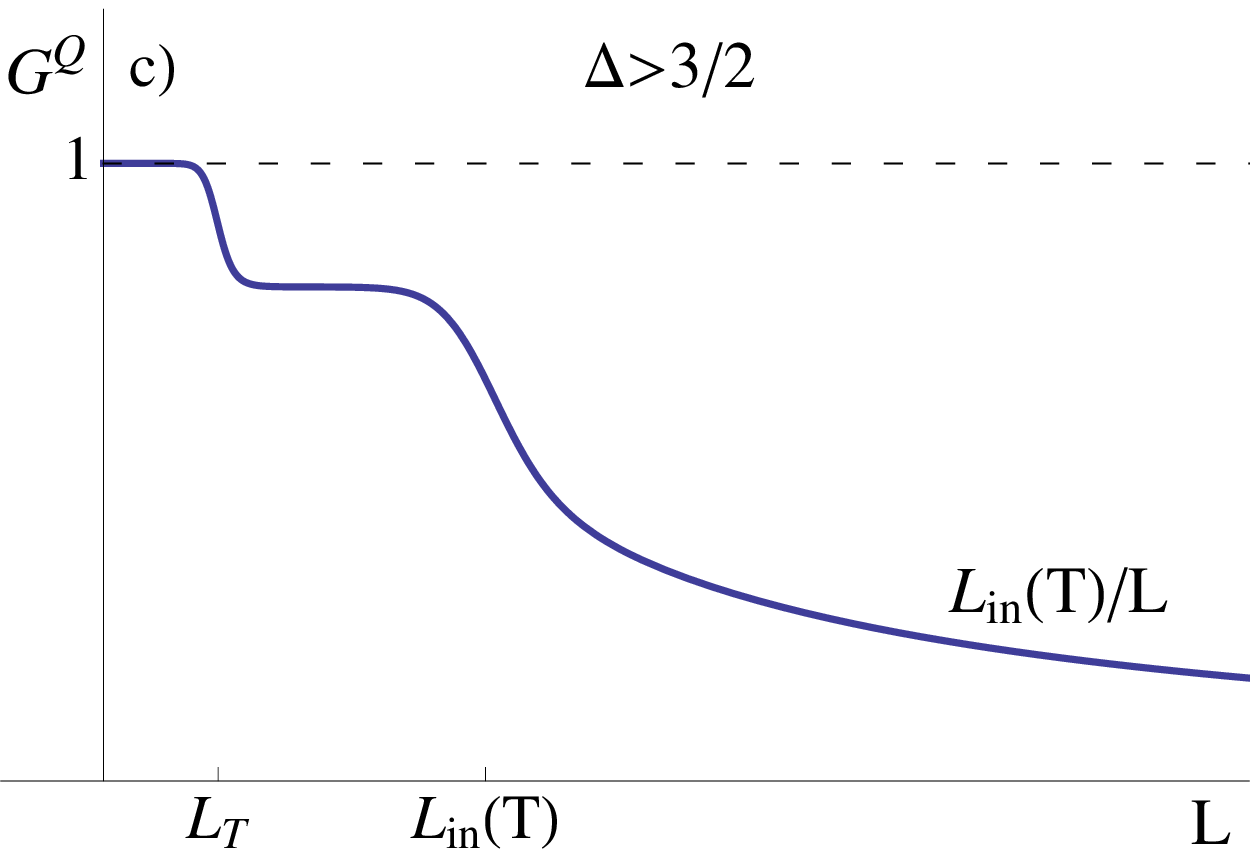}\hspace*{0.5cm}
\includegraphics[width=180pt]{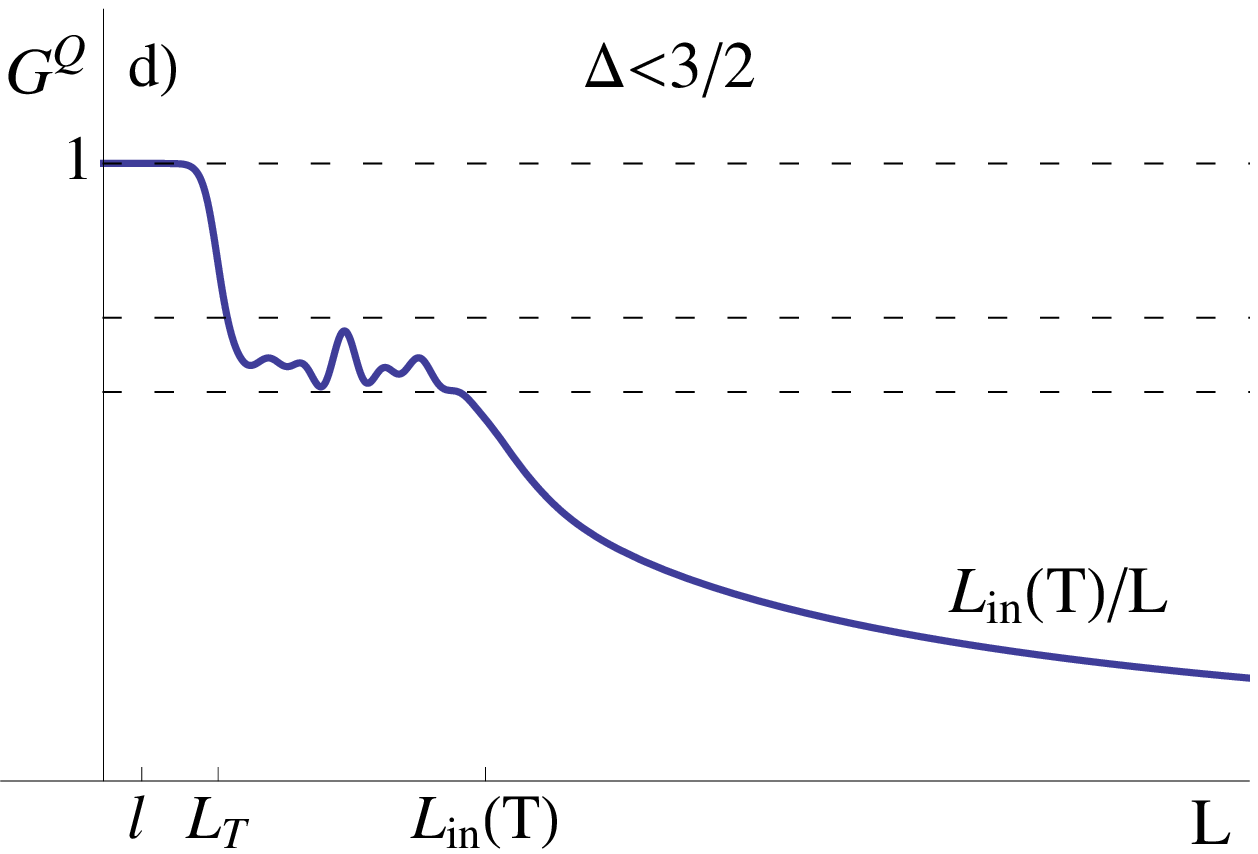}
\caption{Transport properties of a setup with floating $1/3$ mode: Electric ($G$) and thermal ($G^Q$) conductances as functions of the length $L$ of the system.
a) and c):  weak interaction, $\Delta > 3/2$.  
b) and d): strong interaction, $\Delta<3/2$, under the assumption of sufficiently low temperature, $L_T \ll \ell$.   For strong interaction and higher temperatures, $L_T \gtrsim \ell$, the behavior is qualitatively the same as for weak interactions, panels a) and c). }
\label{GEGThGrivnin}
\end{figure}

Figure \ref{GEGThGrivnin} summarises our findings on the length dependence of the electric 
and thermal conductances of a setup with floating $1/3$ 
mode, Fig.~\ref{Fig:SetupFloating}, including both cases of weak and strong interaction. 

\section{Summary and outlook}
\label{s9}

To summarize, we have studied electric and thermal transport properties of a $\nu=2/3$ FQHE junction. We have assumed that the $\nu=2/3$ edge in the middle part of the device is described by counterpropagating 1 and 1/3 modes coupled by interaction and random tunneling, while the leads (coupled to external reservoirs) are characterized by  non-interacting 1 and 1/3 modes, without tunneling between them, see Figs.~\ref{Fig:Setup}, \ref{Fig:Setup-All}.
Our central goal was to explore  the dependence of the electric and thermal two-terminal conductances, $G$ and  $G^Q$ on the   system size $L$ and temperature $T$. We have performed this analysis both for the case of strong interaction between the 1 and 1/ 3 modes (when the low-temperature physics of the interacting segment of the device is controlled by a vicinity of the strong-disorder Kane-Fisher-Polchinski fixed point) and for relatively weak interaction, for which disorder is irrelevant at $T=0$ in the renormalization-group sense. This has allowed us to compare the transport properties in both cases and to understand the similarities and the differences between them.  The main results of this analysis are as follows (see Figs. \ref{Fig:Evolution}, \ref{Fig:Evolution1}, \ref{ThermalCondPlot}):

\begin{enumerate}

\item For a sufficiently small system size $L$, the electric conductance $G(L)$ is close to 4/3 (in units of $e^2/h$), while the thermal conductance  $G^Q$ is close to 2 (in units of $\pi T / 6 \hbar$), independently of the interaction strength. 

\item For large system size, $L > L_{\rm in}$, the system is in an incoherent regime, with $G$ given by 2/3 and $G^Q$ showing Ohmic scaling, $G^Q\propto 1/L$, again for any interaction strength.

\item The hallmark of the strong-disorder fixed point is the emergence of an intermediate range of $L$, in which the electric  conductance shows strong mesoscopic fluctuations (in an interval limited by the values 4/3 and 1/3) and the thermal conductance is $G^Q=1$. 

\item Similar analysis has been performed  also for a modified device where 1/3 mode is not coupled to external reservoirs, see Fig.~\ref{Fig:SetupFloating}. The corresponding results are schematically shown in Fig.~\ref{GEGThGrivnin}.

\end{enumerate}

Below we expand upon a comparison of these results to experimental observations and discuss further {  possible} extensions of our work.

\subsection{Comparison to experiment}

Let us compare our results with existing experimental findings. Almost all measurements of the electric two-point conductance have yielded the value 2/3, see, e.g., Refs.~\cite{Chang92,bid09}. A comparison  with the theoretical results indicate that these samples were in the incoherent regime,  $L > L_{\rm in}(T)$.  
The system sizes in these experiments were $L \ge 30\:\mu$m. We thus conclude that the inelastic length $L_{\rm in}(T)$ was shorter in these devices at temperatures used in the experiment. Let us emphasize that the value 2/3 is characteristic for the incoherent regime both in the case of strong interaction (basin of attraction of Kane-Fisher-Polchinski fixed point) and relatively weak interaction. Therefore, these measurements do not allow one to distinguish between these two {  scenarios}.

The only experiment where values of the electric conductance essentially different from the incoherent value 2/3 were measured, is, to our knowledge, Ref.~\cite{Grivnin14}. Devices studied in that work had only the mode 1 contacted, which is the geometry studied in Sec.~\ref{s8} above. It was found that for system size $L=40\:\mu$m the conductance takes its incoherent value 2/3 at temperature $T\simeq 35\:$mK. On the other hand, samples with   $L=4\:\mu$m and $L=0.4\:\mu$m showed a very different behavior, with conductance values 0.73 and 0.93, respectively. 
Such a departure from the value 2/3 for small systems, $L\lesssim L_{\rm in}(T)$,  is exactly what is expected in the theory.
This suggests that the equilibration length was in the interval between 4 and 40$\:\mu$m, so that the samples of length $L=4\:\mu$m and $L=0.4\:\mu$m were in the coherent regime. (The sample with $L=4\:\mu$m might also be in the crossover.) Unfortunately, the experimental measurements are insufficient to distinguish between three scenarios (see Fig.~\ref{GEGThGrivnin}): 
\begin{enumerate}
\item[(S1)]
strong interaction, $\Delta < 3/2$, and low temperature, $L_T \gg \ell$, 
\item[(S2)]
strong interaction, $\Delta < 3/2$, and higher temperature, $L_T \lesssim \ell$, 
\item[(S3)] weak interaction, $\Delta > 3/2$.
\end{enumerate} 

In a very recent preprint \cite{Banerjee16}, the thermal two-point conductance was measured for several FQHE states in a system of size $L = 150\:\mu$m. The results for the $\nu=2/3$ state were $G^Q=0.33$ and $G^Q=0.25$ for temperature ranges  $10 - 30\:$mK and $30 - 50\:$mK, respectively. These values are six and eight times smaller than the maximal (ballistic) value $G^Q=2$, which yields, in view of 
Eq.~(\ref{GThInc}) for the thermal conductance in the incoherent regime, the following estimates for the equilibration length:
$L_{\rm in} \sim 25\:\mu$m for $T\simeq 20\:$mK and $L_{\rm in} \sim 20\:\mu$m for $T\simeq 40\:$mK.  These values are consistent with the above boundaries for the equilibration length at comparable temperatures. Further, these results for the equilibration length indicate a rather slow variation of $L_{\rm in}$ with temperature, $L_{\rm in} \propto T^{-p}$, with $p \simeq 0.3 - 0.4$. This scaling is clearly inconsistent with the above scenarios S1 [for which $p=2$, see Eq.~(\ref{e6.2})] and S3 [for which $p=2\Delta-2$, see Eq.~({\ref{e6.2b}), is in the range between 1 and 2]. Thus, the experimentally observed value of $p$ points out to the regime S2, in which $p=2\Delta_0-2$, see Eq.~(\ref{e6.2-high}), can take any value between the 0 and 1, depending on $\Delta_0$. We thus conclude that the bare value of the interaction strength corresponds to $\Delta_0 = 1.15 - 1.2$, so that the theory would flow in the Kane-Fisher-Polchinski fixed point at sufficiently low temperature. On the other hand, the temperatures in the experiment \cite{Banerjee16} were apparently not low enough from this point of view: the system was in the regime $L_T \lesssim \ell$. 

The above conclusion on the value of the interaction strength, $\Delta_0 = 1.15 - 1.2$, demonstrates that the very interesting physics of the mesoscopic regime associated with the Kane-Fisher-Polchinski fixed point can be within experimental reach. On the other hand, this remains a highly challenging experimental task, since very low temperatures (below $10\:$mK) are needed, at least for samples with the same strength of random tunneling (i.e., with the same value of the length $\ell$) as in Ref.~\cite{Banerjee16}. {  Alternatively, samples with stronger disorder (i.e., smaller $\ell$) would be favorable for reaching the mesoscopic regime.}  More generally, a systematic experimental study of the length and temperature dependence of the electric and thermal conductances---which would permit also a more systematic comparison of the experiment to the theory---would be of great interest. 

\subsection{What about localization effects?}

It is worth mentioning that we have discarded any localization-type effects in our above analysis. In conventional (non-chiral) one-dimensional disordered systems the Anderson localization plays a major role, and its interplay with the interaction-induced dephasing governs the temperature-dependence of conductivity \cite{gornyi05}. It is therefore natural to ask what are possible implications of Anderson localization for the problem of transport in FQHE edges.  {  As shown in Sec.~\ref{s2.3} on the basis of standard relations between the current emanating from reservoirs to their electrochemical potentials,}
the electric conductance is bounded from below, see Eq.~(\ref{e2.22a}). This implies that the charge transport is necessarily ballistic, so that a strong localization of charge is strictly excluded. It remains to be seen whether quantum-interference corrections of the weak-localization type may be detectable in any of the regimes studied in our work. 

Contrary to the charge transport, the energy transport through disordered $\nu=2/3$ edges may be in principle susceptible to localization effects, since there is an equal number of modes (one) propagating in both directions, which is reflected in the zero lower bound for the thermal conductance, see Eq.~(\ref{e7.1}). We do not expect, however, any essential localization-induced modifications of the analysis of the energy transport in a $\nu=2/3$ FQHE edge with random tunneling (Sec.~\ref{s7}). Indeed, the localization might become strong in the regime where the heat transport is diffusive, i.e., for $L > L_{\rm in}$. The length $L_{\rm in}$ plays a role of the mean free path for the energy transport. For elastic scattering in 1D systems, the backscattering mean free path yields the localization length. However, in the present situation, the scattering is inelastic, i.e., $L_{\rm in}$ serves simultaneously as a dephasing length.  Thus, the effect of the localization in the energy transport in the incoherent regime is expected to be limited to a renormalization of a numerical coefficient $C$ in Eq.~(\ref{GThInc}) by a factor of order unity. 

The effect of localization in the energy transport can be, however, more pronounced if another type of randomness is included: spatial fluctuations in the strength of interaction between the 1 and 1/3 modes. In the absence of tunneling, this type of disorder will lead to localization of bosonic modes (plasmons) with the localization length $\xi_\omega$ scaling as $\xi_\omega \propto 1/\omega^2$ with the frequency $\omega$. 
The characteristic energy of bosons contributing to heat transport is $\omega \sim T$. We thus have to compare $\xi_T$ with the inelastic length $L_{\rm in}(T)$ due to random tunneling. If there is a range of temperatures such that   $\xi_T \ll L_{\rm in}(T)$,  localization will strongly suppress the energy transport in this range. Let us note, however, that this suppression will be only of power-law type (and not exponential), since bosons with low frequencies will escape localization. A detailed study of the interplay of localization of plasmons and their inelastic scattering processes in the context of a disordered $\nu=2/3$ edge in such a regime remains an interesting prospect for future research \cite{fazio98}. 

\subsection{Generalization to other filling factors}

Our analysis can be generalized to transport properties at disordered FQHE edges with counterpropagating modes at other filling fractions; the classification of the corresposnding fixed points was worked out in Ref.~\cite{moore98}. Fractions with two modes propagating in opposite directions are $\nu = n/(nl+1)$, with $n, l > 0$,  where $l$ is odd and $n$ is even. Disorder can be relevant only for the states with $n=2$, i.e., $\nu = 2/3,$ 2/7, 2/11, \ldots. All these fractions are characterized by the existence of a strong-disorder attractive fixed point with $SU(2)$ symmetry in addition to a line of clean fixed points, in analogy with the case of $\nu=2/3$. Thus, the results for systems at these filling factors will be fully analogous to those at $\nu=2/3$ studied in the present paper. For $n>2$ (i.e., $\nu = 4/5,$ 4/13, 6/7, \ldots) disorder is irrelevant, so that only the behavior analogous to that for a  $\nu=2/3$ system in the  case of weak interaction $\Delta > 3/2$ is possible. 

Among systems with $n>2$ modes, a special role is played by fractions which have strong-disorder fixed points with all $n-1$ neutral modes propagating in the direction opposite to the charge mode. These states have filling factors $\nu = n/[n(l+1)-1]$ with odd $l>0$. For the minimal value $l=1$ this yields the series $\nu= 3/7,$ 4/9, \ldots...  The strong-disorder fixed points in these systems (analogous to the SU(2)-symmetric fixed point in the case of $\nu=2/3$)  possess a SU(n) symmetry \cite{kane-fisher95}.  A complete classification of attractive fixed points \cite{moore98} includes also higher-dimensional manifolds of fixed points with lower symmetry. For example, the $\nu=3/5$ disordered edge is characterized by three types of attractive fixed points: (i) an SU(3) symmetric fixed point where all impurity operators are relevant, (ii) lines of SU(2) symmetric fixed points where only one impurity operator is relevant but others are irrelevant, and (iii) two-dimensional parameter space of clean fixed points. The analysis performed in our work can be naturally extended to the $\nu=3/5$ systems in basins of attractions of these different fixed points:

\begin{enumerate}

\item[(i)]
When the bare interactions are such that the theory flows into the SU(3)-symmetric fixed point, the results will be very similar to those  for a $\nu=2/3$ system with $\Delta_0 < 3/2$ (i.e. flowing into the SU(2) fixed point). 

\item[(ii)]
The behavior of the $\nu=3/5$ system with no relevant disorder will be similar to that of the $\nu=2/3$ system in the analogous regime $\Delta > 3/2$. 

\item[(iii)]
In the intermediate situation of a $\nu=3/5$ system corresponding to a line of fixed points with only one disorder operator being relevant, the behavior will combine features of two regimes found in the case of $\nu=2/3$. Specifically, the electric conductance is expected to show a behavior analogous to that shown in Fig.~\ref{Fig:Evolution}, with a regime of strong mesoscopic fluctuations that is a hallmark of the coherent transport at  exactly solvable strong-disorder fixed points. On the other hand, the thermal conductance is expected to show in this mesoscopic regime a plateau with a non-universal value depending on the position on the fixed-point line (in analogy with the right panel of Fig.~\ref{ThermalCondPlot} corresponding to the case of a line of clean fixed points of a $\nu=2/3$ system). In view of the mixed character of this type of fixed points, the equilibration in this case is expected to be characterized by two lengths, analogous to those of Eq.~(\ref{e6.2}) and (\ref{e6.2b}).

\end{enumerate}

The behavior in the incoherent (fully equilibrated) regime is again fully universal, i.e., it does not depend on which of the above three fixed-point types is realized. Specifically, the electric two-terminal conductance is equal to $\nu=3/5$, with exponentially small corrections, as in Eq.~(\ref{e6.8}). The thermal conductance is given asymptotically by the absolute value of the difference between the number of left-moving and right-moving modes, which is $2-1=1$. The approach to this value with increasing $L$ (or temperature) is exponential, in full analogy with the electric conductance. 

Fractions with $n>2$ modes and with counterpropagating neutral modes show even more complex fixed-point structure \cite{moore98}, including mutliple isolated fixed points with non-equivalent properties (in addition to fixed-point lines, planes, etc.). It would be interesting to extend our analysis of the $L$ and $T$ dependences of the electric and thermal conductances to these situations. 

{  Finally, it is worth reminding the reader that we have focused in this paper on the ``minimal model" of the $\nu=2/3$ edge, i.e., on that with a minimal number of modes, $n=2$.  As has been mentioned in Sec.~\ref{s1}, for a smooth confining potential additional pair(s) of counterpropagating 1/3 modes may emerge \cite{chamon94,meir94,wang13}.  It may be interesting to extend our analysis to such models as well. This comment applies also to other filling fractions: the numbers $n$ of modes quoted above refer to corresponding ``minimal models''.}

\section{Acknowledgments}

We thank I. Gornyi, A. Grivnin, M. Heiblum, C.~Nosiglia, J.~Park, and D. Polyakov for useful discussions.  ADM acknowledges the support within the Weston Visiting Professorship at the Weizmann Institute of Science. Y.G. acknowledges support from the CRC 183 Project of the DFG, the DFG grant RO 2247/8-1, and the IMOS Israel-Russia program.

\appendix

 \section{Periodic boundary conditions, mode expansion, and compactification radius for chiral boson fields}
 \label{a2}
 
The fields $\phi_R$ and $\phi_L$ are chiral bosonic fields that correspond to 1/3 and 1 modes, respectively, and satisfy standard commutation relations, see Eqs.~(\ref{e2.11})--(\ref{e2.13}).
 In this Appendix, we present the mode expansion for these fields in a system with periodic boundary conditions. 
 The mode expansion for the field $\phi_R$ reads
\begin{equation}
\phi_R(x)=\frac{2\pi}{L}\frac{N_R}{\sqrt{3}}x-\sqrt{3}\chi_R +\frac{1}{i}\sum_{q> 0}\sqrt{\frac{2\pi}{L q}}\left[e^{iq x}b_q-e^{-iq x}b_q^+\right].
\end{equation}
Here the operators  $b_q^\dagger$, $b_q$ are creation and annihilation operators for modes with non-zero momenta $q = 2\pi n/L$ (where $n$ is integer); they satisfy  the standard bosonic commutation relations, $[b_q,b_{q'}^\dagger]=\delta_{qq'}$. The fields $N_R$ and $\chi_R$ correspond to the zero mode and satisfy the commutation relation
\be
[\chi_R, N_R] = i \,.
\ee
In other words, the operator $e^{i\chi_R}$ raises $N_R$ (the number of $1/3$-quasiparticles) by 1.
In view of periodic boundary conditions and since $N_R$ is integer, the field $\phi_R$ is defined up to an integer multiple of $2\pi/\sqrt{3}$. 
One thus says that $\phi_R$ has $1/\sqrt{3}$ ``compactification radius''.  This corresponds to the fact that a ``good'' tunneling operator is $e^{\pm i \sqrt{3} \phi_r}$.

Similarly, the mode expansion of the left-moving field $\phi_L$ reads
\begin{equation}
 \phi_L(x)=\frac{2\pi}{L}N_L x+\chi_L -\frac{1}{i}\sum_{q< 0}\sqrt{\frac{2\pi}{L |q|}}\left[e^{iq x}b_q-e^{-iq x}b_q^+\right].
\end{equation}
The corresponding commutation relations have the same form as for the right-moving field, $[b_q,b_{q'}^\dagger]=\delta_{qq'}$ and $[\chi_L, N_L] = i$.
The compactification radius is 1.

According to Eq.~(\ref{e2.14}), the neutral mode has the mode expansion with $1/\sqrt{2}$ compactification radius,
\begin{equation}
 \phi_\sigma=\frac{2\pi}{L}\frac{N_\sigma}{\sqrt{2}}x-\sqrt{2}\chi_\sigma+\ldots
\end{equation}
with
\begin{equation}
 N_\sigma=N_R+N_L\,, \qquad \chi_\sigma=(3\chi_R-\chi_L)/2 \,, \qquad [\chi_\sigma, N_\sigma] = i \,.
\end{equation}

\section{Current algebra}
\label{App:CommRelations}

In this  Appendix we elaborate on the hidden symmetry of the neutral mode and show that  the operators 
\begin{eqnarray}
   \label{e4.2a}
 J^z (x)&=& \frac{1}{2\pi \sqrt{2}}\partial_x \phi_\sigma(x) \,,\\
 J^{\pm}(x) \equiv  J^x(x) \pm i J^y(x)&=& \frac{1}{2\pi a}e^{\pm i\sqrt{2}\phi_\sigma(x)} \,.
    \label{e4.3a}
\end{eqnarray}
satisfy the $su(2)_1$ Kac-Moody algebra, Eqs.~(\ref{e4.4}), (\ref{e4.5}). We also demonstrate the  identity
\begin{equation}
\frac{1}{4\pi} \int dx(\partial_x\phi_\sigma)^2=\frac{2\pi}{3}\int dx J^2(x), \qquad J^2(x)=\left[J^x(x)\right]^2+\left[J^y(x)\right]^2+\left[J^z(x)\right]^2
\label{eB3}
\end{equation}
used in the derivation of Eq. (\ref{e4.9}) from Eq. (\ref{e4.1}).

The commutation relations between $J^\pm$ and $J^z$ follow immediately from the commutator of the field $\phi_\sigma$ and the rule valid for arbitrary gaussian fields $A$ and $B$
\begin{equation}
[e^{A}, B]=e^{A}\left[A, B \right].
\end{equation}
To compute the commutator of $J^\pm$ we use  first the 
 equal-time Green's function of the field $\phi_\sigma$,
\begin{equation}
\langle\phi_\sigma(x)\phi_{\sigma}(x^\prime)\rangle=-\ln\frac{2\pi[a-i(x-x^\prime)]}{L},
\label{CorFun}
\end{equation}
to get the 
normal-ordered form of the currents  $J^{\pm}(x)$:
\begin{equation}
J^{\pm}(x)=\frac{1}{ L}:e^{\pm i\sqrt{2}\phi_\sigma(x)}:.
\label{JpmNorm}
\end{equation}
Further, for arbitrary Gaussian fields $A$ and $B$ we have
\begin{equation}
[:e^{A}:, :e^B:]=:e^{A+B}:\left[e^{\langle AB\rangle}-e^{\langle BA\rangle}\right].
\end{equation}
Thus,
\begin{equation}
[J^{+}(x), J^-(x^\prime)]=\frac{1}{(2\pi)^2}:e^{i\sqrt{2}[\phi_\sigma(x)-\phi_\sigma(x^\prime)]}:\left(\frac{1}{\left[a-i(x-x^\prime)\right]^2}-\frac{1}{\left[a+i(x-x^\prime)\right]^2}\right).
\label{eB7}
\end{equation}
The expression in brackets in the right-hand side of Eq. (\ref{eB7})  vanishes in the limit $a\rightarrow 0$ unless $x=x^\prime$. This allows us to expand the normal-ordered exponent in powers of $x-x^\prime$ up to first order: 
\begin{equation}
:e^{i\sqrt{2}[\phi_\sigma(x)-\phi_\sigma(x^\prime)]}:=1+i\sqrt{2}(x-x^\prime)\partial_x\phi_\sigma\equiv1+4\pi i (x-x^\prime)J^z(x) \,.
\end{equation}
Taking now  into account that 
\begin{eqnarray}
\lim_{a\to 0}\left[ \frac{1}{a-i(x-x^\prime)}+\frac{1}{a+i(x-x^\prime)} \right] =2\pi \delta(x-x^\prime),\\[0.2cm]
\lim_{a\to 0}\left[ \frac{1}{\left[a-i(x-x^\prime)\right]^2}-\frac{1}{\left[a+i(x-x^\prime)\right]^2} \right]  = -2\pi i\delta^\prime(x-x^\prime),
\end{eqnarray}
we find
\begin{equation}
[J^{+}(x), J^-(x^\prime)]= -\frac{i}{2\pi}\delta^\prime(x-x^\prime)+2J^z(x)\delta(x-x^\prime),
\end{equation}
which is Eq.~(\ref{e4.5}).

Let us now demonstrate that Eq. (\ref{eB3}) holds. Obviously, the only non-trivial part is to compute $\left[J^x(x)\right]^2+\left[J^y(x)\right]^2$. We start by considering the object
\begin{equation}
J^x(x)J^x(x^\prime)+J^y(x)J^y(x^\prime)=\frac12\left[J^+(x)J^{-}(x^\prime)+J^-(x)J^{+}(x^\prime)\right]
\label{b14}
\end{equation}
that reduces to $\left[J^x(x)\right]^2+\left[J^y(x)\right]^2$ in the limit $x^\prime\rightarrow x$. 
Using Eqs. (\ref{JpmNorm}) and (\ref{CorFun}), we find
\begin{equation}
J^+(x)J^{-}(x^\prime)+J^-(x)J^{+}(x^\prime)=\frac{1}{(2\pi)^2}\left[:e^{i\sqrt{2}[\phi_\sigma(x)-\phi_\sigma(x^\prime)]}:+:e^{-i\sqrt{2}[\phi_\sigma(x)-\phi_\sigma(x^\prime)]}:\right]\frac{1}{\left[a-i(x-x^\prime)\right]^2}.
\label{b15}
\end{equation}
In the limit $x\rightarrow x^\prime$, it is sufficient to expand the normal-ordered exponents up to the second order in $x-x^\prime$:
\begin{equation}
:e^{i\sqrt{2}[\phi_\sigma(x)-\phi_\sigma(x^\prime)]}:+:e^{-i\sqrt{2}[\phi_\sigma(x)-\phi_\sigma(x^\prime)]}:=-2 (x-x^\prime)^2\left(\partial_x\phi_\sigma\right)^2 \,.
\label{b16}
\end{equation}
It follows now from Eqs. (\ref{b14}), (\ref{b15}) and (\ref{b16}) that, in the limit $a\rightarrow 0$,
\begin{equation}
\left[J^x(x)\right]^2+\left[J^y(x)\right]^2=\frac{1}{(2\pi)^2}\left(\partial_x\phi_\sigma\right)^2=2\left[J^z(x)\right]^2.
\end{equation}
Thus, 
\begin{equation}
\frac{2\pi}{3}\int dx J^2(x)=2\pi \int dx \left[J^z(x)\right]^2=\frac{1}{4\pi} \int dx(\partial_x\phi_\sigma)^2 \,,
\end{equation}
which is Eq.~(\ref{eB3}).

\section{Kubo formula and conductance}
\label{App:Conductance}

In this Appendix we derive the results for the parameter $G_{12}$ of an edge  in the two fixed points characterized by $\theta=0$ and $\theta=\pi$, see Sec.~\ref{s5.4}.    The two-terminal conductance $G$ is determined by the values of $G_{12}^{(\mu)}$ of both edges according to Eq.~(\ref{e2.20}). 

We consider first an  edge of our sample with $\theta=0$ and and compute the currents induced in response to a potential $V(x, t)$ 
We assume the applied potential to be constant inside the leads, $\partial_x V(x, t)=0$ for $|x|>L/2$, where it takes values $V(x>L/2)=V_2$ and $V(x<-L/2)=V_1$. The physical charge density $\rho(x)$ and the correction to the Hamiltonian $\delta H$ due to the coupling of the system to the external field  are given by
\begin{eqnarray}
\rho(x)=\frac{1}{2\pi}\left(\partial_x\phi_L+{\cal R}\partial_x\phi_R\right)=\frac{{\cal T}}{2\pi}\partial_x\phi_\rho,\\
\delta H=\frac{{\cal T}}{2\pi}\int dxdt V \partial_x \phi_\rho.
\end{eqnarray}
We will be interested in the current in the right lead, $x> L/2$ as defined  from the continuity equation
\begin{equation}
\hat{j}=-\frac{1}{2\pi}v_L\partial_x \phi_L+\frac{{\cal R}v_R}{2\pi}\partial_x\phi_R-\frac{{\cal T}^2}{2\pi}V_2\,.
\label{currentBasic}
\end{equation}
The last term in the expression for current is the manifestation of the chiral anomaly. 

The Kubo formula gives 
\begin{equation}
j(x,\omega)=-\frac{{\cal T}^2}{2\pi} V(x,\omega)-\frac{{\cal T}}{2\pi}\int_{-L/2}^{L/2}\partial_y V(y, \omega)\int\frac{d\omega^\prime}{2\pi}\frac{\left[\hat{j}(x), \phi_\rho(y)\right]_{\omega^\prime}}{\omega-\omega^\prime +i0}
\label{Kubo}
\end{equation}
with
\begin{equation}
\left[\hat{j}(x, \omega), \phi_\rho(y, \omega^\prime)\right]\equiv 2\pi \delta(\omega+\omega^\prime)
\left[\hat{j}(x), \phi_\rho(y)\right]_{\omega}.
\end{equation}
The fields $j(x, \omega)$ and $\phi_\rho(y, \omega)$ can be expressed in terms of the incoming left and right fields $\Phi_L$ and $\Phi_R$. Specifically, we find for $\phi_\rho(y, \omega)$ with $|y|<L/2$ 
\begin{equation}
\phi_\rho(y, \omega)=\frac{{\cal T}e^{-i(y/v_\rho+\delta_{\rho L})\omega}}{1-{\cal R}^2e^{2i\omega\tau}}\Phi_{L}(0, \omega)+\frac{{\cal T}{\cal R}e^{-i(y/v_\rho+\delta_{\rho R})\omega}}{1-{\cal R}^2e^{2i\omega\tau}}\Phi_R(0, \omega), 
\end{equation}
where 
\begin{equation}
\delta_{\rho L}=\frac{L}{2v_L}-\frac{L}{2v_\rho}\,, \qquad \delta_{\rho R}=\frac{L}{2v_R}-\frac{L}{2v_\rho}-\frac{L}{v_\sigma}.
\end{equation}
Further, the current $j(x, \omega)$ in the right lead ($x>L/2$) is given by
\begin{eqnarray}
\hat{j}(x, \omega)&=&\frac{1}{2\pi}\frac{i\omega {\cal T}^2 {\cal R}e^{i(x/v_R+\delta_{RR})\omega}}{1-{\cal R}^{2}e^{2i\omega\tau}}\Phi_R(0, \omega)
+\frac{1}{2\pi}
\frac{i\omega {\cal R}^2 e^{i(x/v_R+\delta_{RL})\omega}(e^{2i\omega\tau}-1)}
{1-{\cal R}^{2}e^{i2\omega\tau}}\Phi_L(0, \omega)+\nonumber\\&&+\,\frac{i\omega}{2\pi}e^{-ix\omega/v_L}\Phi_L(0, \omega) -\frac{{\cal T}^2}{2\pi}V_2 \,,
\label{current}
\end{eqnarray}
where
\begin{equation}
\delta_{RR}=\frac{L}{v_\sigma}-\frac{L}{v_R}\,,\qquad \delta_{RL}=-\frac{L}{2v_L}-\frac{L}{2v_R}.
\end{equation}
In the limit $x\gg  L$ the integral (\ref{Kubo}) is dominated by the pole $\omega^\prime=\omega$, with the last term in the current operator (\ref{current}) not contributing. Evaluating the integral in the limit $\omega\tau\ll 1$ we get
\begin{equation}
j=-\frac{1}{2\pi}{\cal T}^2 V_2-\frac{1}{2\pi}{\cal R}^2(V_2-V_1)\equiv \frac{1}{2\pi}\left(-V_2+\frac{1}{3}V_1\right).
\label{Eq:jRes1}
\end{equation}
Comparison of Eq. (\ref{Eq:jRes1}) to Eq. (\ref{e2.19}) shows that an edge in the stable $\theta=0$ fixed point is characterized by $G_{12}=0$.  

Let us now discuss the case of an edge with $\theta=\pi$. In terms of the modes $\tilde{\phi}^R$ and $\tilde{\phi}_L$ introduced in Eq. (\ref{Eq:sigmarhotoRtLt}), the current operator at $x> L/2$ reads [note the change of sign in front of the right-moving field compared to Eq.(\ref{currentBasic})]
\begin{equation}
\hat{j}=-\frac{1}{2\pi}v_L\partial_x \tilde\phi_L-\frac{{\cal R}v_R}{2\pi}\partial_x\tilde\phi_R-\frac{{\cal T}^2}{2\pi}V_2 \,.
\end{equation}
Taking into account the scattering matrix at the edge of the right lead  (see Fig. \ref{Fig:ScatteringPi}), we find
\begin{equation}
\phi_\rho(y, \omega)=\frac{{\cal T}e^{-i(y/v_\rho+\delta_{\rho L})\omega}}{1+{\cal R}^2e^{2i\omega\tau}}\Phi_{L}(0, \omega)-\frac{{\cal T}{\cal R}e^{-i(y/v_\rho+\delta_{\rho R})\omega}}{1+{\cal R}^2e^{2i\omega\tau}}\Phi_R(0, \omega)
\end{equation}
and
\begin{eqnarray}
\hat{j}(x, \omega) &=& -\frac{1}{2\pi}\frac{i\omega {\cal T}^2 {\cal R}e^{i(x/v_R+\delta_{RR})\omega}}{1+{\cal R}^{2}e^{2i\omega\tau}}\Phi_R(0, \omega)
-\frac{1}{2\pi}
\frac{i\omega {\cal R}^2 e^{i(x/v_R+\delta_{RL})\omega}(1+e^{2i\omega\tau})}
{1+{\cal R}^{2}e^{i2\omega\tau}}\Phi_L(0, \omega)
\nonumber \\
&+& \frac{i\omega}{2\pi}e^{-ix\omega/v_L}\Phi_L(0, \omega) -\frac{{\cal T}^2}{2\pi}V_2 \,.
\label{current1}
\end{eqnarray}

Evaluation of the current according to Eq.~(\ref{Kubo}) gives now
\begin{equation}
j=-\frac{1}{2\pi}{\cal T}^2 V_2+\frac{1}{2\pi}\frac{{\cal T}^2 {\cal R}^2}{1+{\cal R}^2}(V_2-V_1)=-\frac{1}{2\pi}\frac{{\cal T}^2}{1+{\cal R}^2} V_2-\frac{1}{2\pi}\frac{{\cal T}^2{\cal R}^2}{1+{\cal R}^2} V_1\equiv \frac{1}{2\pi}\left(-\frac12 V_2-\frac{1}{6}V_1\right) \,.
\label{Eq:jRes2}
\end{equation}
Equation (\ref{Eq:jRes2}) shows that an edge with $\theta=\pi$ is characterized by $G_{12}=1/2$.

\begin{figure}
\centering
\includegraphics[width=180pt]{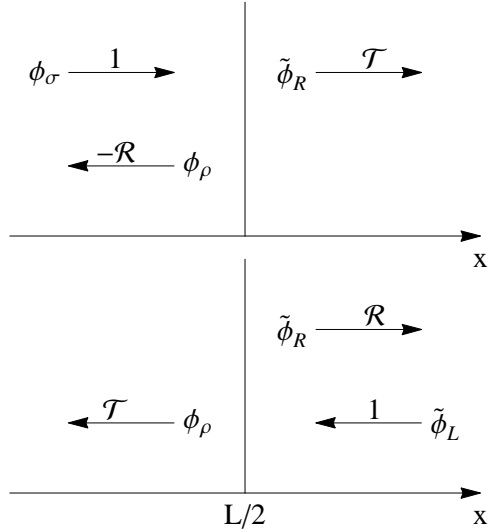}
\caption{Scattering at the boundary with the right lead for the fixed point $\theta=\pi$, see Eq. (\ref{Eq:sigmarhotoRtLt}).}
\label{Fig:ScatteringPi}
\end{figure}

\section{Length scales}
\label{App:Length-scales}

For completeness, we present in this Appendix (largely following Ref.~\cite{kfp}) the derivation of disorder-induced length scales  that determine borders of various transport regimes. 
We consider separately the cases of strong ($\Delta<3/2$) and weak ($\Delta<3/2$) interaction between the 1 and 1/3 modes.

\subsection{Strong interaction ($\Delta<3/2$)}

If the bare interaction is sufficiently strong, $\Delta_0 < 3/2$, the disorder is RG-relevant and the theory flows towards a disordered fixed point with $\Delta = 1$. Let $W_0 \ll 1$ be the bare dimensionless strength of disorder (i.e.,  the bare disorder-induced scattering rate $1/\tau_0$ in units of the ultraviolet energy cutoff).

For simplicity, let us exclude the case when the bare value $\Delta_0$ is in a close vicinity of $3/2$. Then, the renormalization of interaction is not particularly important at the initial state of RG, and the disorder increases according to 
\be
\label{e-length1}
\frac{dW}{d \cal L} = (3 - 2\Delta_0) W \,, \qquad {\cal L} = \ln (L/a) \,,
\ee
i.e., $W(L) \sim W_0 (L/a)^{3-2\Delta_0}$.
Thus, the disorder becomes strong $W\sim 1$ at the length scale $\ell$ determined by the condition $W_0 (\ell / a)^{3-2\Delta_0} \sim 1$, where $a$ is the ultraviolet length cutoff. This yields
\be
\label{e-length2}
\ell \sim a W_0^{-1/(3-2\Delta_0)}.
\ee
The physical meaning of $\ell$ is as follows. At distances $L$ smaller than $\ell$ the disorder is of no importance, and the system is described in terms of two decoupled bosonic modes determined by the bare interaction strength (characterized by the parameter $\Delta_0$). On the other hand, for $L \gg \ell$, these modes are strongly mixed by disorder, and the RG flow drives the system towards $\Delta=1$ \cite{kfp}. 

Exactly at $\Delta=1$ the eigenmodes of the system are the 2/3 and the neutral modes, which are completely decoupled, and thus do not equilibrate. 
On the other hand, when the temperature is nonzero (and the initial value $\Delta_0$ is different from unity), the RG flow towards $\Delta=1$ is stopped by temperature in the vicinity of the point $\Delta=1$. Thus, the eigenmodes remain weakly coupled by disorder, which leads to an inelastic scattering, establishing the equilibration between them  at a length scale $L_{\rm in}(T)$. 
The derivation of the length $L_{\rm in}$ in this regime that is presented below closely follows Ref.~\cite{kfp}. The coupling between the neutral and the 2/3 modes is described by a perturbation of the action of the $\Delta=1$ theory:
\be
\label{e-length3}
S_{\rm pert} \sim V_0^{\rm in} \int dx d\tau \partial_x\phi_\rho \cos(\sqrt{2}\phi_\sigma) u(x) \,,
\ee
where $u(x) = {\rm tr}U(x)\sigma_z U^{-1(x)}$ originates from the random SU(2) rotation $U(x)$ that is needed to gauge out the disorder,   see Sec.~\ref{s4}. The initial value of the coupling $V_0^{\rm in}$ (at the scale $\sim \ell$ which serves as an ultraviolet starting point for the strong-disorder stage of the RG ) is 
\be
\label{e-length4}
V_0^{\rm in} \simeq \sqrt{\frac{\Delta_0-1}{2}} (v_\rho + v_\sigma).
\ee
The initial value of the dimensionless scattering rate determined by this term is $W_0^{\rm in} \sim \Delta_0 - 1$. This means that at high temperature (such that the associated $L_T$ is of order $\ell$) the inelastic relaxation length due to the term (\ref{e-length3}) is of the order of $\ell/(\Delta_0 - 1)$. 
The RG equation for $W_{\rm in}$ is determined by the scaling dimension $\delta=2$ of the operator in Eq.~(\ref{e-length4}) at the attractive fixed point $\Delta = 1$, yielding, after disorder averaging,
\be
\label{e-length5}
\frac{dW^{\rm in}}{d \cal L} = (3 - 2\delta) W \equiv - W \,.
\ee
Thus, we get for scales $L$ larger than $\ell$ the renormalized value
\be
\label{e-length6}
W^{\rm in} (L) \sim W_0^{\rm in} \frac{\ell}{L}.
\ee
This renormalization stops at the scale $L_T \sim v_\sigma/T$, where the dimensionless strength of the perturbation is $W^{\rm in} (L_T) \sim W_0^{\rm in} \ell / L_T \ll 1$.   For scales $L$ larger than $L_T$, the renormalization becomes trivial, 
\be 
\label{e-length7}
W^{\rm in} (L) \sim (L/L_T) W^{\rm in} (L_T) \,, \qquad L > L_T \,.
\ee 
The length scale at which the renormalized value $W^{\rm in} (L)$ reaches unity is the inelastic scattering (equilibration) length $L_{\rm in} (T)$, i.e.,
$W^{\rm in} (L_{\rm in} (T)) \sim 1$.  Using Eqs.~(\ref{e-length7}) and (\ref{e-length8}), we get
\be 
\label{e-length8}
L_{\rm in}(T) \sim \frac{L_T^2}{\ell W_0^{\rm in}} \sim \frac{1}{\Delta_0-1} \frac{1}{\ell} \left( \frac{v_\sigma}{T} \right)^2 \,,
\ee
as was found in Ref.~\cite{kfp}. 

{
As is clear from the above analysis, the result (\ref{e-length8}) holds when the temperature is sufficiently low, so that $L_T \gtrsim \ell$. In the opposite case of higher temperatures, $L_T \ll \ell$, the renormalization (\ref{e-length1}) is stopped at the scale $L_T$ (rather than at the scale $\ell$), 
\be
\label{e-length8a}
W(L_T) \sim W_0 \left( \frac{L_T}{a}\right)^{3-2\Delta_0}  = \left( \frac{L_T}{\ell}\right)^{3-2\Delta_0}\,.
\ee
This yields the initial value of the dimensionless inelastic scattering rate $W_0^{\rm in}$ at scale $L_T$:
\be
\label{e-length8b}
W_0^{\rm in} \sim (\Delta_0-1 )\left( \frac{L_T}{\ell}\right)^{3-2\Delta_0}\,,
\ee
and thus, in full analogy with Eq.~(\ref{e-length7}),
\be
\label{e-length8c}
W^{\rm in} (L) \sim \frac{L}{L_T} W^{\rm in} (L_T) \sim (\Delta_0-1 )\left( \frac{L_T}{\ell}\right)^{3-2\Delta_0} \frac{L}{L_T}\,, \qquad L > L_T\,.
\ee
Equating this expression for $W^{\rm in} (L)$ to unity, we get
the equilibration length
\be
\label{e-length8d}
L_{\rm in}(T) \sim \frac{\ell}{\Delta_0-1} \left( \frac{\ell}{L_T}\right)^{2-2\Delta_0} \,.
\ee
Clearly, the expressions (\ref{e-length8}) and (\ref{e-length8d}) match when $L_T \sim \ell$. }

\subsection{Weak interaction ($\Delta < 3/2$)}

For weak interaction, $\Delta_0 < 3/2$, the RG flow  (\ref{e-length1}) corresponds to weakening of disorder with increasing length scale. This renormalization stops at the thermal length $L_T \sim v_{1/3} / T$, where the disorder is equal to
\be
\label{e-length9}
W(L_T) \sim W_0 \left( \frac{L_T}{a}\right)^{3-2\Delta} \,.
\ee
The renormalization of interaction is of no particular importance, $\Delta \simeq \Delta_0$.
For larger scales $W$ increases linearly with $L$, in full analogy with Eq.~(\ref{e-length7}). The scale at which it reaches unity is the equilibration length $L_{\rm in}(T)$, which is thus given by
\be
L_{\rm in}(T) \sim W_0^{-1} a  \left( \frac{L_T}{a} \right)^{2\Delta-2}\sim W_0^{-1} a  \left(\frac{v_{1/3}}{a T} \right)^{2\Delta-2} \,,
\label{e-length10}
\ee
in agreement with Ref.~\cite{kfp}.

\end{document}